\documentclass[
 amsmath,amssymb,
 aps,
prb,
floatfix,
]{revtex4-2}

\usepackage{multirow}
\usepackage{graphicx}

\begin{document}

\title{Strictly Localized States on the Socolar Dodecagonal Lattice}
\author{M. Akif Keskiner}
\email{akif.keskiner@bilkent.edu.tr}
\author{M.\"O. Oktel}
\email{oktel@bilkent.edu.tr}
\affiliation{Department of Physics, Bilkent University, Ankara, 06800, TURKEY}

\date{\today}

\begin{abstract}
Socolar dodecagonal lattice is a quasicrystal closely related to the better-known Ammann-Beenker and Penrose lattices. The cut and project method generates this twelve-fold rotationally symmetric lattice from the six-dimensional simple cubic lattice. We consider the vertex tight-binding model on this lattice and use the acceptance domains of the vertices in perpendicular space to count the frequency of strictly localized states. We numerically find that  these states span $f_{\mathrm{Num}}\simeq 7.61$ \% of the Hilbert space. We give 18 independent localized state types and calculate their frequencies. These localized state types provide a lower bound of $f_{\mathrm{LS}} =\frac{10919-6304\sqrt{3}}{2} \simeq 0.075854$, accounting for more than $99 \%$ of the zero-energy manifold. Numerical evidence points to larger localized state types with smaller frequencies, similar to the Ammann-Beenker lattice. On the other hand, we find sites forbidden by local connectivity to host localized states. Forbidden sites do not exist for the Ammann-Beenker lattice but are common in the Penrose lattice. We find a lower bound of $f_{\mathrm{Forbid}}\simeq 0.038955$ for the frequency of forbidden sites. Finally, all the localized state types we find can be chosen to have constant density and alternating signs over their support, another feature shared with the Ammann-Beenker lattice. 
\end{abstract}

\maketitle

\section{Introduction}

Quasicrystals are formed by non-periodic yet highly symmetric arrangements of atoms. While their initial synthesis \cite{she84} was done by carefully controlled alloying, they have since been observed to form through natural processes \cite{bin09}. More recently,  defect-free and highly controllable synthetic systems with quasicrystalline order have been demonstrated. These meta-quasicrystals in photonic systems\cite{var13}, polaritons\cite{tan14}, cold-atoms \cite{vie19,sin15} and synthetic surfaces \cite{col17} promise direct access to elementary excitations of quasicrystals which goes beyond the structural tools such as x-ray scattering \cite{jan97}. 

The high degree of symmetry in quasicrystals makes it possible to understand structural properties in great detail. For example, the x-ray diffraction images of quasicrystals are now well characterized in terms of the quasicrystal lattice and the unit cell decorations \cite{lev86,soc86}. However, this symmetry does not yield a  tool like Bloch's theorem to constrain the excitation spectrum. Except for some examples in one dimension \cite{ost83,koh83,jag21}, the excitation spectrum of  quasicrystalline systems is not well understood. Eigenstates not only contain the extended and localized state possibilities of disordered systems but can also be critically self-similar \cite{sut86}. The energy spectrum can be singularly continuous with multifractal properties \cite{pie95,mjk17}.  
The connection between the energy spectrum and the symmetries of the quasicrystal is not obvious. In the absence of a general theory, large-scale numerical calculations \cite{rie98,tsu91,zja00} or in depth-understanding of specific eigenstates \cite{sut86,rpg98,kka14} are valuable.

A common occurrence in the quasicrystal spectrum is the existence of strictly localized states (LS) \cite{ara88}. LS are identically zero beyond a finite domain and are also called confined states or compact localized states. LS were first identified in numerical diagonalization of tight-binding models of the Penrose lattice \cite{koh86,oda86,cho85}. Local destructive interference, similar to Aharonov-Bohm cages \cite{vid98}, confines the LS wavefunction to a finite region. LS have zero energy for bipartite lattices \cite{sutbip86}, and their properties have been studied for two well-known models, the Penrose (PL) \cite{ara88,kts17,mok20} and Ammann-Beenker lattices (ABL) \cite{kog20,okt21}. In both cases, almost ten percent of the Hilbert space is spanned by the LS. However, there are significant differences between the LS in the two models. In the ABL, all sites host LS,  but a substantial portion of the sites in the PL are forbidden by local connectivity to be in the support of an LS. The LS in the PL are formed by only six LS types, while the ABL seems to have an infinite number of LS types. While the choice for LS type expansion is not unique, the known ABL LS types can all be chosen to be rotationally symmetric around an eight-fold symmetric vertex. Furthermore, all LS type wavefunctions in the ABL can be selected to be of constant density. The interference leading to localization is provided only by the fluctuating sign of the wave function. There is no choice for LS types in the PL with constant density. Finally, the LS in the PL show remarkable robustness \cite{day20}. A uniform magnetic field applied to the system leaves the total LS fraction in the PL invariant, while the LS fraction in the ABL quickly decreases with the applied field. 

These striking differences between the two commonly used models of quasiperiodic order make it worthwhile to explore the LS in other quasicrystal systems. It is essential to ask if the PL or the ABL display a more generic behavior and which properties in their definition lead to such qualitative differences. Another quasicrystal lattice closely related to the PL and ABL is the Socolar dodecagonal lattice (SDL) \cite{soc89}. SDL can be constructed by projection from a six-dimensional simple cubic lattice, while PL and ABL are projected from 5 and 4-dimensional simple cubic lattices. All three lattices have scaling symmetry with simple inflation-deflation rules, can be constructed with local matching rules, and admit simple decorations which line up to form a quasiperiodic grid called an Ammann pattern. 

A recent paper considered the vertex tight-binding model on the SDL \cite{kog21}. The LS fraction was numerically calculated as $0.076$, and eleven LS types were identified. However, the contributions of the LS types were significantly short of the numerical result. In this paper, we consider the same model but use the perpendicular space accounting method \cite{mok20,okt21} to count the frequencies of the LS types. Our numerical result for the total LS fraction agrees with Ref.\cite{kog21}. However, we find that the LS type frequencies significantly differ from what was reported. We calculate the contribution of the  LS types reported in  Ref.\cite{kog21} as $f_{\mathrm{LS}} \simeq 0.0719$. We identify seven more LS types bringing the total covered to $ f_{\mathrm{LS}}=\frac{10919-6304\sqrt{3}}{2}
\simeq 0.0758$. We show the perpendicular space images for the support of all LS types and prove their independence. We conjecture that there are LS types of lower frequencies related to the types we identified and the total frequency sum gives $f_{\mathrm{Conj}}=\frac{-1089+629\sqrt{3}}{6}\simeq0.076660$. We inspect the local density of states (LDOS) formed by the LS and find forbidden sites where the local density of states for LS is zero. We give four arguments on the lattice which lead to forbidden sites. These arguments show that at least  $3.8955 \%$ of all sites are forbidden. The calculation of LDOS in perpendicular space suggests that there may be more forbidden sites, but we have not found real space arguments to prove that they cannot host LS. While the presence of forbidden sites is similar to the PL, the existence of large LS types with low frequencies makes the spectrum more similar to the ABL. All LS types have wavefunctions that can be chosen to have a constant density similar to the ABL. We also find that the LS fraction decreases quickly with an applied magnetic field. Overall our results indicate that the properties of the ABL zero energy manifold may be more commonly observed in quasicrystals, and the PL may have other unidentified properties which restrict the behavior of its LS.  

The paper is organized as follows: We introduce the SDL giving its projective definition in the next section. The following section \ref{sec:Numerical} details our numerical method and results for the local density of states and total LS fraction. We identify LS types and discuss their properties in section \ref{sec:LS types} and section \ref{sec:Forbidden} details the forbidden sites arguments. Finally, we compare our results for SDL to ABL and PL cases and discuss outstanding questions in \ref{sec:Conclusion}.

\section{Cut and project definition of the Socolar dodecagonal lattice}
\label{sec:Model}

\begin{figure}[!htb]
    \centering
    \includegraphics[trim=8mm 8mm 8mm 8mm,clip,width=0.48\textwidth]{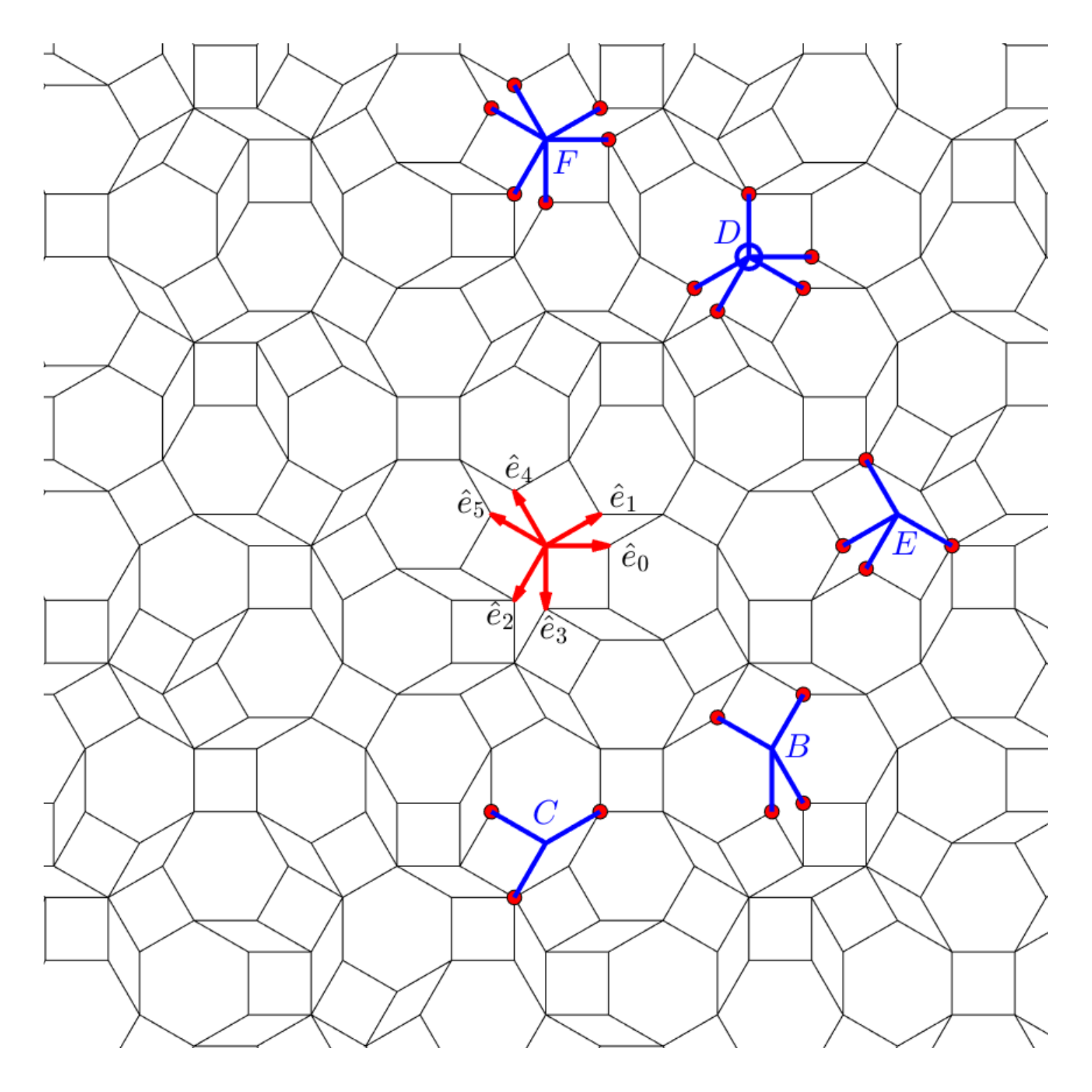}
    \caption{Socolar dodecagonal lattice has three kinds of tiles, a square, a hexagon, and a rhombus with $\pi/6$ angle. All bonds are parallel to the six star vectors $\hat{e}_m$ shown in red. Each vertex can be classified by its connections to the nearest neighbors. There are only five vertex types B, C, D, E, and F. One example of each is highlighted in the figure.}
    \label{fig:Fig01_RealSpaceLattice}
\end{figure}

Socolar constructed the SDL as an analog of the PL with twelve-fold rotational symmetry  \cite{soc89}. The same construction method applied to eight-fold symmetry gives the ABL. Thus, these three lattices share remarkable properties beyond quasiperiodic order. All three are self-similar under simple deflation rules, and simple decorations of tiles generate matching rules for the lattice construction. Following the work of De Bruijn \cite{bru81,bru81b}, all three lattices can be constructed as duals of grids made of equally spaced lines. Grids of four-fold symmetry generate the ABL, five-fold symmetry generates the PL and six-fold symmetric "hexagrids" generate the SDL. Here the most important distinction for the SDL appears as the hexagrid has to be chosen so that three lines that make $2 \pi/3$ angles with each other always meet at a single point. The triple intersection points give rise to a third tile shape, in contrast to just two for the ABL and PL. The SDL defines a quasiperiodic tiling of regular hexagons, squares, and $\pi/6$ angle rhombuses. (See Fig.\ref{fig:Fig01_RealSpaceLattice}).       

SDL can be generated using the dual grid method, the deflation method, or the quasiperiodically spaced Ammann decoration method. However, as we use perpendicular space properties
to explore the LS, we first repeat the definition through the cut-project approach. We start with a six-dimensional space($\mathbb{R}^{6}$) filled with a grid of unit cubes. We can define any interior point of each cube indexed with six integers $k_{0},...,k_{5}$ in terms of the orthogonal unit vectors $\hat{u}_{n}=(\delta_{0n},...,\delta_{5n})$ as 
\begin{equation}
    \vec{x}=\sum_{n=0}^{5}x_{n}\hat{u}_{n}
\end{equation}
with $k_{n}-1<x_{n}<k_{n}$ for $n=0,...,5$. This 6-dimensional space  can be also spanned by the following orthogonal vectors defined in terms of the complex number $\zeta=e^{\frac{i\pi}{6}}$:
\begin{equation} \label{eq2}
\begin{split}
\vec{a}_{1} & = \sum_{n=0,1,4,5}\Re({\zeta}^{n})\hat{u}_{n}-\sum_{n=2,3}\Re({\zeta}^{n})\hat{u}_{n}, \\
 \vec{a}_{2} & = \sum_{n=0,1,4,5}\Im({\zeta}^{n})\hat{u}_{n}-\sum_{n=2,3}\Im({\zeta}^{n})\hat{u}_{n}, \\
 \vec{a}_{3} & = \sum_{n=0,3,4}\Re({\zeta}^{-n})\hat{u}_{n}-\sum_{n=1,2,5}\Re({\zeta}^{-n})\hat{u}_{n}, \\
 \vec{a}_{4} & = \sum_{n=0,3,4}\Im({\zeta}^{-n})\hat{u}_{n}-\sum_{n=1,2,5}\Im({\zeta}^{-n})\hat{u}_{n}, \\
 \vec{a}_{5} & = (1,0,1,0,1,0) = \hat{u}_0+\hat{u}_2+\hat{u}_4,\\
 \vec{a}_{6} & = (0,1,0,1,0,1) = \hat{u}_1+\hat{u}_3+\hat{u}_5,
\end{split}
\end{equation}
with $\vec{a}_{p}.\vec{a}_{q}=3\delta_{pq}$. We also define $\xi=2+\sqrt{3}.$ An intercept vector $\vec{\gamma}=\sum_{n=0}^{5}\gamma_{n}\hat{u}_{n}$, is chosen so that $\gamma_{0}+\gamma_{2}+\gamma_{4}=\gamma_{1}+\gamma_{3}+\gamma_{5}=0$. The vertices of the SDL are given by $\vec{k}.\vec{a}_{1}\hat{x}+\vec{k}.\vec{a}_{2}\hat{y}$, where $\vec{k}=(k_{0},...,k_{5})\in\mathbb{Z}^{6}$, whose open cube has a non-empty intersection with the plane described by the following equations
\begin{equation} \label{eq3}
\begin{split}
(\vec{x}-\vec{\gamma}).\vec{a}_{3} & =0,\\ 
(\vec{x}-\vec{\gamma}).\vec{a}_{4} & =0,\\
(\vec{x}-\vec{\gamma}).\vec{a}_{5} & =0,\\
(\vec{x}-\vec{\gamma}).\vec{a}_{6} & =0.
\end{split}
\end{equation}

\begin{figure}[!htb]
    \centering
    \includegraphics[trim=8mm 8mm 8mm 8mm,clip,width=0.34\textwidth]{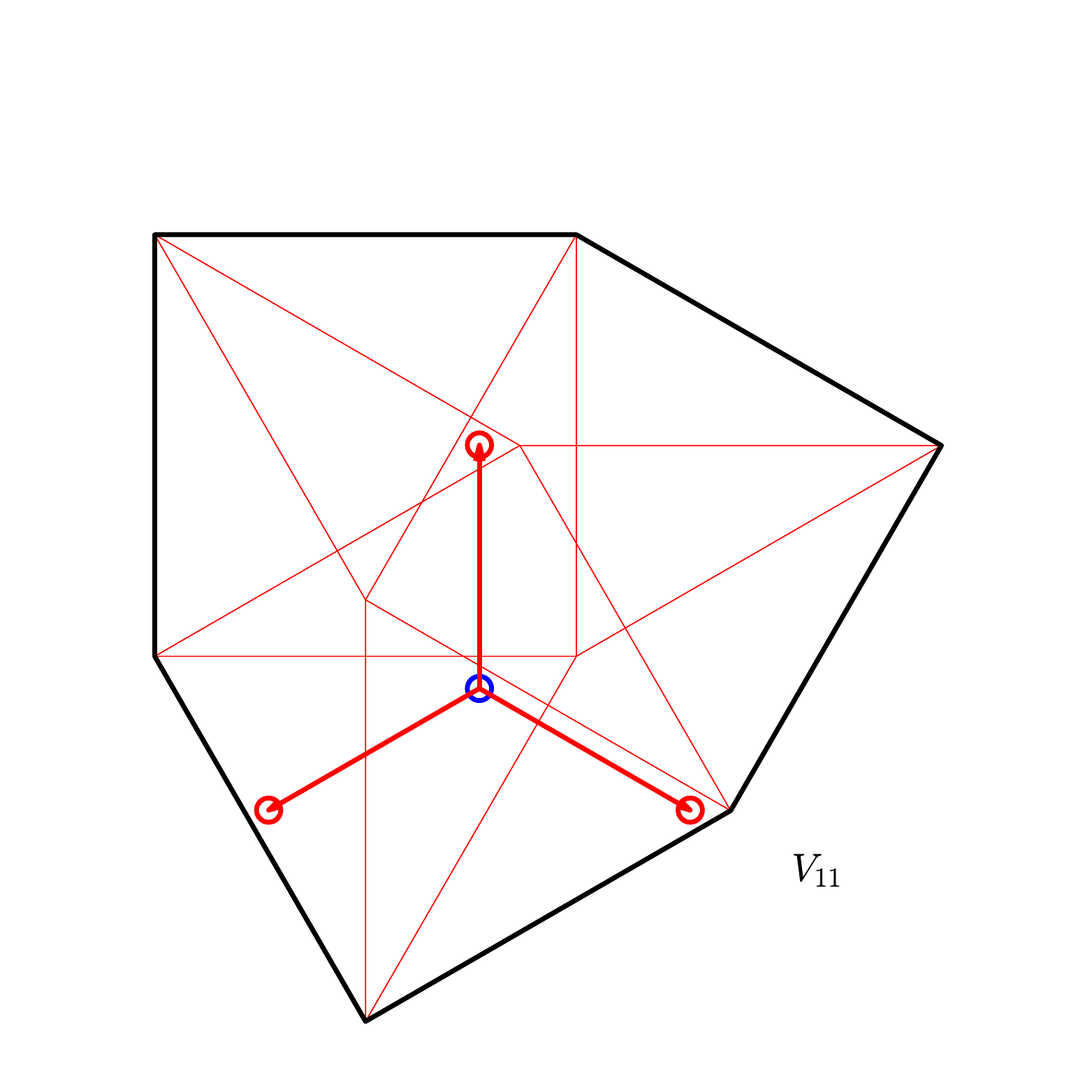}
    \includegraphics[trim=8mm 8mm 8mm 8mm,clip,width=0.34\textwidth]{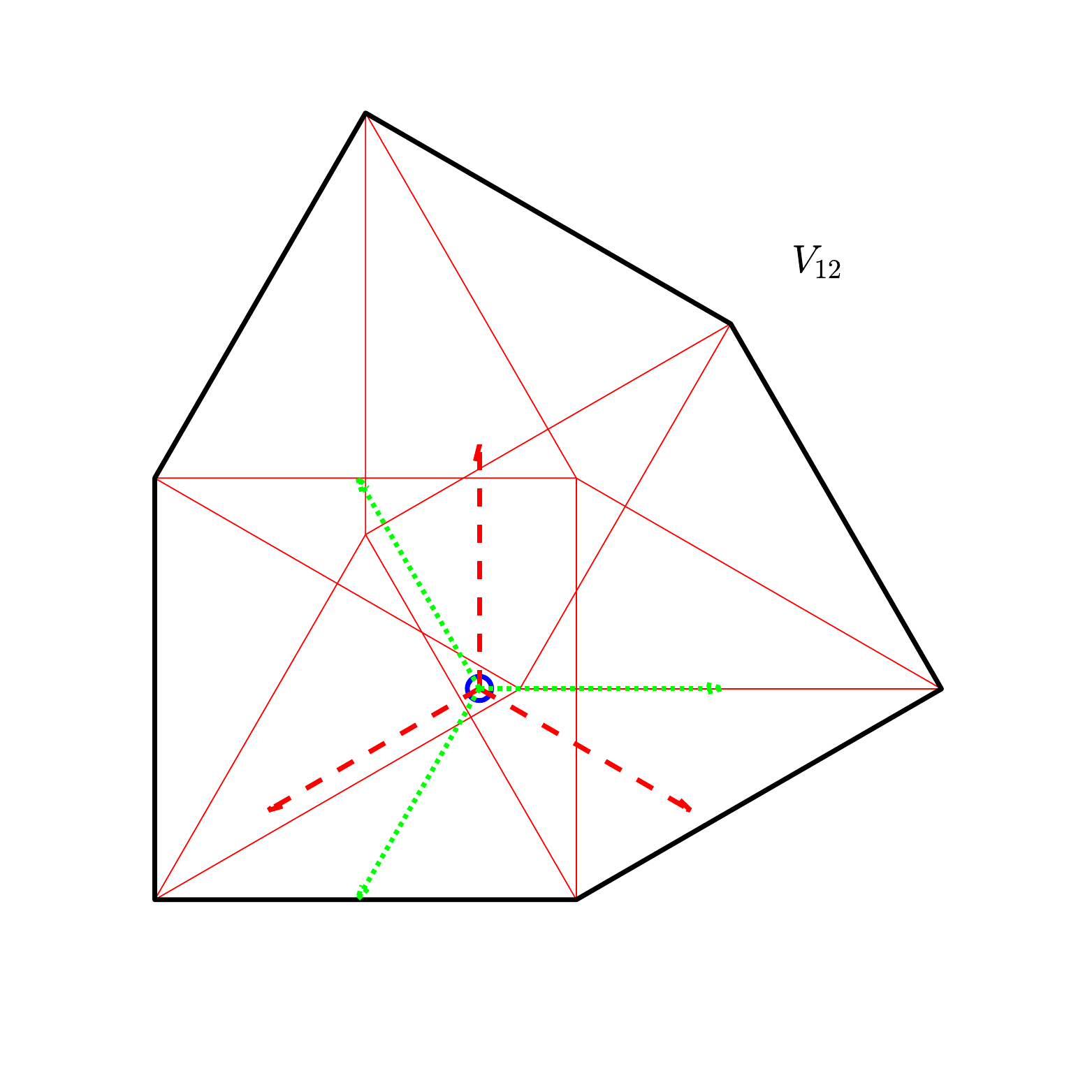}\\
    \includegraphics[trim=8mm 8mm 8mm 8mm,clip,width=0.34\textwidth]{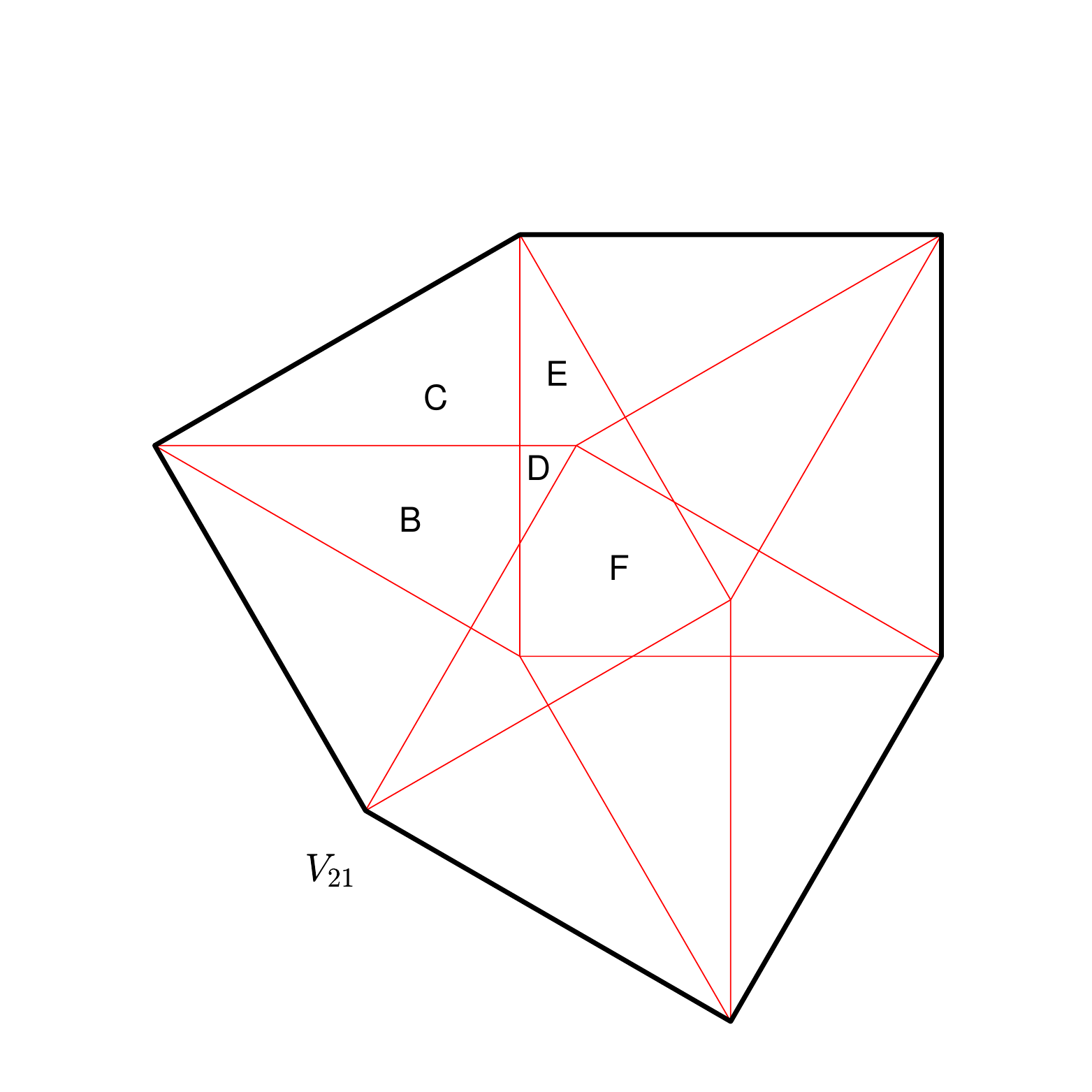}
    \includegraphics[trim=8mm 8mm 8mm 8mm,clip,width=0.34\textwidth]{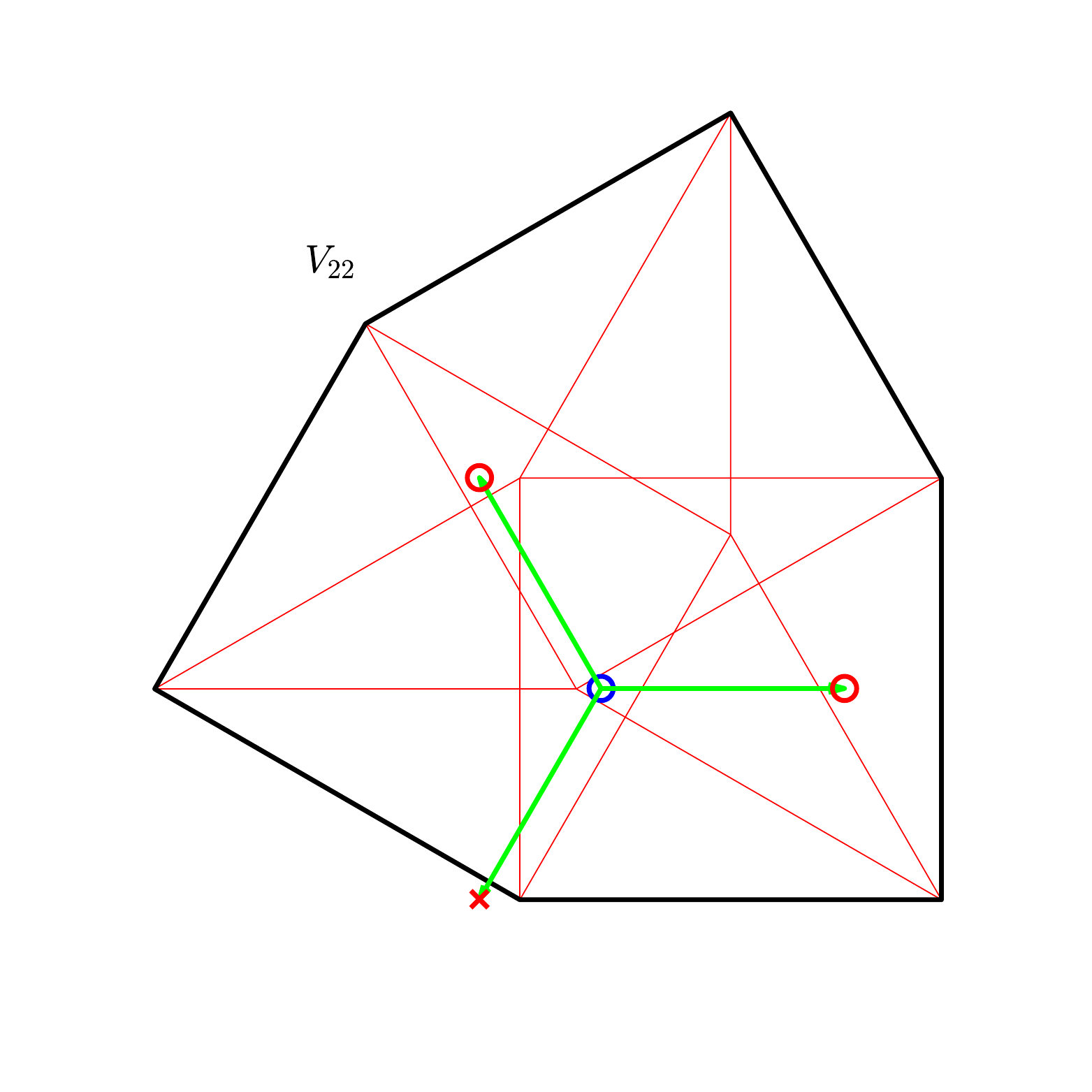}
       
    \caption{The perpendicular space of the SDL consists of four hexagons, $V_{ij}$. $V_{11}, V_{22}$ form the even sublattice, and sites in them can only have a bond connecting to the odd sublattice formed by $V_{12}, V_{21}.$ The D vertex highlighted in Fig.\ref{fig:Fig01_RealSpaceLattice} has the perpendicular space space image shown in $V_{12}$, the six possible bond directions taking this point to $V_{11},V_{22}$ are shown. Only five of six remain inside their respective hexagons, making this point a D vertex. $V_{21}$ is marked with regions belonging to vertex types. Unmarked regions can be deduced by symmetry. }
    \label{fig:Fig02_PerpSpaceLattice}
\end{figure}

If a point in the six-dimensional lattice $\vec{R}=\sum_{n=0}^5k_{n}\hat{u}_{n}$, has an open unit cube satisfying Eq.\ref{eq3}, we can express it in terms of the $a$-vectors as
\begin{equation} \label{eq4}
\begin{split}
\vec{R} & =\frac{1}{3}\big\{(\vec{k}.\vec{a}_{1})\vec{a}_{1}+(\vec{k}.\vec{a}_{2})\vec{a}_{2}+(\vec{k}.\vec{a}_{3})\vec{a}_{3}+(\vec{k}.\vec{a}_{4})\vec{a}_{4}+(\vec{k}.\vec{a}_{5})\vec{a}_{5}+(\vec{k}.\vec{a}_{6})\vec{a}_{6}\big\}\\ 
 & =\frac{1}{3}\big\{x^{\parallel}\vec{a}_{1}+y^{\parallel}\vec{a}_{2}+\tilde{x}\vec{a}_{3}+\tilde{y}\vec{a}_{4}+x^{Li}\vec{a}_{5}+y^{Li}\vec{a}_{6}\big\}.
\end{split}
\end{equation}
The point $(x^{\parallel},y^{\parallel})$ is the real space projection of $\vec{R}$, which is defined in 2-dimensional physical space, and the point $(\tilde{x},\tilde{y},x^{Li},y^{Li})$ is the perpendicular space projection of $\vec{R}$, which is defined in 4-dimensional perpendicular space. Nevertheless, $\vec{R}.\vec{a}_{5}=k_{0}+k_{2}+k_{4}=i \in \mathbb{Z}$ and $\vec{R}.\vec{a}_{6}=k_{1}+k_{3}+k_{5}=j \in \mathbb{Z}$. Both $i$ and $j$ can take only the values  1 or 2\cite{soc89}. Therefore, the perpendicular space projections of all lattice points do not fill the 4-dimensional perpendicular space, but fall on  four two-dimensional planes. Furthermore, the perpendicular space projections  lie only within four hexagons in these planes which we designate as $V_{ij}=\big\{ V_{11}, V_{12}, V_{21}, V_{22}\big\}$ (See Fig.\ref{fig:Fig02_PerpSpaceLattice}).

The projection from the six-dimensional space to the real space means that a positive move in the $n^{th}$ direction $k_n \rightarrow k_n+1$ changes real space position by a star vector $\hat{e}_n$. The star vectors are 
\begin{eqnarray}
\hat{e}_n&=&\cos(n \pi/6) \hat{i} + \sin(n \pi/6) \hat{j}, \: n=0,1,4,5 \\ \nonumber
\hat{e}_n&=&-\cos(n \pi/6) \hat{i} -\sin(n \pi/6) \hat{j},\: n=2,3.
\end{eqnarray}
as shown in Fig.\ref{fig:Fig01_RealSpaceLattice}. The same move is also reflected in perpendicular space, however through a different set of vectors $\hat{\tilde{e}}_n$ where the projection rules dictate
\begin{eqnarray}
\label{Eq:PerpSpaceStarVectors}
    \hat{\tilde{e}}_0=\hat{e}_0, \:  \hat{\tilde{e}}_2=\hat{e}_4, \:  \hat{\tilde{e}}_4=\hat{e}_2, \\ \nonumber
 \hat{\tilde{e}}_3=\hat{e}_3, \:  \hat{\tilde{e}}_1=\hat{e}_5, \:  \hat{\tilde{e}}_5=\hat{e}_1. 
\end{eqnarray}
The combination of perpendicular space allowed regions and star vectors makes it possible to describe the local environment of the lattice. Consider a point inside the perpendicular space hexagon $V_{12}$ as shown in Fig.\ref{fig:Fig02_PerpSpaceLattice}. Any vertex can, in principle, have $12$ bonds emanating from it in $\pm\hat{e}_m$ directions for $m=0,...,5$. However, if a point has a perpendicular space projection inside $V_{12}$, its first index $i$ can only change to $2$. Thus, vectors $+\hat{e}_0,+\hat{e}_2,+\hat{e}_4$ are allowed, while the $i$ index forbids their negatives. Similarly, the second index, $j$, allows for $-\hat{e}_1,-\hat{e}_3,-\hat{e}_5$, and forbids a positive move along these three directions. These allowed vectors are shown on the hexagon $V_{12}$ in Fig.\ref{fig:Fig02_PerpSpaceLattice}.  Correct index change is not enough for a move to be allowed. Also, the result of the move must lie in the new $V_{ij}$ hexagon. For the  point  shown in Fig.\ref{fig:Fig02_PerpSpaceLattice}, all three moves along  $-\hat{e}_1,-\hat{e}_3,-\hat{e}_5$ result in a point inside $V_{11}$, while only two out of the three moves along $+\hat{e}_0,+\hat{e}_2,+\hat{e}_4$ lie inside $V_{22}$. Thus, we conclude that this vertex is connected to five neighbors, with bond directions along $-\hat{e}_1,-\hat{e}_3,-\hat{e}_5,+\hat{e}_0,+\hat{e}_2$. Notice that the correspondence in Eq.\ref{Eq:PerpSpaceStarVectors} allows the real space star vector $\hat{e}_2$ rather than $\hat{e}_4$.

The perpendicular space construction makes it easy to deduce the local environment of any vertex if its perpendicular space position is known. Any vertex has  at most six and at least three bonds. Following Ref.\cite{soc89}, we classify vertices into five vertex types up to rotations and reflections. One example of each is highlighted in Fig.\ref{fig:Fig01_RealSpaceLattice}. Vertex types C, D, and F, have 3, 5, and 6 bonds, respectively. Two vertex types, B and E, have four bonds. As the number and orientation of the bonds are uniquely determined by the perpendicular space position of the vertex, it is possible to split the perpendicular space hexagons into smaller regions corresponding to each vertex type. This division is shown on $V_{21}$ in Fig.\ref{fig:Fig02_PerpSpaceLattice}, which can be extended to all unmarked regions and other hexagons by symmetry. As the projection from the six-dimensional lattice to the perpendicular space is linear, all the areas inside the hexagons are filled densely and uniformly with projected points. Consequently, area sizes in perpendicular space reflect the frequencies of local environments. 

Consider the central hexagon corresponding to the F vertices. The area of that small hexagon divided by the area of $V_{ij}$ gives the frequency of $F$ vertices as $f_{F}=7-4\sqrt{3}\simeq0.0718$. Similarly other vertex types have frequencies $f_B=\frac{3}{4}(9-5\sqrt{3})\simeq0.255$, $f_C=\frac{3}{4}(\sqrt{3}-1)\simeq 0.549$, $f_D=\frac{3}{4}(11\sqrt{3}-19)\simeq0.0394$, and $f_E=\frac{1}{4}(9-5\sqrt{3})\simeq0.0849$ . The perpendicular space area method for calculating local environment frequencies can be extended beyond the first neighbors. We use this method to count the LS types in the following sections.

The perpendicular space picture is also helpful in describing the symmetries of the SDL. First, notice that points in $V_{11}$ and $V_{22}$ are only connected to points in $V_{12}, V_{21}$ but not to each other. Thus the lattice is bipartite, composed of two sublattices. We refer to the collection of points in $V_{11}, V_{22}$ as the even sublattice and $V_{12}, V_{21}$ as the odd sublattice. Also, notice that the perpendicular space hexagon shapes can be obtained from each other by rotations. Under a counter-clockwise rotation of $\pi/6$ we have $V_{22}\rightarrow V_{12}\rightarrow V_{11} \rightarrow V_{21} \rightarrow V_{22}$. Any local environment has a thirty-degree rotated copy in the lattice. However, the sublattices are exchanged for the rotated copy. Similarly, any local environment must have a sixty-degree rotated copy with the same sublattice assignments. However, the two copies have exchanged $V_{11} \leftrightarrow  V_{22}$ and $V_{12} \leftrightarrow  V_{21}$.

We define a tight-binding model on the SDL as
\begin{equation}
\label{eq: Hamiltonian}
    {\cal H}=-\sum_{<ij>} |\vec{R}_i\rangle \langle \vec{R}_j |.
\end{equation}
where the sites $<i,j>$ are connected by a bond. We focus on the eigenstates that have zero density beyond a finite lattice region. Such LS appear at zero energy for bipartite lattices. If an energy $E$ is an eigenvalue, the bipartite property of the lattice ensures that $-E$  is also an eigenvalue. Similarly, changing the sign of the wave function on only one of the sublattices can be used to choose zero energy states to be confined to only one of the sublattices. In the following, we present calculations and examples for LS that are non-zero only on the odd sublattice of $V_{12}, V_{21}$, understanding that everything can be mapped to the even sublattice by a thirty-degree rotation.

\begin{figure}[!htb]
    \centering
    \includegraphics[clip,width=0.38\textwidth]{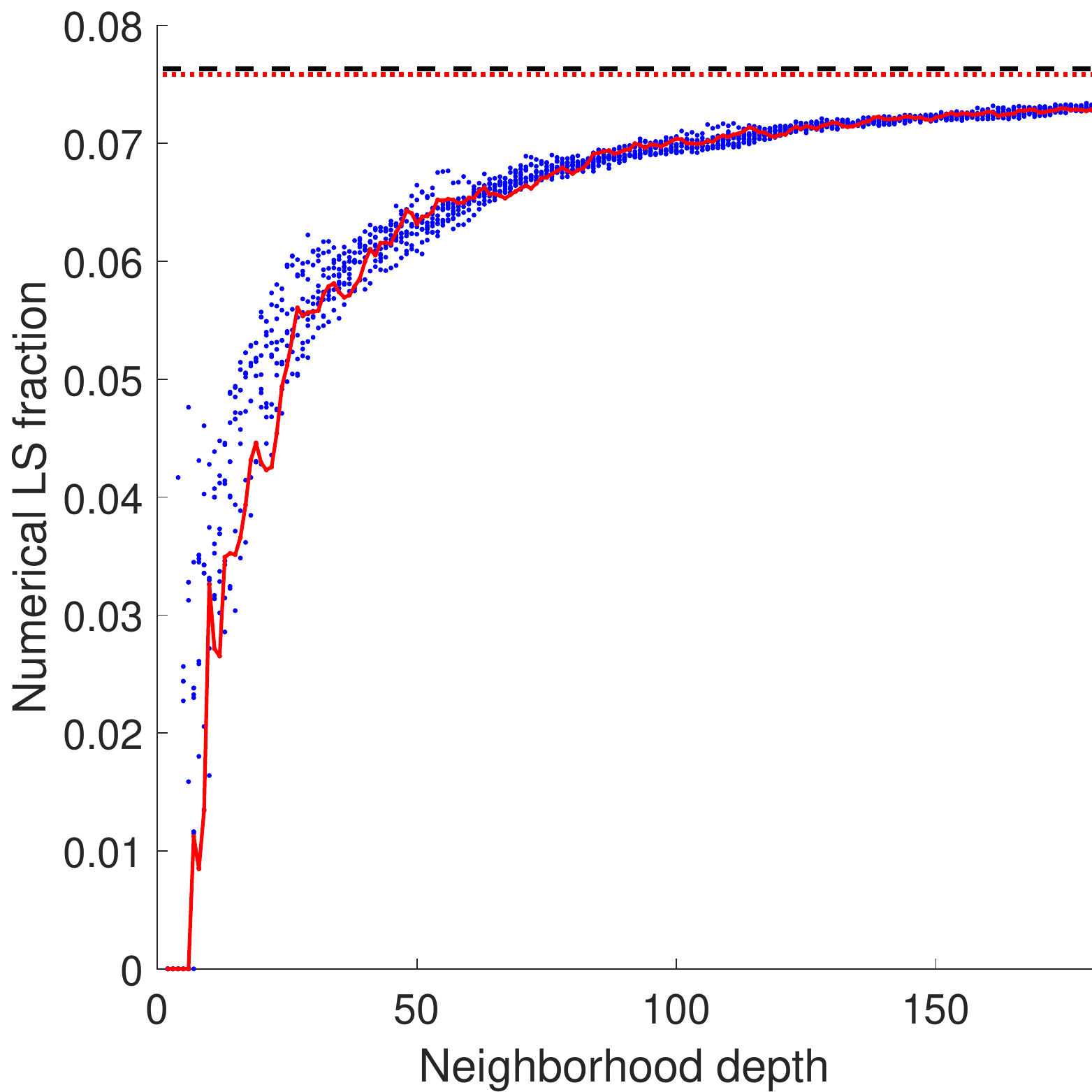}
     \includegraphics[clip,width=0.38\textwidth]{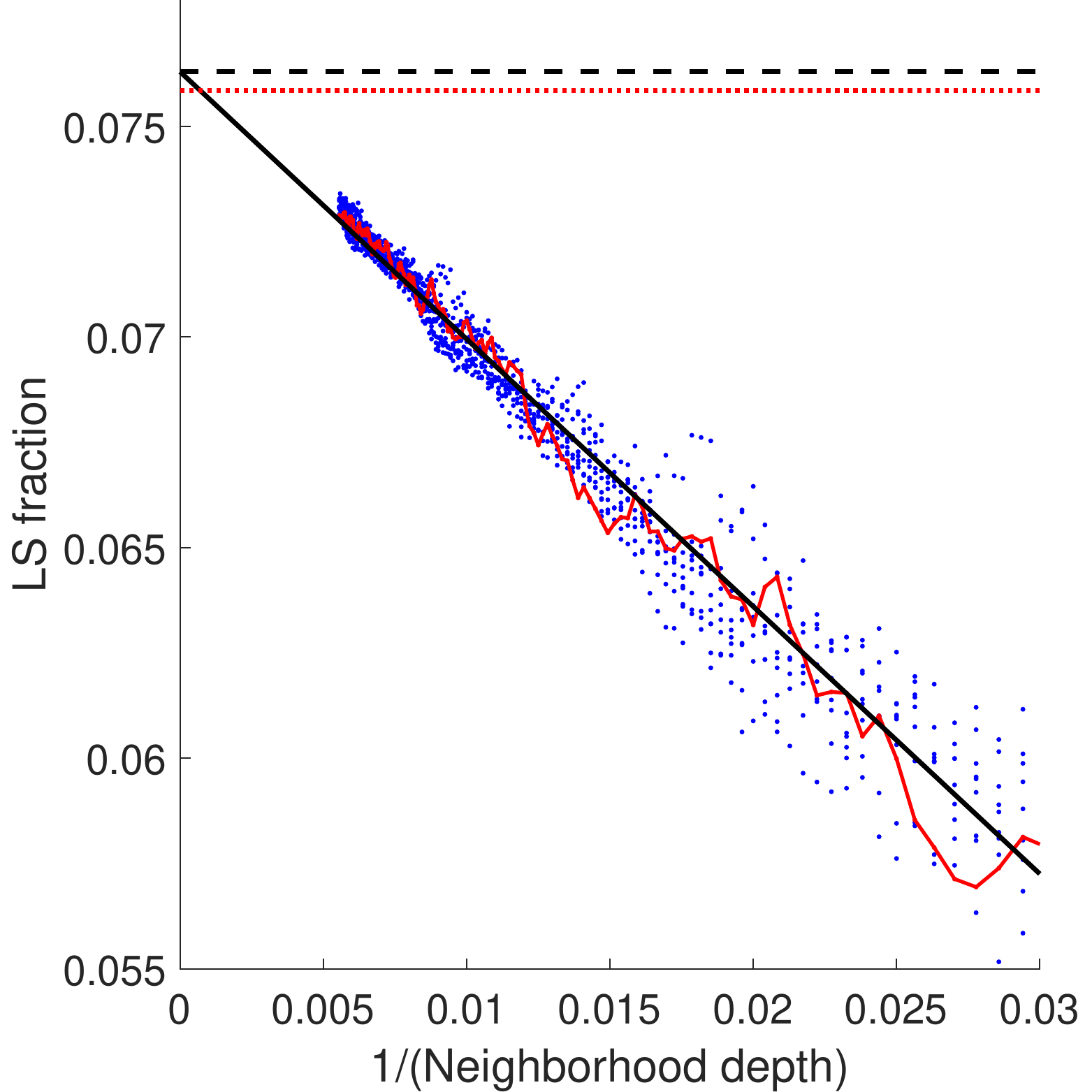}
    \caption{(a) Numerically calculated LS fraction as a function of neighborhood depth. We used ten randomly chosen initial points to generate the neighborhood, each providing one data point at each depth. For one initial point, data points are linked by the red line to show variation with depth. The Black dashed line is our best estimate for the infinite lattice's LS fraction, and the red dashed line is the lower bound provided by LS types. (b) The same data plotted as a function of inverse neighborhood depth, showing the linear fit used to estimate the infinite lattice LS fraction.}
    \label{fig:Fig03_NumericalLSFraction}
\end{figure}

\section{Numerical LS fraction and the zero energy local density of states}
\label{sec:Numerical}

We construct large SDL domains by using the algorithm presented in the previous section. We start by specifying the initial point's perpendicular space coordinates and calculating its nearest neighbors' real and perpendicular space positions. Repeating the same process on the nearest neighbors generates the second neighbors, and iteration can be repeated up to the desired neighborhood depth $D$. This process also records the Hamiltonian with open boundary conditions within this neighborhood. The most extensive domains we use have a neighborhood depth of $D=180$, containing approximately 66 000 sites.   

This construction method has the advantage that all the sites on the boundary belong to the same sublattice. If the initial site is in the even sublattice and the neighborhood depth $D$ is even, then all the sites on the boundary are in the even sublattice. Recall that LS can be chosen to lie only in one sublattice. Hence, if we find any LS in the odd sublattice in this finite domain, it is guaranteed that the same LS exists for the infinite SDL. There will be no extra LS generated due to the open boundary conditions. To be more specific, we can write the Hamiltonian in a block form by grouping the odd and even lattice sites
\begin{equation}
  \label{eq:MatrixC}
    {\cal H}=\begin{bmatrix}
0 & {\cal C} \\
{\cal C}^T & 0
\end{bmatrix},
\end{equation}
where ${\cal C}$ acts on only the odd sublattice sites.

We do not use matrix diagonalization to find the eigenvalues of the above Hamiltonian. Instead, we numerically find the dimension of the null space of the sparse matrix ${\cal C}$. Available QR decomposition routines efficiently handle this calculation. The size of the null space of the matrix ${\cal C}$ divided by the total number of sites in the odd sublattice gives us a numerical estimate $f_{\mathrm{Est}}(D)$ of the LS fraction \cite{okt21}.

In Fig.\ref{fig:Fig03_NumericalLSFraction}, we plot the numerically obtained LS fraction as a function of neighborhood depth D. We repeat the calculation for ten  randomly chosen initial points. While the LS fraction depends on the perpendicular space position of the initial point for small neighborhood depths, the results quickly converge to a narrow band for $D>50$. The numerical LS fraction increases with increasing neighborhood depth. Our method captures all the LS which lie entirely inside the boundary but misses any LS which are present in the infinite lattice but cross the boundary. Thus, the lower numerical LS fraction is a boundary effect. We can utilize this to obtain more information about the infinite system.  As we expect the estimated LS fraction to be deficient due to the boundary, the numerical calculations for large enough domains should follow
\begin{equation}
    f_{\mathrm{Est}}(D) \simeq f_{\mathrm{Num}} - \frac{C_1}{D}.
\end{equation}
We plot the same data as a function of inverse neighborhood depth in Fig.\ref{fig:Fig03_NumericalLSFraction}, where one can observe that the above scaling form provides a good fit. By fitting a line to the data, we can extract an estimate for the LS fraction of the infinite system as
\begin{equation}
    f_{\mathrm{Num}}\simeq 0.0761.
\end{equation}
This value agrees with the value reported in Ref.\cite{kog21}, which is obtained with larger lattices. It is also close to the analytical lower bound calculated by counting LS types below. However, considering the spread of the data coming from different neighborhoods, the numerical result has an uncertainty of $\pm 0.0006$.

The numerical calculation yields information beyond the dimension of the null space. The QR decomposition also results in an orthonormal vector set that spans the null space. As all the LS are degenerate, the basis set found from the numerical diagonalization does not resemble the LS types discussed in the next section. Nevertheless, a basis independent physical quantity can be calculated from this set. We define the local density of states (LDOS) from LS as
\begin{equation}
    \rho(\vec{R}_i)=\sum_m |\langle \vec{R}_i | \Psi_m \rangle |^2,
\end{equation}
where $|\Psi_m\rangle$ satisfy both ${\cal H}|\Psi_m>=0$ and $\langle\Psi_m|\Psi_n\rangle=\delta_{n,m}$. Because we use the null space of ${\cal C}$, the wavefunctions forming the LS manifold are automatically localized to only one sublattice. 

\begin{figure}[!htb]
    \centering
    \includegraphics[trim=8mm 8mm 8mm 8mm,clip,width=0.48\textwidth]{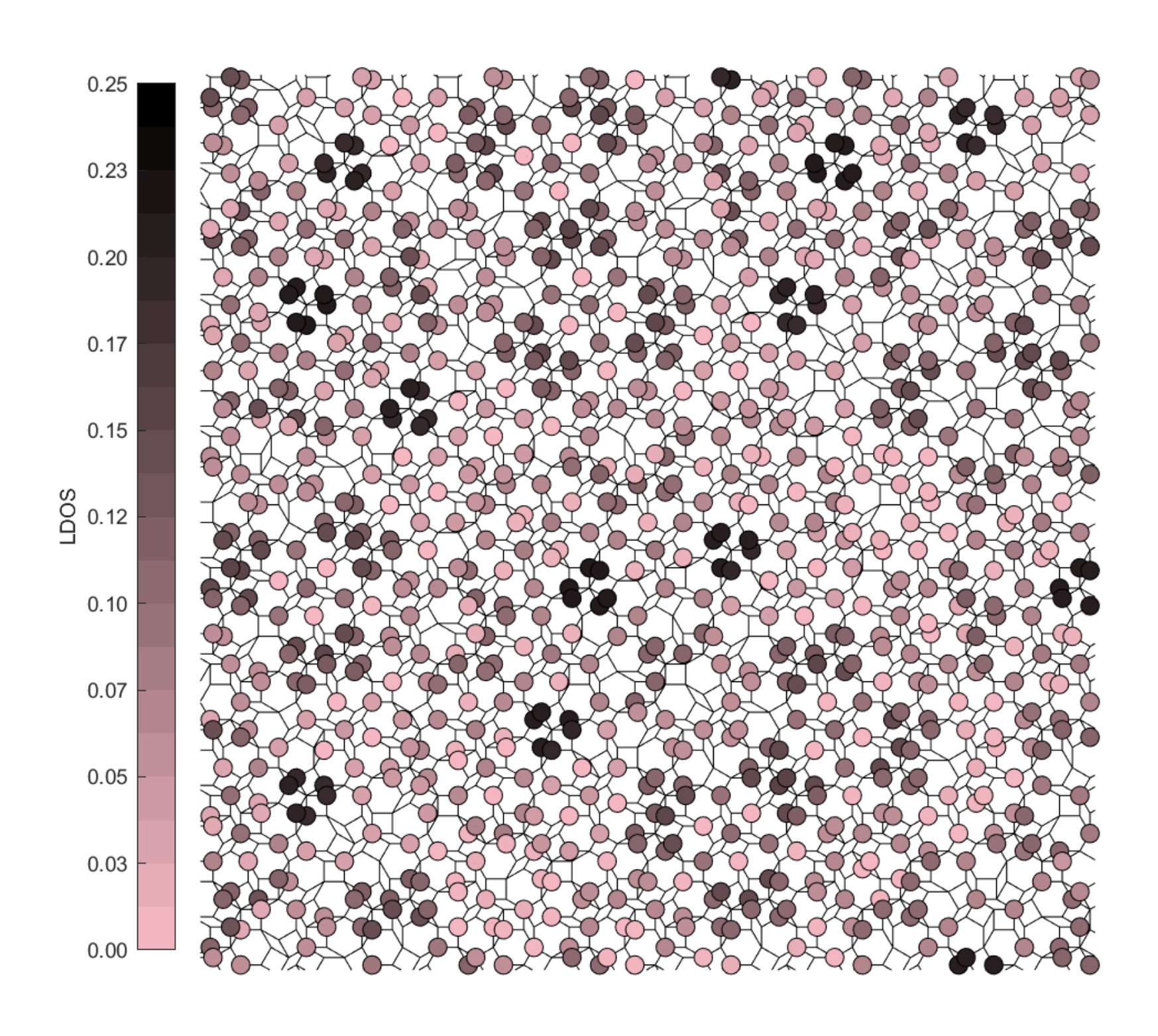}
    \caption{LDOS on the odd sublattice. Notice that some sites have zero LDOS, although they are in the odd sublattice. The LDOS is highly correlated with the local environment, as explained by the existence of LS types. The highest LDOS value is $\rho_{max}\simeq0.2132$.}
    \label{fig:Fig04_LDOS}
\end{figure}

A representative result for LDOS is displayed in Fig.\ref{fig:Fig04_LDOS}. First, we notice that, unlike the PL, there are no large regions devoid of LS. Most of the sites host LS. Still, it is crucial to notice that some sites have zero LDOS. In the ABL, no such forbidden sites are found. We also see that local connectivity is strongly correlated with LDOS. The sites with the largest LDOS have $\rho_{max}\simeq 0.2132$ and are always nearest neighbors of $F$ vertices of the even sublattice. Yet not all nearest neighbors of $F$ sites have maximum LDOS. This correlation is explained with the LS types and their overlaps in the next section.

Recent experiments in synthetic lattices, whether in cold atom systems, polaritons, or assembled surfaces, feature local probes of density. We believe that general properties of LDOS, such as the existence of forbidden sites or maximum of the LDOS, can be observed with local probes such as scanning tunneling microscopy \cite{col17} or site-resolved density measurements of cold atoms \cite{vie19}.  

\section{Localized State Types}
\label{sec:LS types}

\begin{figure}[!htb]
    \centering
    \includegraphics[trim=8mm 8mm 8mm 8mm,clip,width=0.48\textwidth]{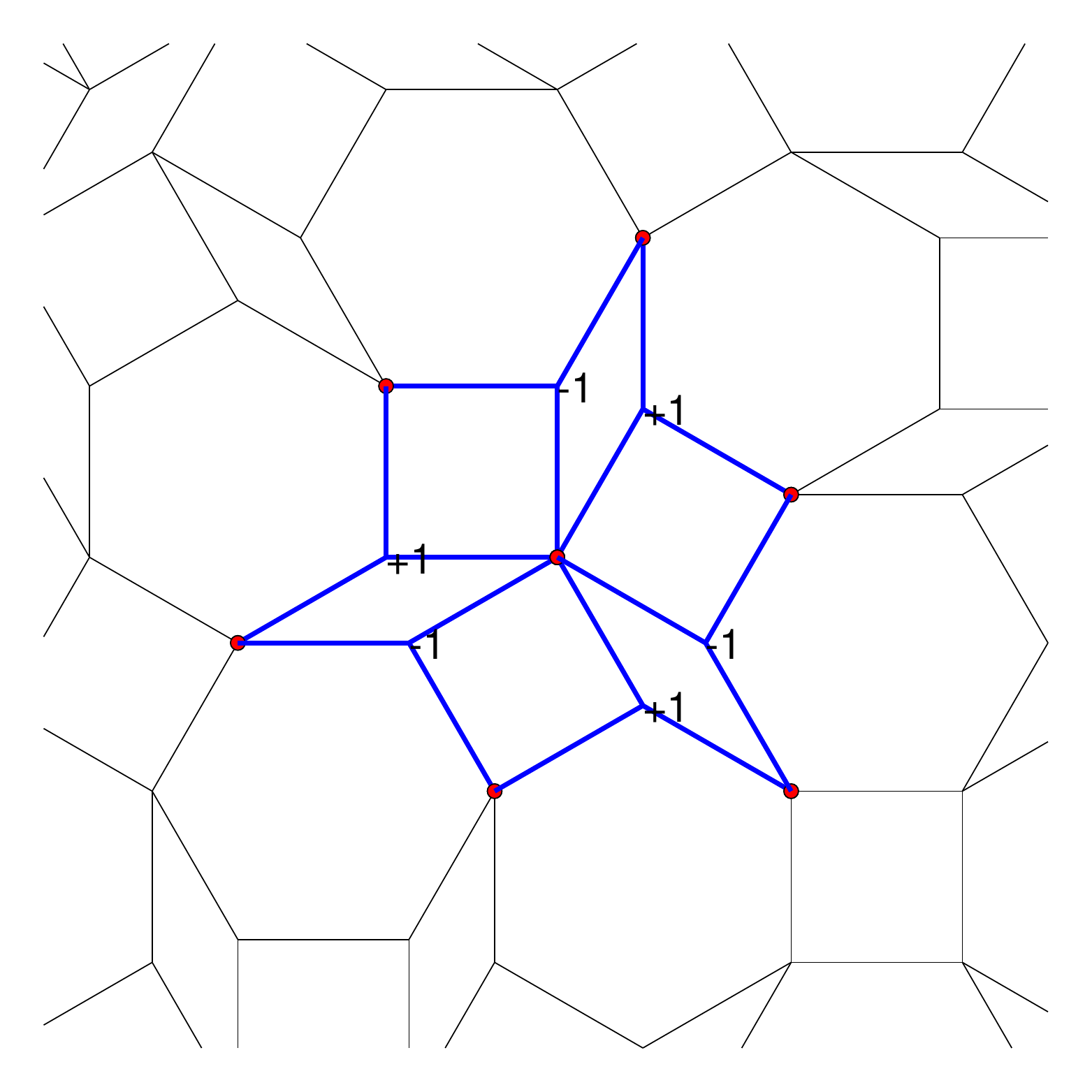}
    \caption{The LS type-A1 has six C sites around an F vertex. Notice that the sums of the wavefunctions linked to the nearest neighbor sites marked in red are zero.}
    \label{fig:TypeA1_RealSpace}
\end{figure}

\begin{figure}[!htb]
    \centering
    \includegraphics[trim=8mm 8mm 8mm 8mm,clip,width=0.43\textwidth]{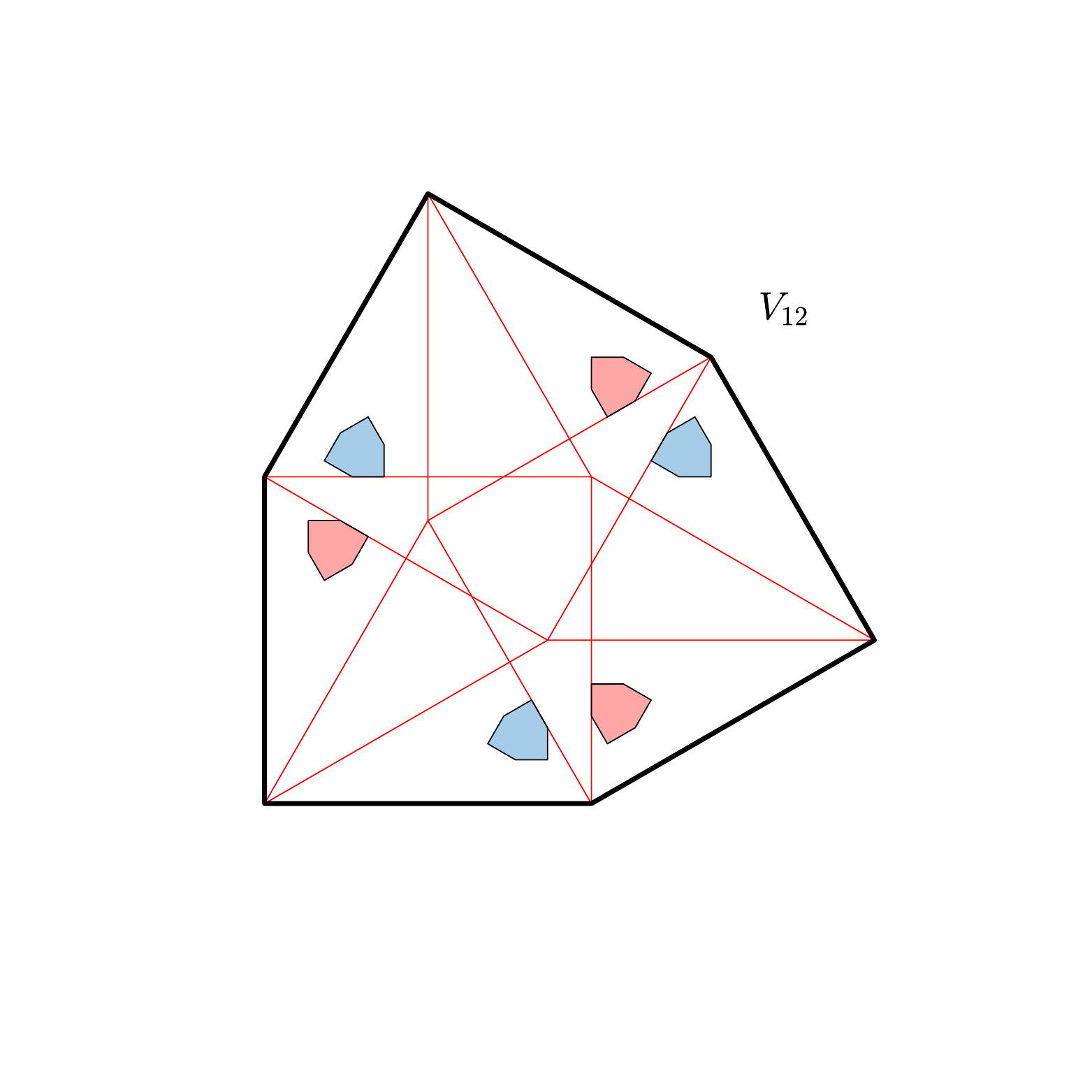}
        \includegraphics[trim=8mm 8mm 8mm 8mm,clip,width=0.43\textwidth]{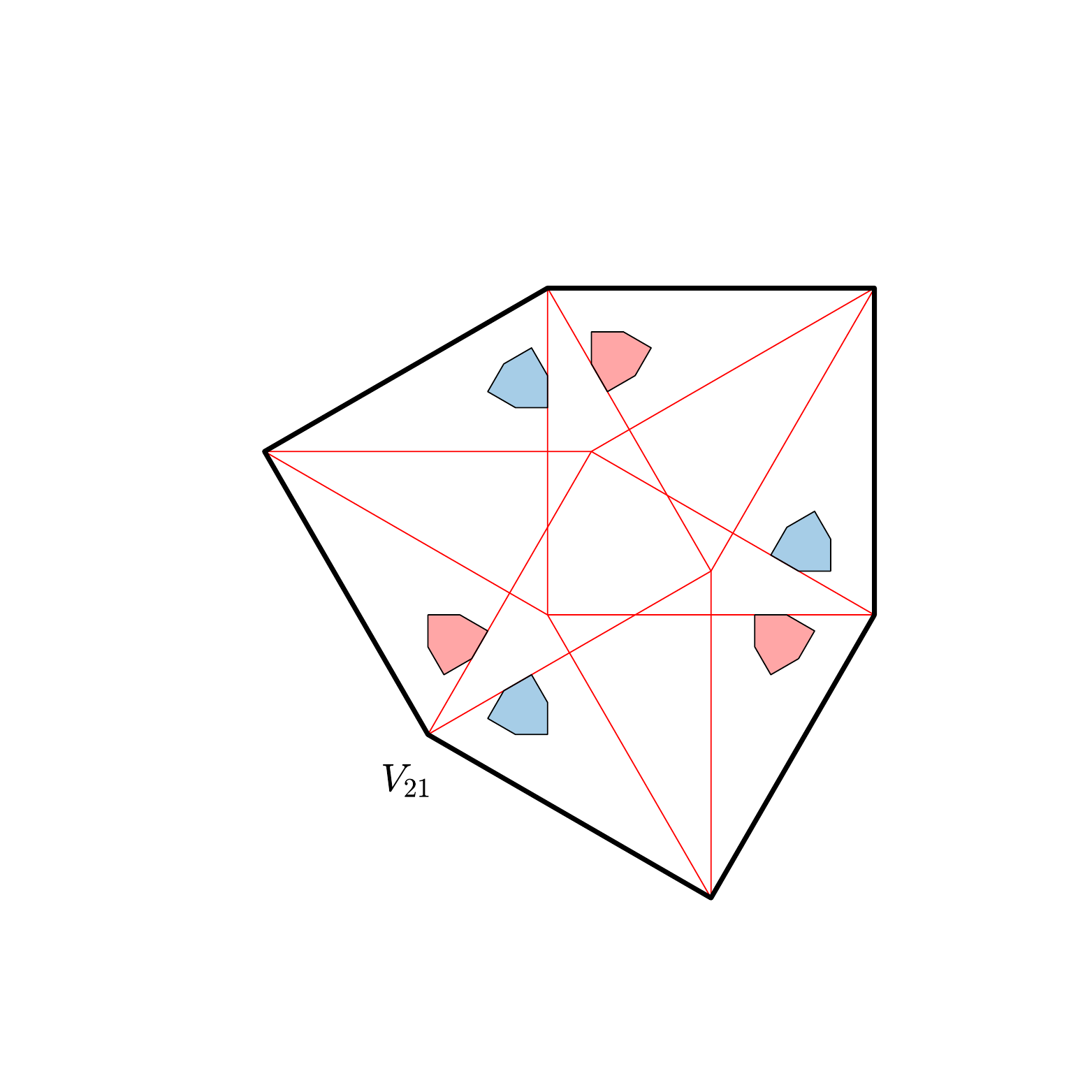}
    \caption{The allowed regions in $V_{12}$ and $V_{21}$ for the type-A1 state in Fig.\ref{fig:TypeA1_RealSpace} are shown in blue. Three of the sites in the support are in one hexagon and the remaining three in the other. Another type-A1 LS can be constructed by rotating Fig.\ref{fig:TypeA1_RealSpace} by $\pi/3$. Allowed regions for the rotated state are shown in red.}
    \label{fig:TypeA1_PerpSpace}
\end{figure}
 
\begin{figure}[!htb]
    \centering
    \includegraphics[trim=8mm 8mm 8mm 8mm,clip,width=0.43\textwidth]{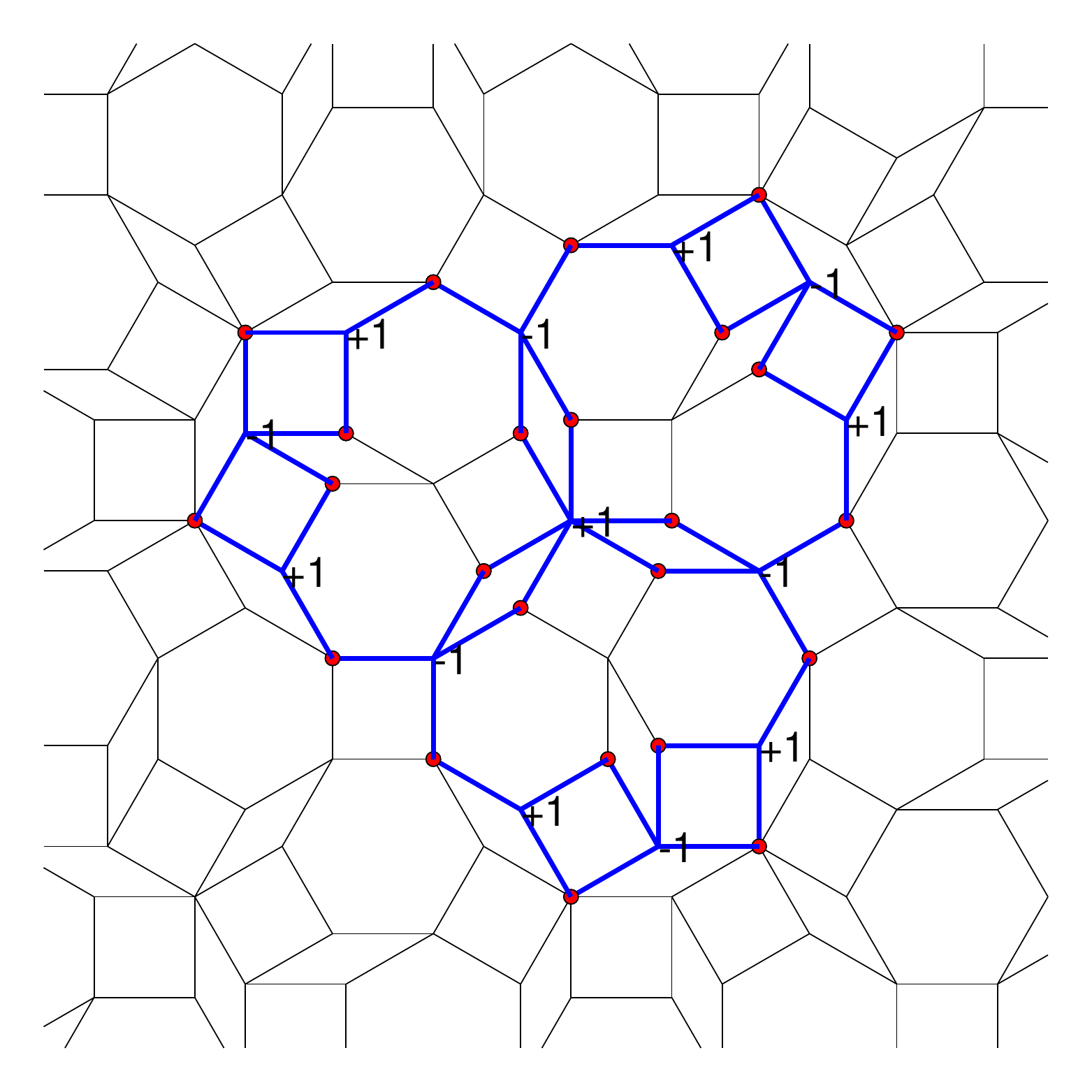}
    \includegraphics[trim=8mm 8mm 8mm 8mm,clip,width=0.43\textwidth]{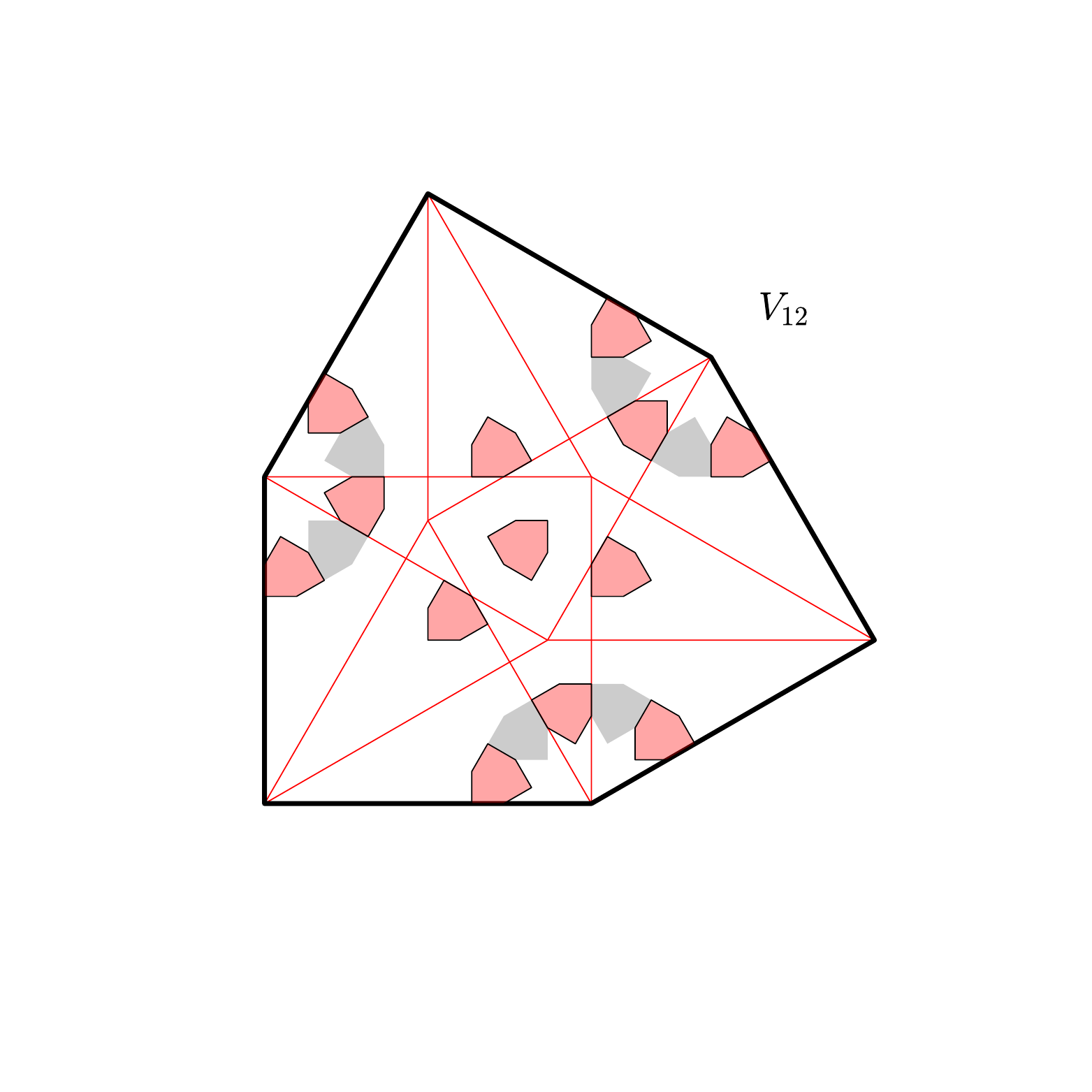}
    \caption{LS type-A2 in real space and its allowed regions in $V_{12}$. Perpendicular space regions covered by previous LS types are shown as the grey background.}
    \label{fig:TypeA2_RealSpace}
\end{figure}

 \begin{table}[hbt!]
\begin{tabular}{||c| c  ||} 
 \hline
 LS Type & Areas of polygons in terms of $\xi$   \\ [0.5ex] 
 \hline\hline
  $V_{ij}$ & $\frac{3}{2}(1+\xi)$ \\ 
 \hline
 A1,A2 & $\frac{3}{4}(\xi^{-2}+\xi^{-3})$ \\ 
 \hline
 A3,A4 &  $\frac{3}{4}(\xi^{-4}+\xi^{-5})$  \\
 \hline
 B1,B2&  $\frac{1}{2}(2\xi^{-2}+\xi^{-4})$   \\
 \hline
B3,B4 &  $\frac{1}{2}(2\xi^{-4}+\xi^{-6})$  \\
 \hline
 C1,C2 &  $\frac{3}{4} \xi^{-3}$   \\ 
 \hline
  C3,C4 &  $\frac{3}{2}\xi^{-4}$   \\
 \hline
 C5 &  $\frac{3}{2}\xi^{-5}$   \\ 
 \hline 
 C6,C7,C8,C9 &  $\frac{3}{4}\xi^{-5}$   \\ 
 \hline 
 C10 &  $\frac{1}{4}(\xi^{-5}+\xi^{-6})$  \\[1ex] 
 \hline

\end{tabular}

 \caption{Perpendicular space areas for the allowed areas of vertices for the 18 LS types. The ratio of these areas to the area of $V_{12}$, combined with symmetry factors yield the LS type frequencies.}
  \label{tbl:PolygonAreas}
\end{table}

\begin{table}[hbt!]

\begin{tabular}{||c| c c c||} 
 \hline
 LS Type & Frequency & &  \\ [0.5ex] 
 \hline\hline
 A1 & $\frac{\xi^{-3}}{2}~~=$ & $\frac{26-15\sqrt{3}}{2}$ & $\sim$~~0.009619 \\ 
 \hline
 A2 &  $\frac{\xi^{-3}}{2}~~=$& $\frac{26-15\sqrt{3}}{2}$& $\sim$~~0.009619 \\
 \hline
 A3 &  $\frac{\xi^{-5}}{2}~~=$ & $\frac{362-209\sqrt{3}}{2}$ & $\sim$~~0.000691 \\
 \hline
 A4 &  $\frac{\xi^{-5}}{2}~~=$& $\frac{362-209\sqrt{3}}{2}$& $\sim$~~0.000691 \\
 \hline
 B1 &  $\frac{3\xi^{-3}+\xi^{-4}}{4}~~=$ & $\frac{175-101\sqrt{3}}{4}$ & $\sim$~~0.015717 \\ 
 \hline
  B2 &  $\frac{3\xi^{-3}+\xi^{-4}}{4}~~=$ & $\frac{175-101\sqrt{3}}{4}$ & $\sim$~~0.015717 \\
 \hline
 B3 &  $\frac{3\xi^{-5}+\xi^{-6}}{4}~~=$ & $\frac{2437-1407\sqrt{3}}{4}$ & $\sim$~~0.001128 \\ 
 \hline 
 B4 &  $\frac{3\xi^{-5}+\xi^{-6}}{4}~~=$ & $\frac{2437-1407\sqrt{3}}{4}$ & $\sim$~~0.001128 \\ 
 \hline 
 C1 &  $\frac{\xi^{-3}+\xi^{-4}}{4}~~=$ & $\frac{123-71\sqrt{3}}{4}$ & $\sim$~~0.006098 \\ 
 \hline
  C2 &  $\frac{\xi^{-3}+\xi^{-4}}{4}~~=$ & $\frac{123-71\sqrt{3}}{4}$ & $\sim$~~0.006098 \\ 
 \hline
 C3 &  $\frac{\xi^{-4}+\xi^{-5}}{2}~~=$ & $\frac{459-265\sqrt{3}}{2}$ & $\sim$~~0.003268 \\ 
 \hline
 C4 &  $\frac{\xi^{-4}+\xi^{-5}}{2}~~=$ & $\frac{459-265\sqrt{3}}{2}$ & $\sim$~~0.003268 \\ 
 \hline
  C5 &  $\frac{\xi^{-5}+\xi^{-6}}{2}~~=$ & $\frac{1713-989\sqrt{3}}{2}$ & $\sim$~~0.000876 \\ 
 \hline
  C6 &  $\frac{\xi^{-5}+\xi^{-6}}{4}~~=$ & $\frac{1713-989\sqrt{3}}{4}$ & $\sim$~~0.000438 \\ 
 \hline
 C7 &  $\frac{\xi^{-5}+\xi^{-6}}{4}~~=$ & $\frac{1713-989\sqrt{3}}{4}$ & $\sim$~~0.000438 \\ 
 \hline
 C8 &  $\frac{\xi^{-5}+\xi^{-6}}{4}~~=$ & $\frac{1713-989\sqrt{3}}{4}$ & $\sim$~~0.000438 \\ 
 \hline
 C9 &  $\frac{\xi^{-5}+\xi^{-6}}{4}~~=$ & $\frac{1713-989\sqrt{3}}{4}$ & $\sim$~~0.000438 \\ 
 \hline
  C10 &  $~~~~\frac{\xi^{-6}}{2}~~~~~~=$ & $\frac{1351-780\sqrt{3}}{2}$ & $\sim$~~0.000185 \\ 
 \hline
 Total &$\frac{23\xi^{-4}+24\xi^{-5}}{2}=$   & $\frac{10919-6304\sqrt{3}}{2}$ & $\sim$~~0.075855 \\ [1ex] 
 \hline

\end{tabular}

 \caption{The eighteen LS types and their frequencies.}
  \label{TBL:LSFrequencies}
\end{table}

\begin{figure}[!htb]
    \centering
    \includegraphics[trim=8mm 8mm 8mm 8mm,clip,width=0.48\textwidth]{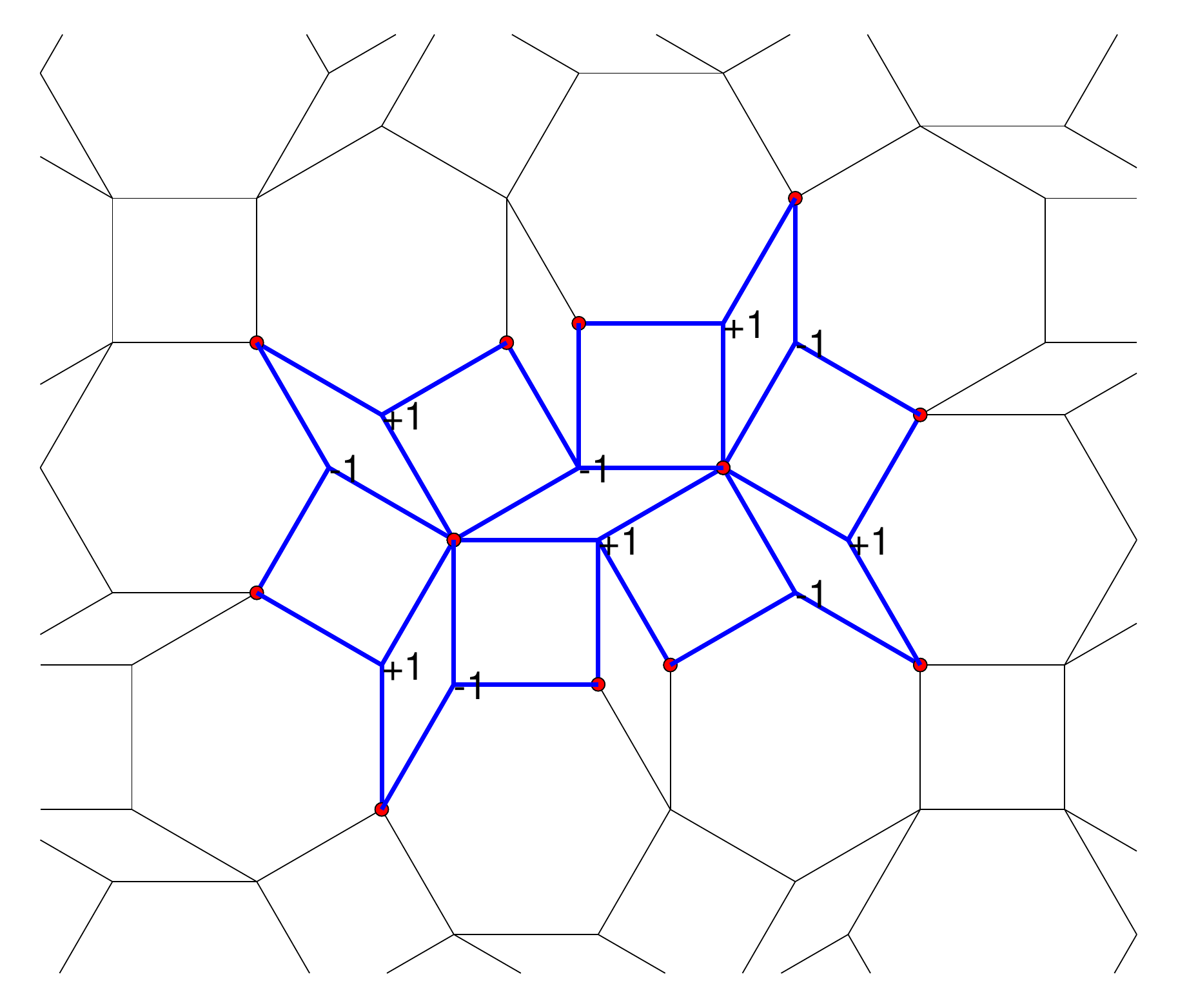}
    \includegraphics[trim=8mm 8mm 8mm 8mm,clip,width=0.48\textwidth]{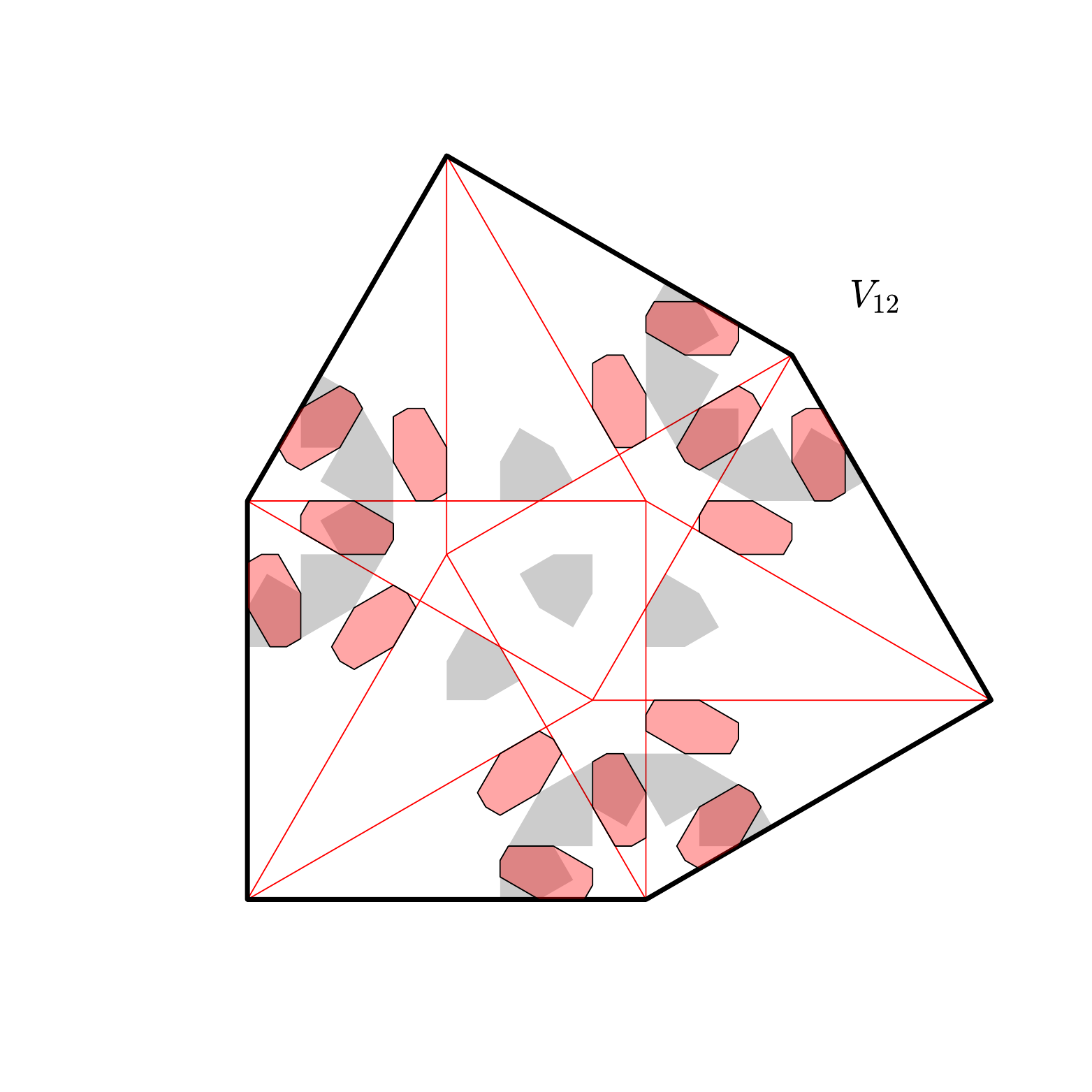}
    \caption{LS type-B1 in real space and corresponding allowed regions for all three independent orientations in $V_{12}$. Although there is overlap with previous LS types, at least one allowed hexagon lies entirely in an uncovered region, proving independence from the LS types above.}
    \label{fig:TypeB1_RealSpace}
\end{figure}

\begin{figure}[!htb]
    \centering
    \includegraphics[trim=8mm 8mm 8mm 8mm,clip,width=0.48\textwidth]{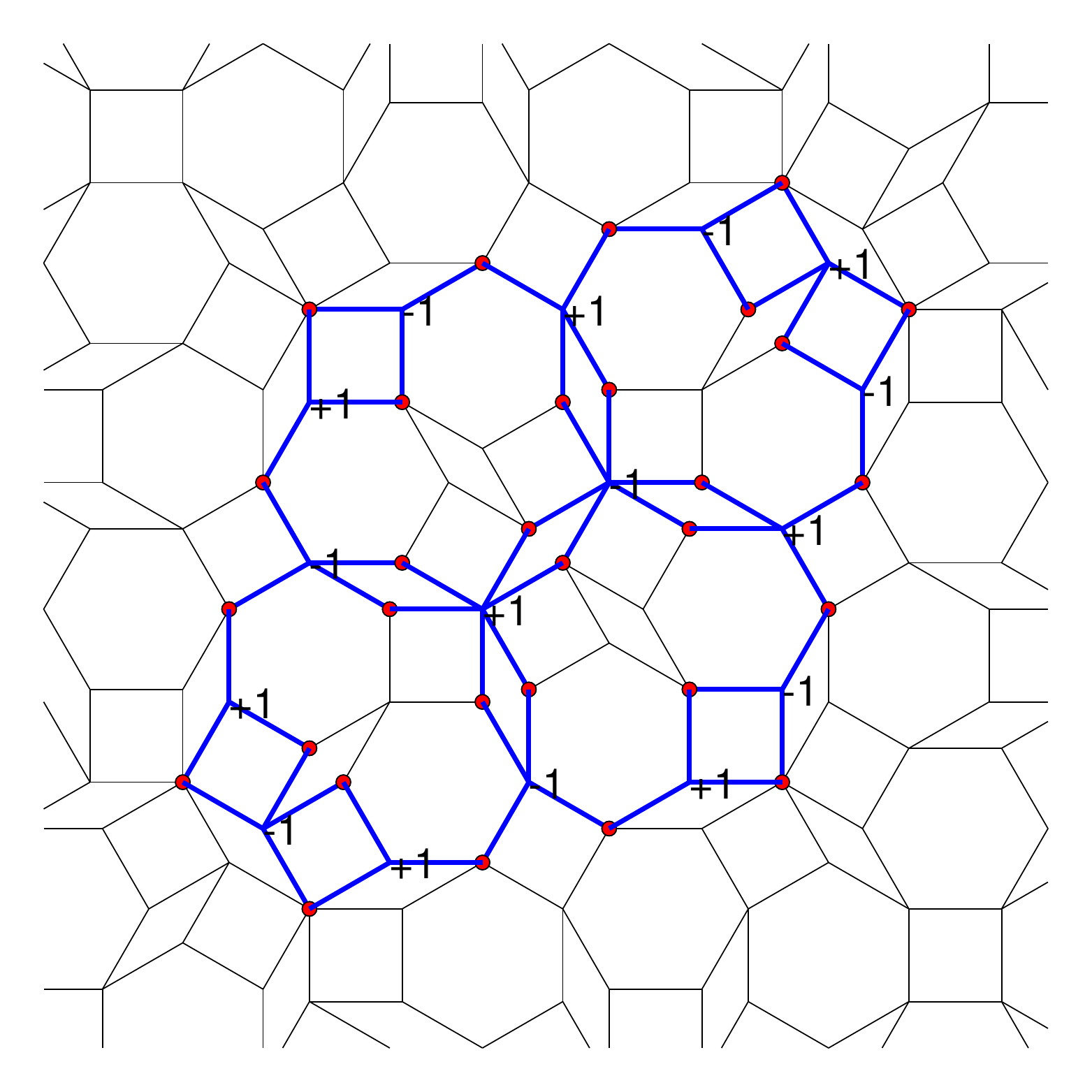}
    \includegraphics[trim=8mm 8mm 8mm 8mm,clip,width=0.48\textwidth]{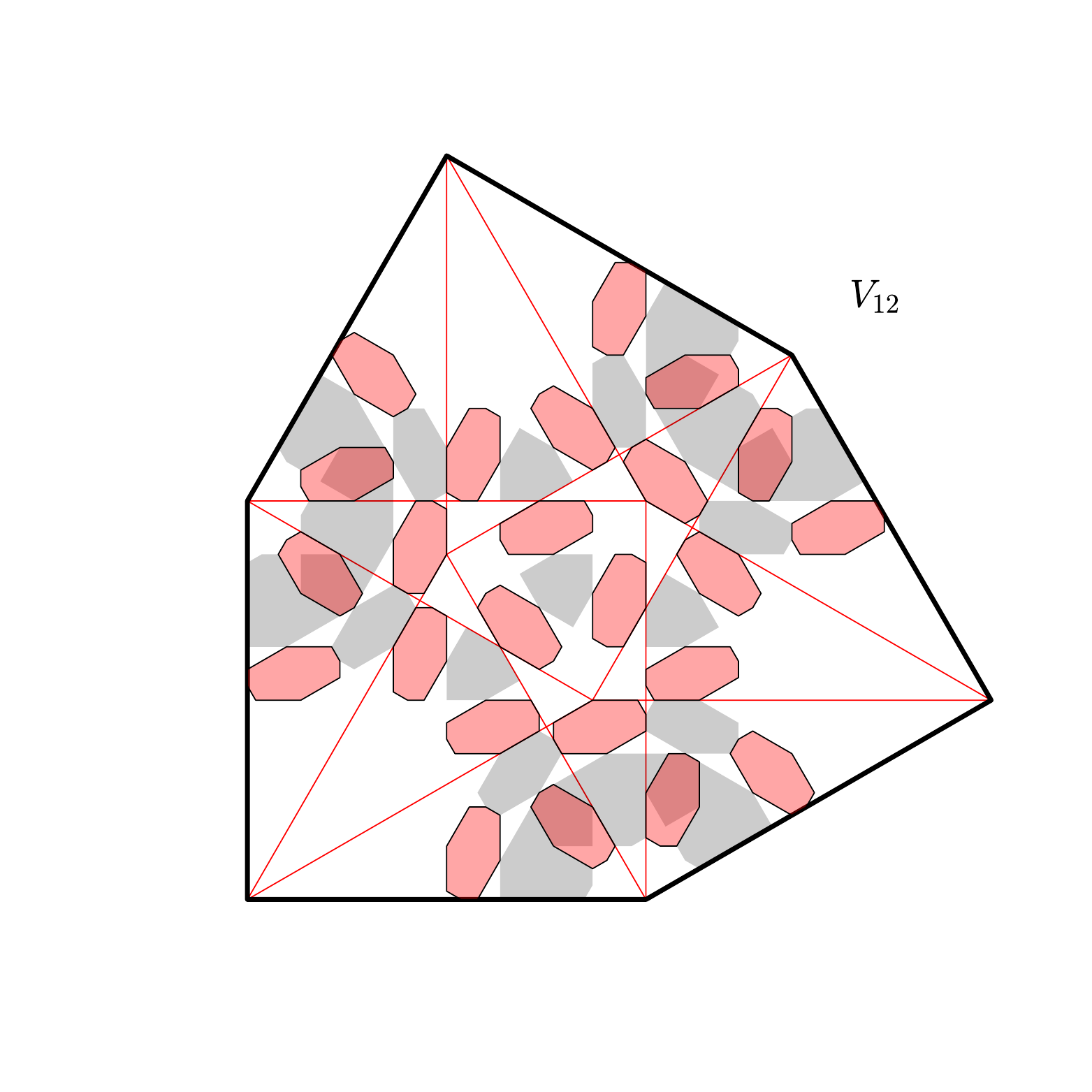}
    \caption{Type-B2 in real and perpendicular space. Its allowed region area and frequency are the same as type-B1.}
    \label{fig:TypeB2_RealSpace}
\end{figure}

\begin{figure}[!htb]
    \centering
    \includegraphics[trim=8mm 8mm 8mm 8mm,clip,width=0.48\textwidth]{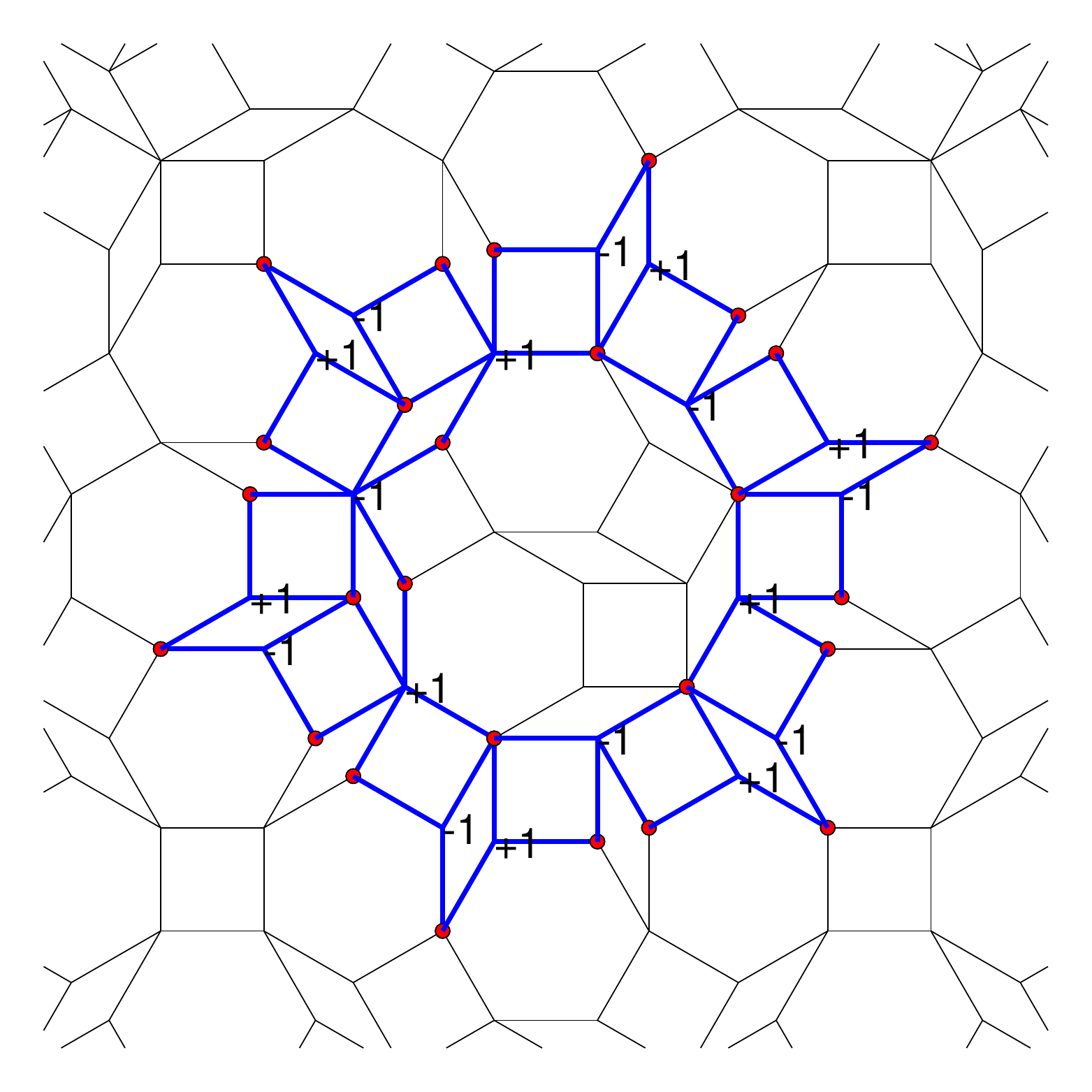}
     \includegraphics[trim=8mm 8mm 8mm 8mm,clip,width=0.48\textwidth]{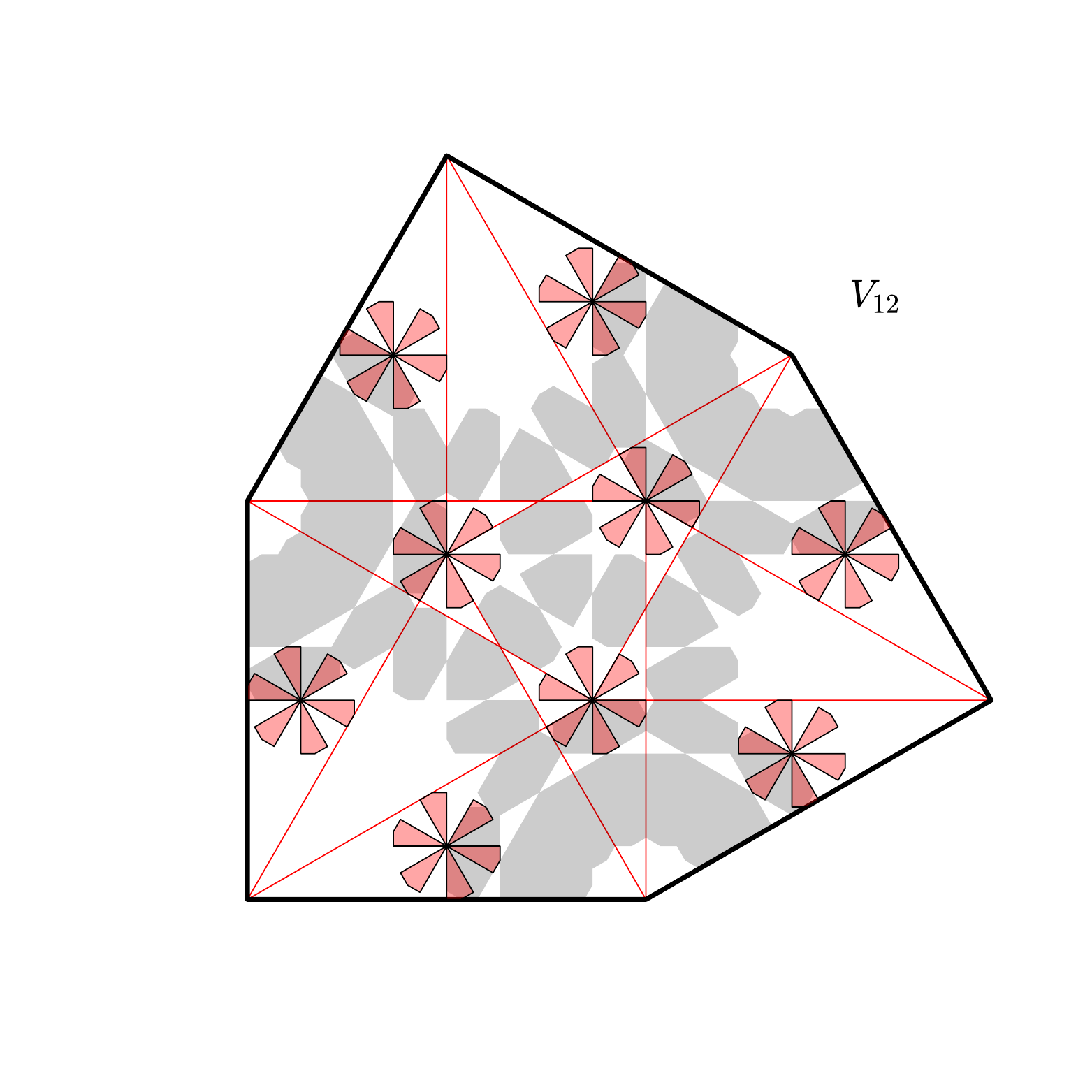}
    \caption{Type-C1 LS and allowed regions for its six independent rotated copies.This LS type has frequency $f_{C1}=~\frac{\xi^{-3}+\xi^{-4}}{2}\simeq0.006098$}
    \label{fig:TypeC1_RealSpace}
\end{figure}

\begin{figure}[!htb]
    \centering
    \includegraphics[trim=8mm 8mm 8mm 8mm,clip,width=0.48\textwidth]{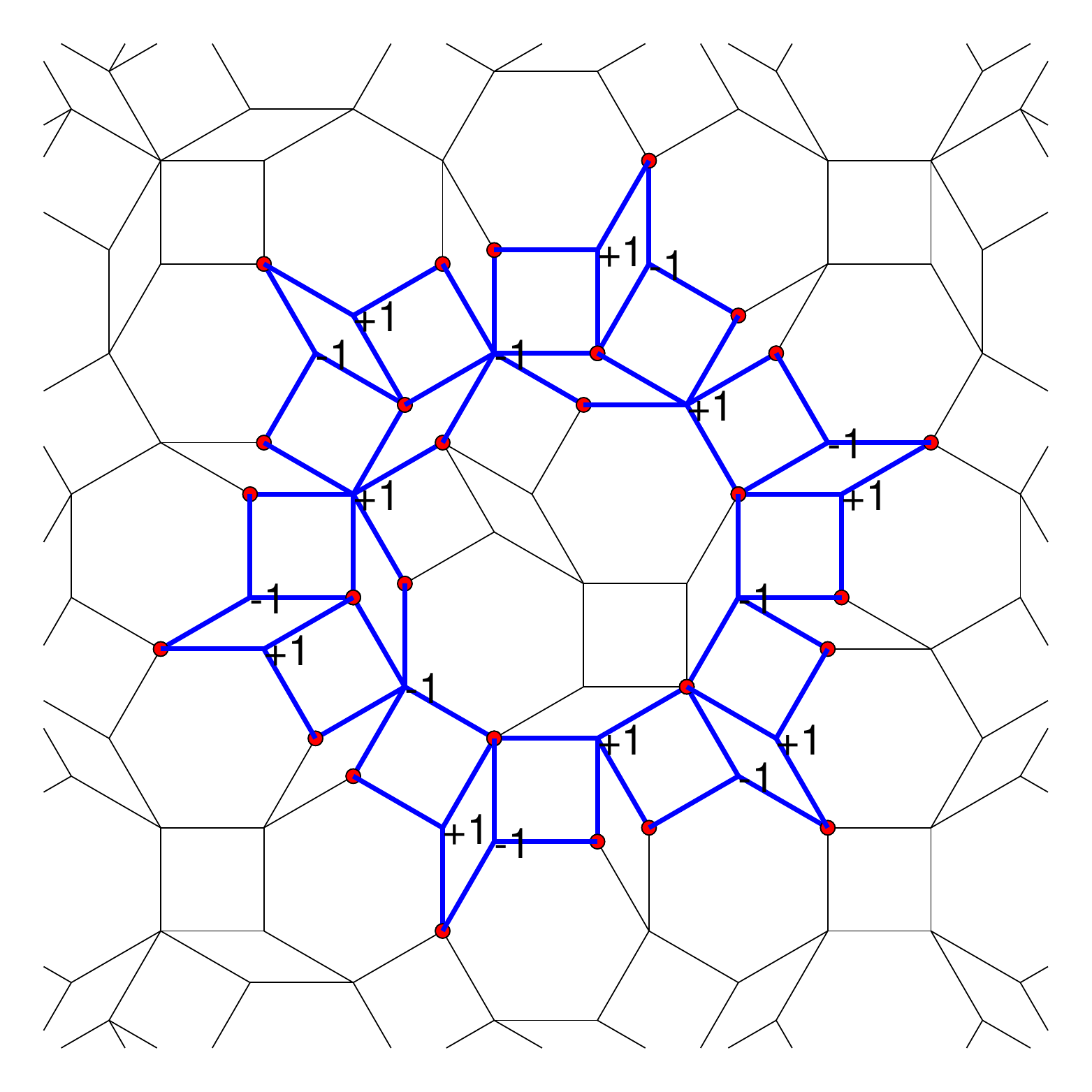}
    \includegraphics[trim=8mm 8mm 8mm 8mm,clip,width=0.48\textwidth]{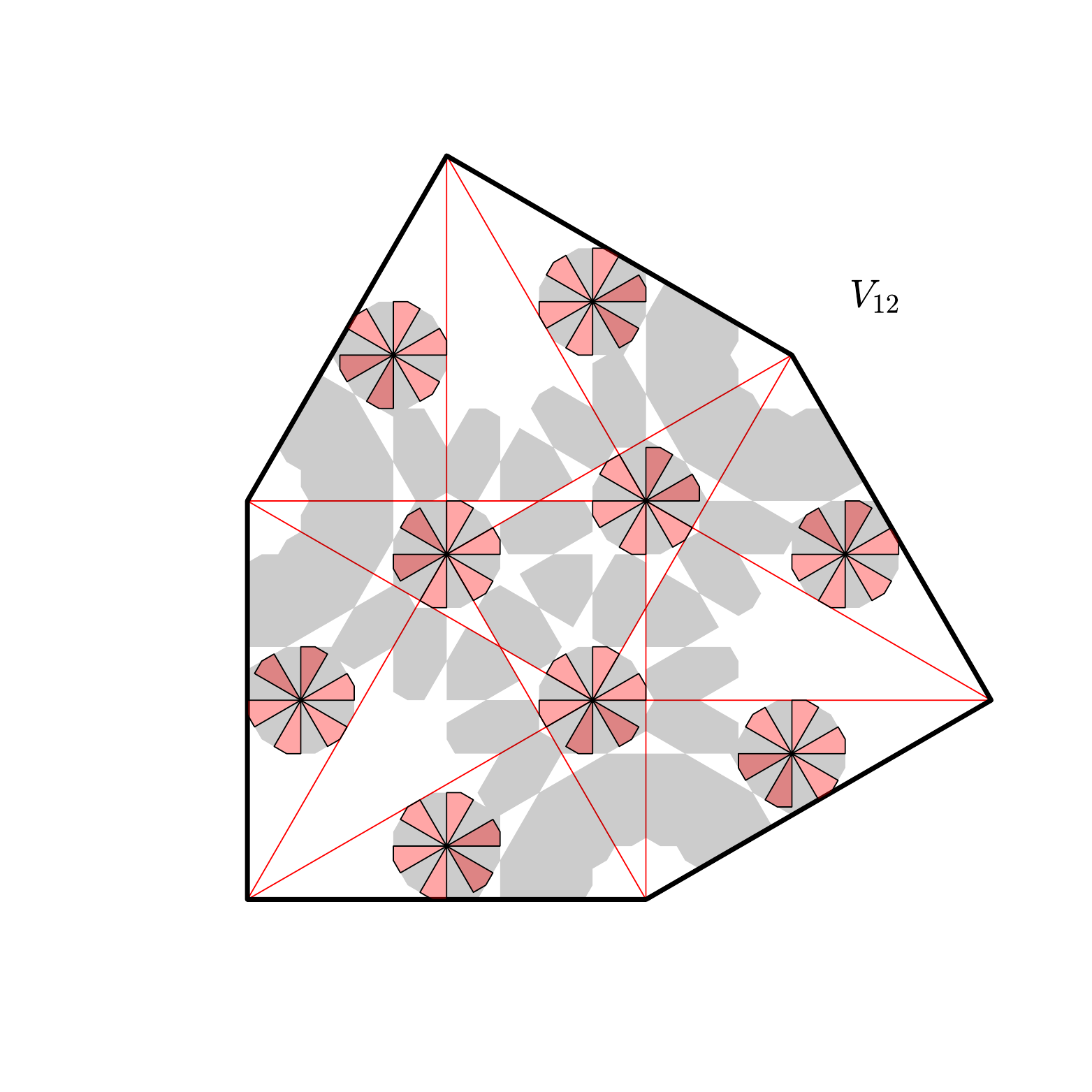}
    \caption{Type-C2 LS has allowed areas adjacent to type-C1. The two LS types have the same frequency.}
    \label{fig:TypeC2_RealSpace}
\end{figure}

\begin{figure}[!htb]
    \centering
    \includegraphics[trim=8mm 8mm 8mm 8mm,clip,width=0.48\textwidth]{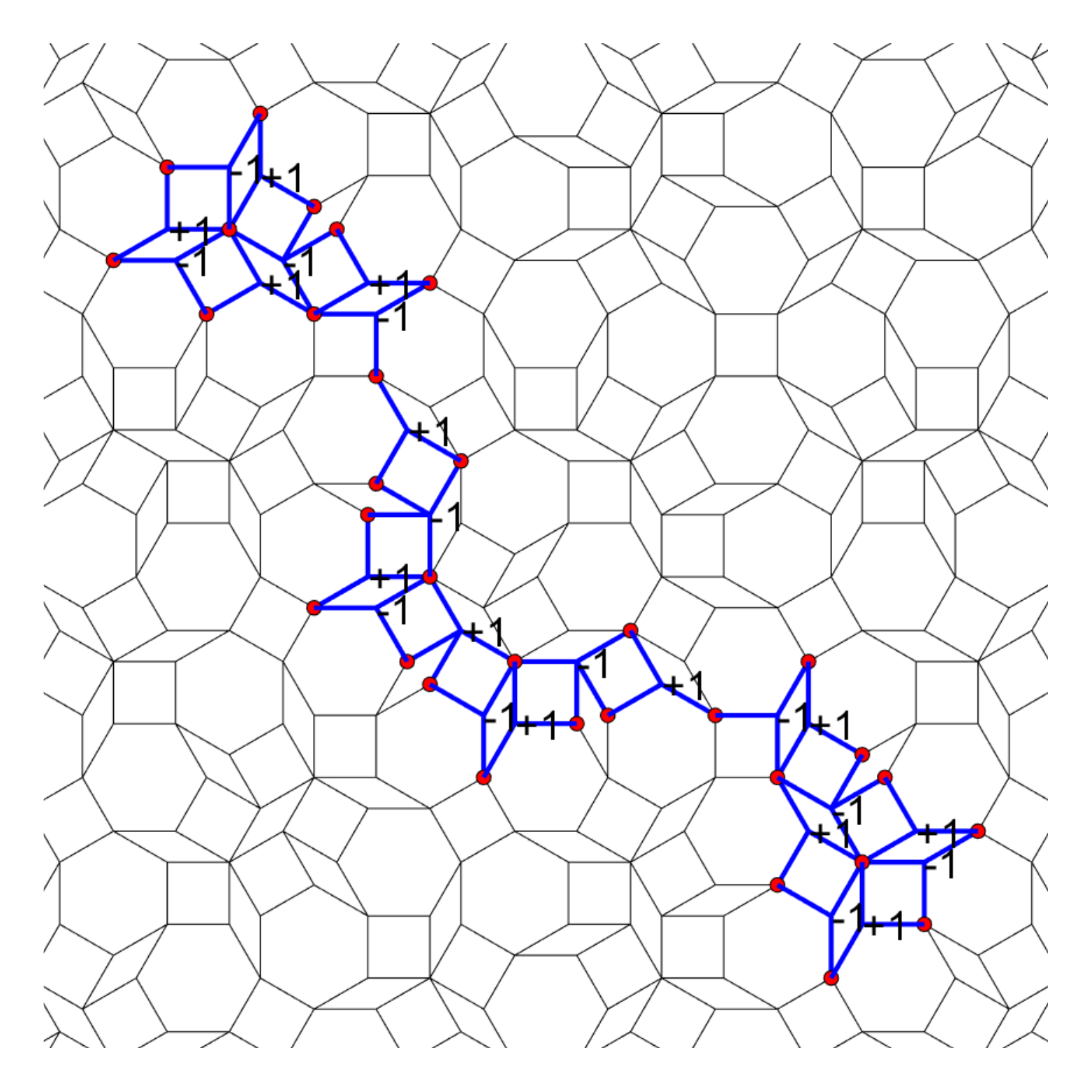}
    \includegraphics[trim=8mm 8mm 8mm 8mm,clip,width=0.48\textwidth]{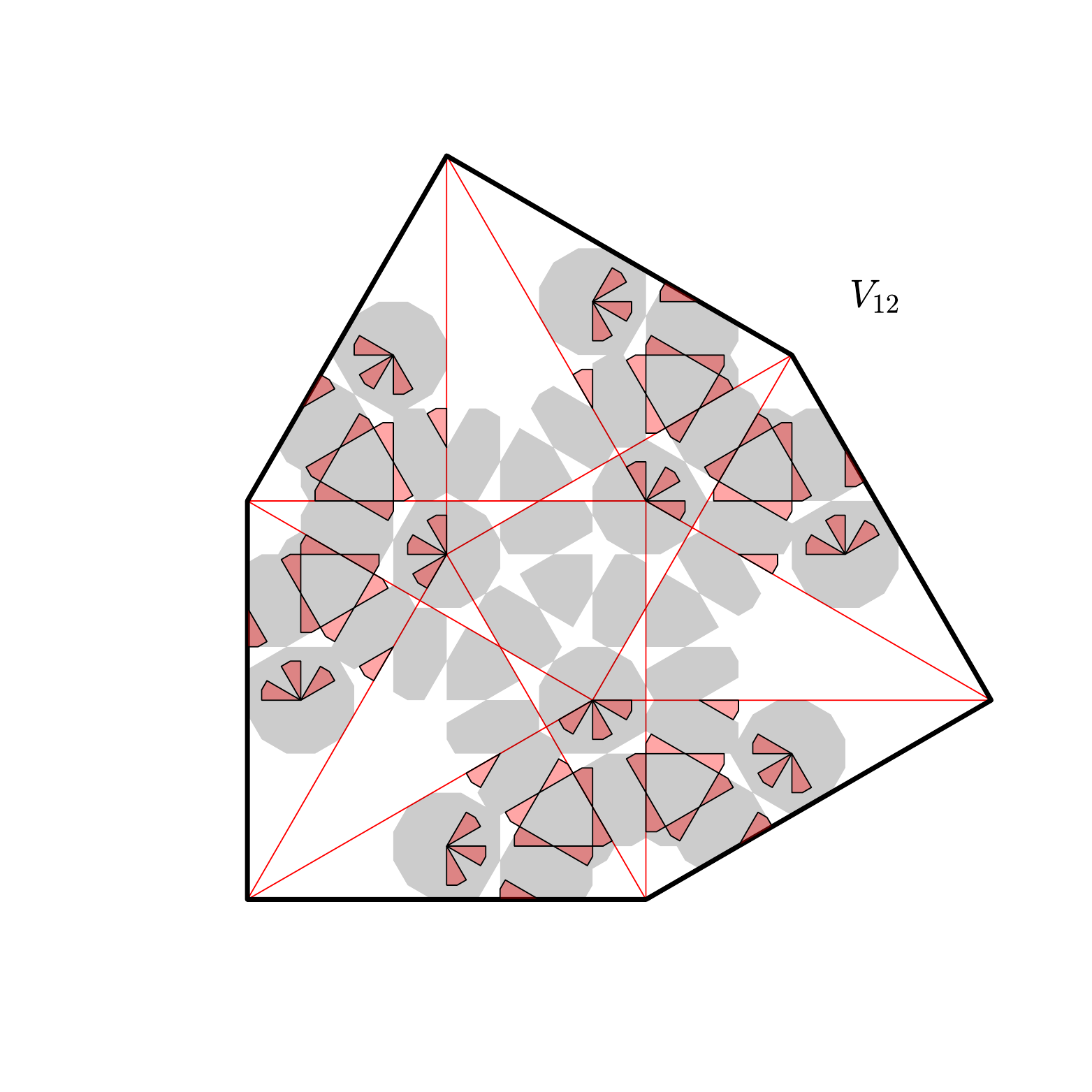}
    \caption{LS type-C3. While the allowed area is similar in shape to types C1 and C2, its scaled down. The frequency is $\frac{\xi^{-4}+\xi^{-5}}{2}\simeq0.003268$.}
    \label{fig:TypeC3_RealSpace}
\end{figure}

\begin{figure}[!htb]
    \centering
    \includegraphics[trim=8mm 8mm 8mm 8mm,clip,width=0.48\textwidth]{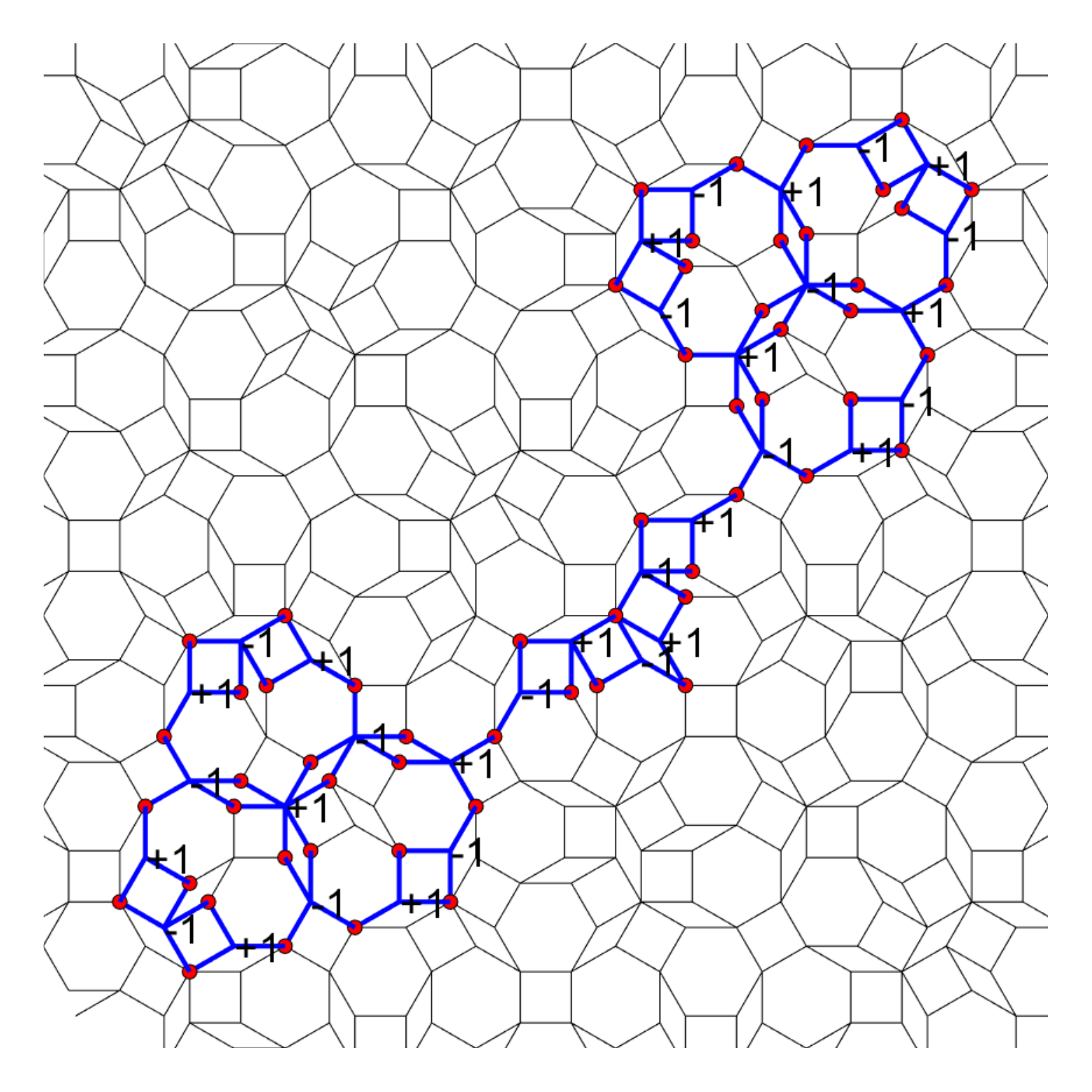}
    \includegraphics[trim=8mm 8mm 8mm 8mm,clip,width=0.48\textwidth]{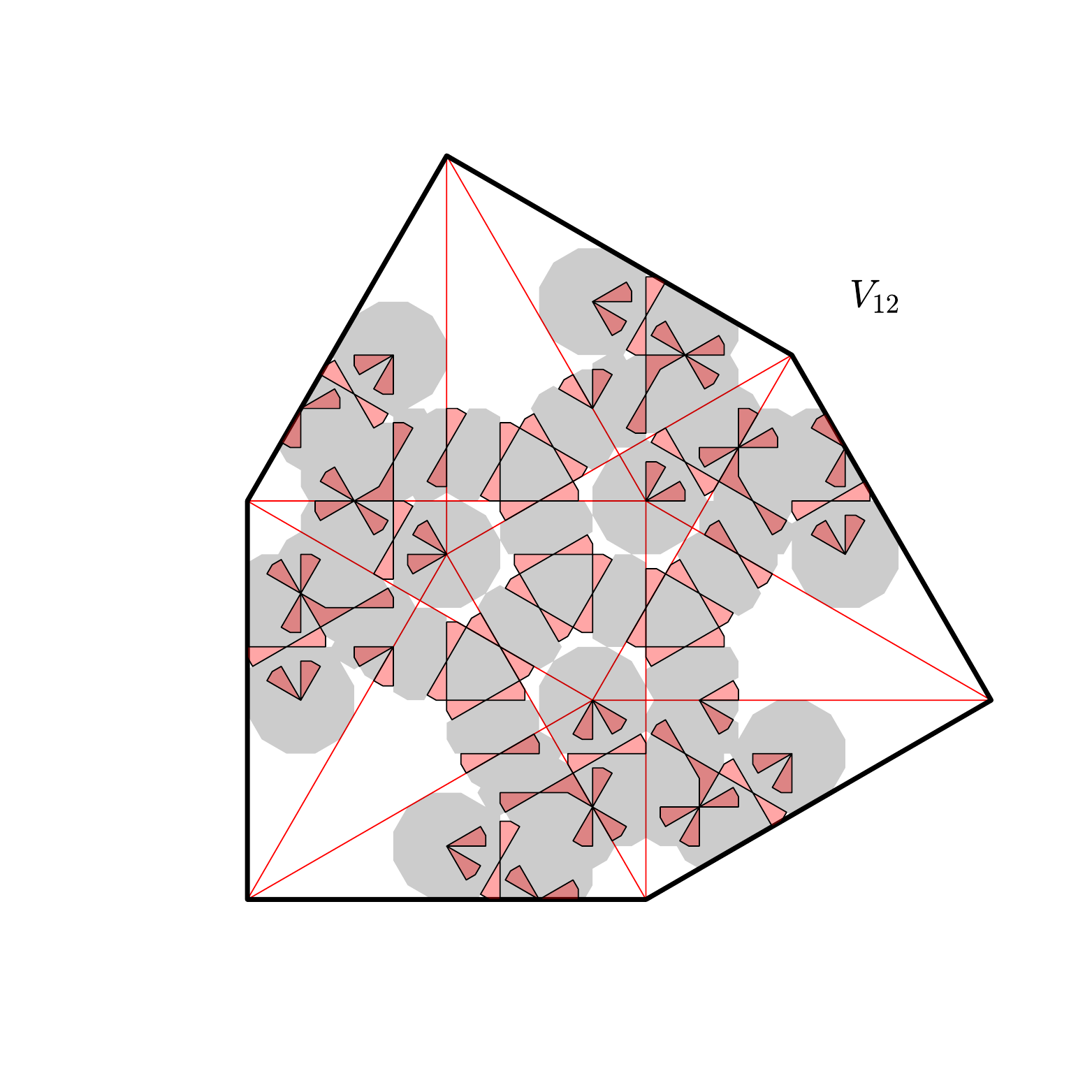}
    \caption{LStype C4 has the same frequency as C3, and its allowed areas fill in the empty regions between C3 allowed areas.}
    \label{fig:TypeC4_RealSpace}
\end{figure}

Without Bloch's theorem, there is no simple way to label the eigenstates of quasicrystals. This problem is most clearly seen in the LS, where almost 8\% of all the states are degenerate and even energy cannot be used to index the states. However, the LS are confined in a finite region, and we can describe how many independent zero energy states are translated copies of specific LS types. If an LS is identified in a limited area of the lattice, it must have infinitely many copies throughout the infinite lattice. This is ensured by Conway's theorem \cite{gar77} showing that any finite pattern is repeated throughout the quasicrystal. One must then ask what percentage of the eigenstates correspond to the translated and rotated copies of the same LS. We designate all the rotated and translated copies of an LS as the same LS type.

We can designate any LS as an LS type. However, we aim to find LS types with the highest frequency while remaining independent from other LS types. For the PL, just six LS types are enough to span the whole zero energy manifold \cite{ara88,kts17,mok20}. Recent ABL results \cite{kog20,okt21} indicate that an infinite number of LS types organized into generations are required. The first few generations of these types were shown to account for most of the numerically obtained degeneracy. Orthogonality of LS types is not easily assured; however, their independence is more straightforward to prove. If one of the vertices of the support of an LS type is not in the support of any other LS type, then independence follows trivially. 

The frequency of an LS type is found by counting the occurrences of its support throughout the lattice. This counting is generally done by using the scaling symmetries of the lattice, which gets increasingly more complex as the support region for the LS type grows. Another way of calculating the frequencies is based on calculating the perpendicular space acceptance region for the vertices on which the LS type is defined. We successfully applied this method to the PL, the ABL, and local isomorphism classes of pentagonal quasicrystals, which lack simple scaling symmetries \cite{mok20,okt21,okt22}. Here we use the same procedure for the SDL.

In a recent paper \cite{kog21}, Koga introduced 11 LS types for the SDL. One surprising result of this paper was that the total frequency of the calculated LS types accounts for only 84.2\% of the numerically obtained LS fraction. Here, we also find the same 11 LS types. However, our results for their frequencies are different. We believe the frequencies reported in Ref.\cite{kog21} are incorrect, possibly due to the complexity of the scaling for the SDL. Our results for the same 11 types give us a lower bound $f_{\mathrm{LS}}\simeq 0.0719$, which is almost 94.5\% of the numerical result. We find a further 7 LS types, which bring out the total to  $f_{\mathrm{LS}}=\frac{10919-6304\sqrt{3}}{2}\simeq0.075855$, which is 99.7\% of the numerical result. 
Considering the projections of the acceptance regions in perpendicular space, we believe that infinitely many LS types are required to span the zero energy manifold. A similar scenario was found in the ABL. We give both the real space configurations and the perpendicular space acceptance domains for 18 LS types. The first eight types have high frequency and are presented in this section, while the remaining ten are given in the appendix. 

As the first LS type, consider the state in Fig.\ref{fig:TypeA1_RealSpace}. We call this state type-A1 LS. Its support has 6 C vertices, and wavefunction has an alternating $\pm1$ sign. We plot the acceptance region for all six vertices in Fig\ref{fig:TypeA1_PerpSpace} in blue. As can be seen, three of the vertices in the support lie in $V_{12}$ and the other three in $V_{21}$. The LS support has rotational symmetry under $2 \pi/3$ rotation; however, recall that such a rotation exchanges the perpendicular space hexagons $V_{12}\leftrightarrow V_{21}$. Thus, a $2\pi/3$ rotated version of the type-A1 ls remains independent, as can be seen from the areas plotted in red in Fig. \ref{fig:TypeA1_PerpSpace}. The two can also be distinguished by observing that the central $F$ vertex lies in $V_{11}$ in one case and $V_{22}$ in the other. For all the LS types,  symmetry is restored once all rotated copies of the same state are considered. Hence, for the remaining LS types, we only plot the allowed areas in $V_{12}$. A simple rotation gives the allowed areas for $V_{21}$.  

The frequency of LS type-A1 is the ratio of the areas of one of the allowed hexagons in Fig.\ref{fig:TypeA1_PerpSpace} and  $V_{12}$. This ratio is $f_{A1}=\xi^{-3}/2=\frac{26-15\sqrt{3}}{2}\simeq0.00962$. We further checked this frequency by numerically counting $F$ sites with six $C$ vertex neighbors. Thus, our results indicate that the frequency reported in Ref.\cite{kog21}, $0.0052$, is incorrect.

As the next LS type, consider the real space and perpendicular space images for type-A2 displayed in Fig.\ref{fig:TypeA2_RealSpace}. Type-A2 LS has 13 sites in its support, but allowed regions for any of these vertices have the same area as type-A1 LS. Hence Type-A2 has the same frequency as Type-A1. Furthermore, no site in the support of type-A1 can also belong to the support of type-A2. This is most easily observed by comparing the perpendicular space allowed regions for the two types. To facilitate such comparisons, we plot the total allowed areas belonging to all previous LS types in gray on all subsequent perpendicular space images.

In figure Fig.\ref{fig:TypeB1_RealSpace} we give display type-B1 LS. First, notice that the support is symmetric under $\pi$ rotation. Hence copies of the same state rotated by $\pi/3$ as well as $2\pi/3$ are independent of the original. The second important thing to notice is an overlap between the support of type-B1 and type-A2. Six out of the ten sites in the support of type-B1 can also be in the support of a type-A2 LS. Thus, two such states are not orthogonal. However, notice that four of the sites in the support are not covered by previous LS types. As a result, the B1 type is independent of A1 and A2. The independence is most easily seen in the perpendicular space figure. Even if a single allowed area covers a new portion of the perpendicular space, the independence of the new LS type is established. This reasoning is enough to demonstrate the independence of all the remaining states in this paper. 

The LS fraction for type-B1 and type-B2 is the same $f_{B1}=f_{B2}=\frac{3\xi^{-3}+\xi^{-4}}{4}=\frac{175-101\sqrt{3}}{4}\simeq0.01572$. The definition of the type-B2 state is given in Fig.\ref{fig:TypeB2_RealSpace}. Once again, independence is established by the newly covered areas, such as those in the $F$ vertex regions. One can also explore the interplay between independence and orthogonality by considering the LDOS of a point shared by type-A1 and type-B2 states. If these two states were orthogonal, LDOS would be a sum of their densities. As the type-A1 wavefunction is normalized by $1/\sqrt{6}$, and type-B2 by $1/4$, the LDOS at a common point would be $\rho_{max}=\frac{1}{6}+\frac{1}{16}=\frac{11}{48}\simeq 0.2292$. This value is higher than the maximum value from the numerical calculation $\rho_{max}\simeq 0.2132$. As these states are not orthogonal, their overlap $\langle B2|A1\rangle=\frac{1}{2\sqrt{6}}$ reduces the LDOS just due to these two types to $  9/46\simeq 0.1956$. Other LS types contribute to the LDOS at this point to raise its value to the numerically observed peak. 

Next, we give four states with no rotational symmetry in real space, types C1 to C4. Their real space and perpendicular space images are given in Figs.\ref{fig:TypeC1_RealSpace},\ref{fig:TypeC2_RealSpace},\ref{fig:TypeC3_RealSpace},\ref{fig:TypeC4_RealSpace}. Now all six copies obtained by rotations are independent. Hence, the frequencies for type-C1 and type-C2 are $f_{C1}=f_{C2}=\frac{\xi^{-3}+\xi^{-4}}{4}=\frac{123-71\sqrt{3}}{4}\simeq0.00609$ and, for types C3 and C4 are $f_{C3}=f_{C4}=\frac{\xi^{-4}+\xi^{-5}}{2}=\frac{459-265\sqrt{3}}{2}\simeq0.00327$.  

The total LS frequency of the types given in this section is $0.0694$ which corresponds to $91.2\%$ of the numerically observed LS fraction. We identified 10 more LS types as given in the appendix. We give the allowed area for a single vertex of each  LS type in table \ref{tbl:PolygonAreas}. These values combined with the symmetry factors and the total area of the perpendicular space give the frequency of LS types, listed in table \ref{TBL:LSFrequencies}. 
The LS type frequencies  sum up to $f_{\mathrm{LS}}=\frac{10919-6304\sqrt{3}}{2}\simeq0.07585$ a lower bound which is quite close to our numerical estimate. 

We used small (up to 35 neighbors) system sizes to identify new LS types. Instead of QR decomposition, we use Gauss-Jordan elimination to put the ${\cal C}$ matrix into a reduced row echelon form. While this is not numerically efficient, it gives a set of LS where wavefunctions are rational numbers. This approach is limited by the cluster size we use, and the existence of other LS types is almost assured by comparing large cluster null space results with the total areas covered by the LS types we identified. We expect that many, possibly infinitely many, independent small frequency LS types to exist in the SDL, similar to the  ABL. Another interesting point about the LS types is that they can all be chosen to have a constant density over their support. Such a choice was possible for the ABL but not for the PL. It is unclear why the LS would be organized in a specific form for the two lattices or which property of the PL prevents the same situation.

 \begin{table}[hbt!]
\begin{tabular}{||c| c |c| c|c|c|c|c|c|c||} 
 \hline
 LS Type & Frequency & LS Type& Frequency& LS Type& Frequency& LS Type& frequency& LS Type& Frequency\\ [0.5ex] 
 \hline\hline
 $A^{0}$ & $\xi^{-3}$ & $B^{0}$ &$\frac{3\xi^{-3}+\xi^{-4}}{2}$& $C_{x}^{0}$ &$\frac{\xi^{-3}+\xi^{-4}}{2}$ & $C_{y}^{0}$ &${\xi^{-4}+\xi^{-5}}$ & $C_{z}^{0}$ &${\xi^{-5}+\xi^{-6}}$\\ 
 \hline
 $A^{1}$ &  $\xi^{-2}{\xi^{-3}}$& $B^{1}$ &$\xi^{-2}\frac{3\xi^{-3}+\xi^{-4}}{2}$& $C_{x}^{1}$ &$\xi^{-2}\frac{\xi^{-3}+\xi^{-4}}{2}$ & $C_{y}^{1}$ &$\xi^{-2}(\xi^{-4}+\xi^{-5})$ & $C_{z}^{1}$ &${\xi^{-2}(\xi^{-5}+\xi^{-6}})$ \\
 \hline
 $A^{2}$ &  $\xi^{-4}{\xi^{-3}}$& $B^{2}$ &$\xi^{-4}\frac{3\xi^{-3}+\xi^{-4}}{2}$& $C_{x}^{2}$ &$\xi^{-4}\frac{\xi^{-3}+\xi^{-4}}{2}$ & $C_{y}^{2}$ &$\xi^{-4}(\xi^{-4}+\xi^{-5})$ & $C_{z}^{2}$ &$\xi^{-4}({\xi^{-5}+\xi^{-6}})$ \\
 \hline
$A^{3}$ &  $\xi^{-6}{\xi^{-3}}$& $B^{3}$ &$\xi^{-6}\frac{3\xi^{-3}+\xi^{-4}}{2}$& $C_{x}^{3}$ &$\xi^{-6}\frac{\xi^{-3}+\xi^{-4}}{2}$ & $C_{y}^{3}$ &$\xi^{-6}(\xi^{-4}+\xi^{-5})$ & $C_{z}^{3}$ &$\xi^{-6}({\xi^{-5}+\xi^{-6}})$ \\
 \hline
 ... &  ... & ...& ... &... & ... &...&...&...&... \\ 
 \hline
$A^{n}$ &  $\xi^{-2n}{\xi^{-3}}$& $B^{n}$ &$\xi^{-2n}\frac{3\xi^{-3}+\xi^{-4}}{2}$& $C_{x}^{n}$ &$\xi^{-2n}\frac{\xi^{-3}+\xi^{-4}}{2}$ & $C_{y}^{n}$ &$\xi^{-2n}(\xi^{-4}+\xi^{-5})$ & $C_{z}^{n}$ &$\xi^{-2n}({\xi^{-5}+\xi^{-6}})$ \\ 
 \hline 
... &  ... & ...& ... &... & ... &...&...&...&... \\ 
 \hline
  $\sum_{0}^{\infty}$ &  $\frac{\xi^{-1}}{\xi^{2}-1} $& $\sum_{0}^{\infty}$ &  $\frac{3\xi^{-1}+\xi^{-2}}{2(\xi^{2}-1)}$ &$\sum_{0}^{\infty}$ &  $\frac{\xi^{-1}+\xi^{-2}}{2(\xi^{2}-1)}$ &$\sum_{0}^{\infty}$& $\frac{\xi^{-2}+\xi^{-3}}{\xi^{2}-1}$&$\sum_{0}^{\infty}$& $\frac{\xi^{-3}+\xi^{-4}}{\xi^{2}-1}$\\  [1ex] 
 \hline

\end{tabular}
 
 \caption{The scaling argument leading to the conjectured LS frequency $ f_{\mathrm{Conj}}= \frac{3\xi^{-1}+2\xi^{-2}+2\xi^{-3}+\xi^{-4}}{\xi^{2}-1}\simeq 0.076660$. }
 
 \label{tbl:InfiniteSum}
\end{table}

The perpendicular space images of LS types suggest a hierarchical organization. The allowed perpendicular space region for a single vertex of  type-A3 is just the allowed region for the type-A1 state scaled by  $\xi$. Consequently, their frequencies differ by a factor of $\xi^2$. The SDL has deflation symmetry, where $\xi$ is the scaling factor between the original lattice and its deflation \cite{soc89}. It is not unreasonable to conjecture that an infinite number of LS types are generated by deflation \cite{kog20}. We present such a conjecture in table \ref{tbl:InfiniteSum}.  

The support of any LS type defines a domain in the real space lattice. The deflation applied to this domain creates a larger domain and possibly new LS types. We specify five different domains. In the domain of $A^{0}$, the LS types A1 and A2 are defined. Applying deflation to this domain generates the domain $A^{1}$, where LS types A3 and A4 occur. The LS types B1 and B2 live in the $B^{0}$ domain. LS types  C1, C2, define $C_{x}^{0}$, C3, C4 similarly define $C_{y}^{0}$. Our last domain, $C_{z}^{0}$, contains the LS types C6, C7, C8, and C9. If these are the only domains that are not generated by deflation, and deflations of only these domains can generate all LS types, we get $ f_{\mathrm{Conj}}= \frac{3\xi^{-1}+2\xi^{-2}+2\xi^{-3}+\xi^{-4}}{\xi^{2}-1}\simeq 0.076660$ in the thermodynamic limit. This value is close to our numerical estimate.

\section{Forbidden Sites}
\label{sec:Forbidden}

\begin{figure}[!htb]
    \centering
    \includegraphics[clip,width=0.43\textwidth]{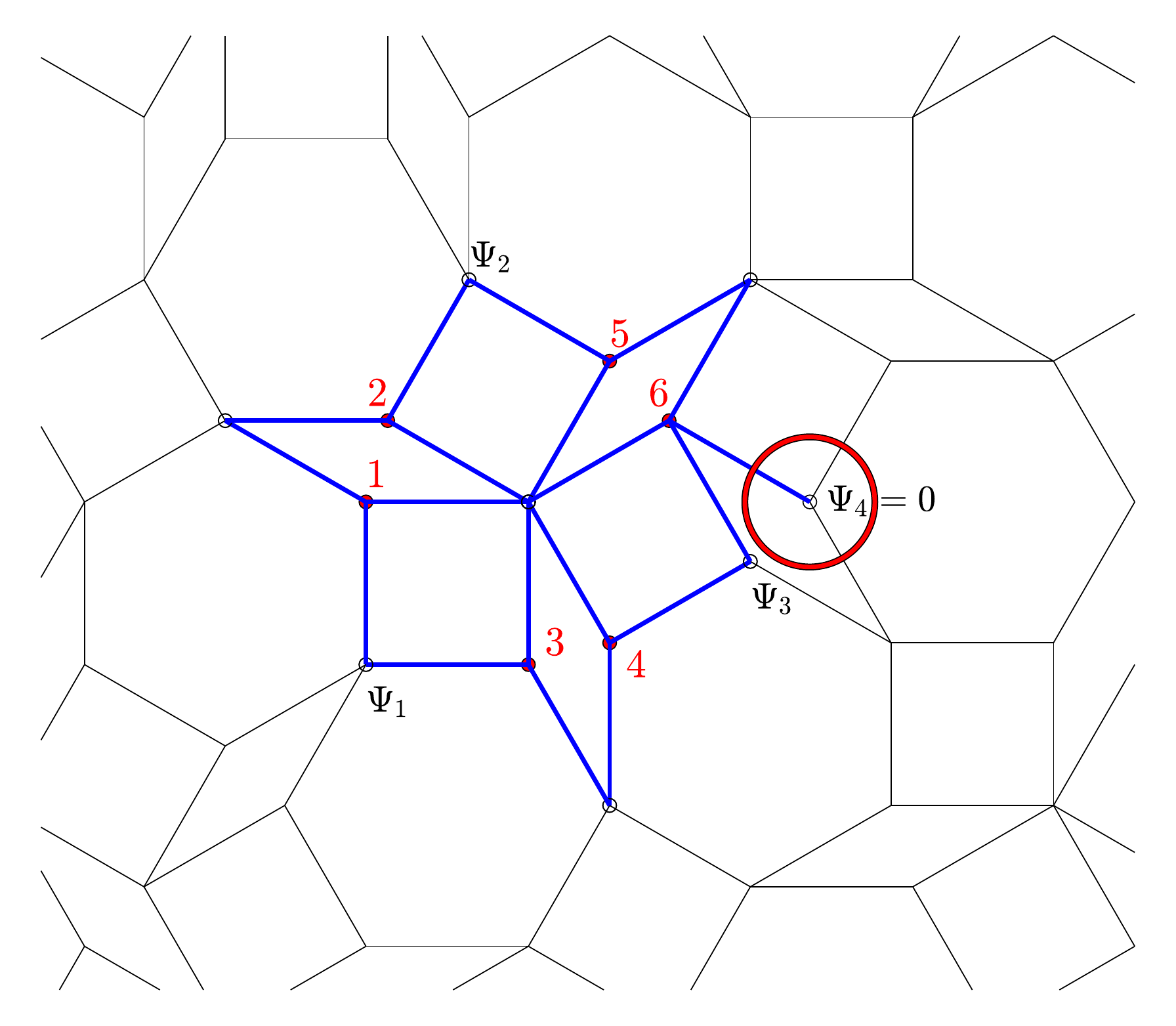}
    \includegraphics[clip,width=0.43\textwidth]{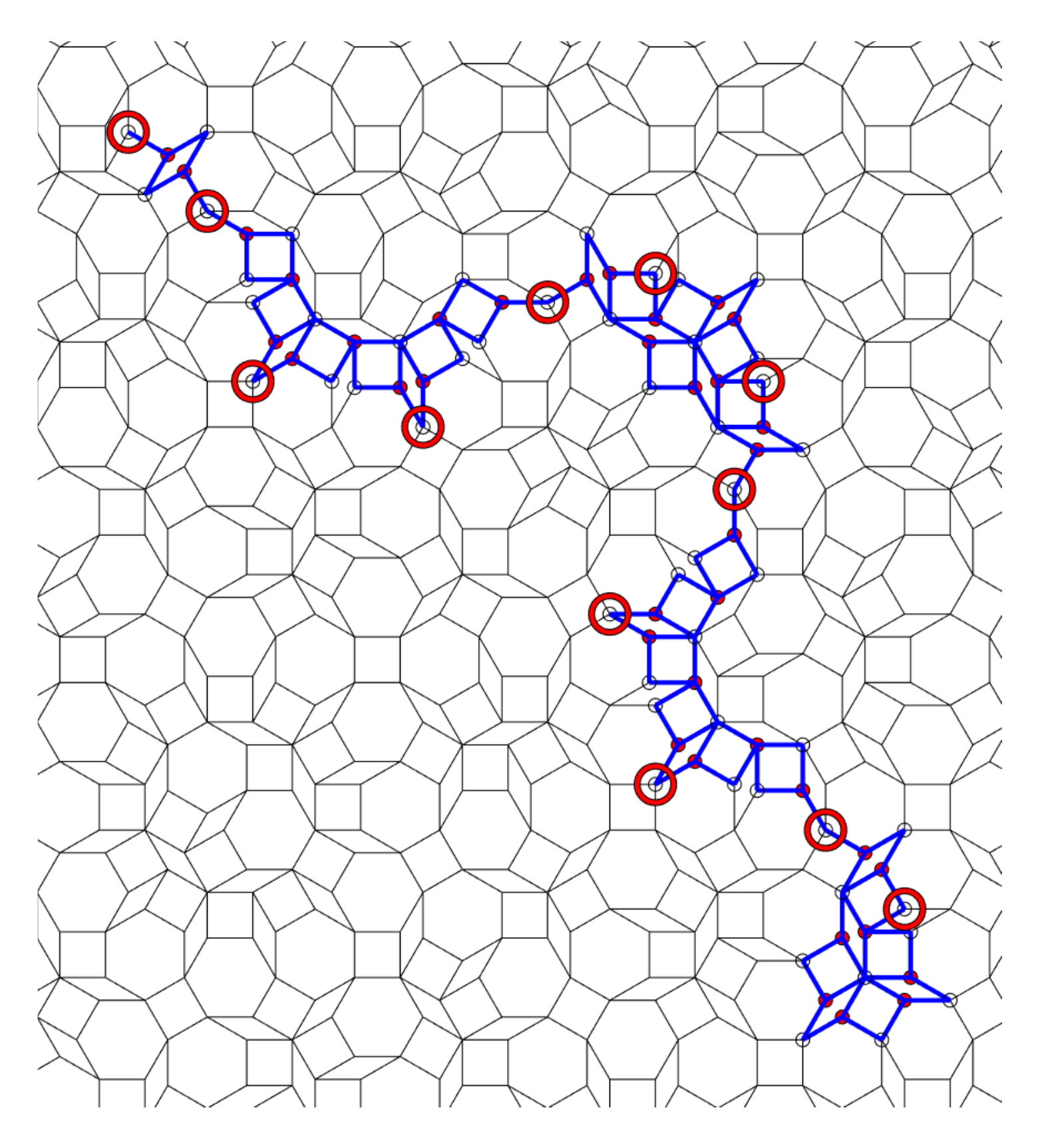}
    \caption{(a)The local configuration forbids the site encircled in red. It is easy to see that $\Psi_1=\Psi_2=\Psi_3$ which makes $\Psi_4=0$. (b) Up to eleven more sites (encircled in red) can be forbidden following the first site.}
    \label{fig:ForbiddenRealSpace1}
\end{figure}
\begin{figure}[!htb]
    \centering
    \includegraphics[clip,width=0.48\textwidth]{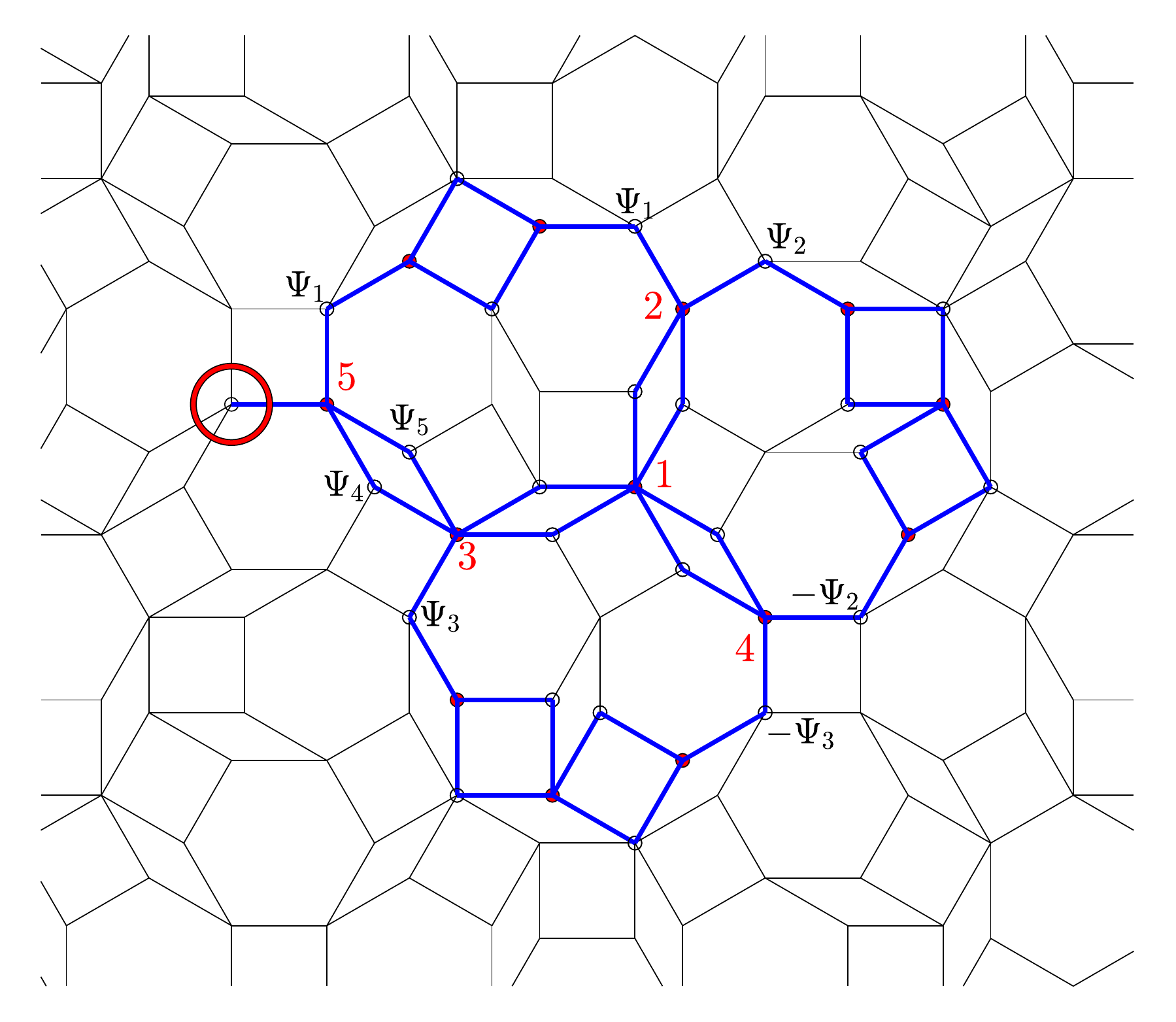}
      \includegraphics[clip,width=0.48\textwidth]{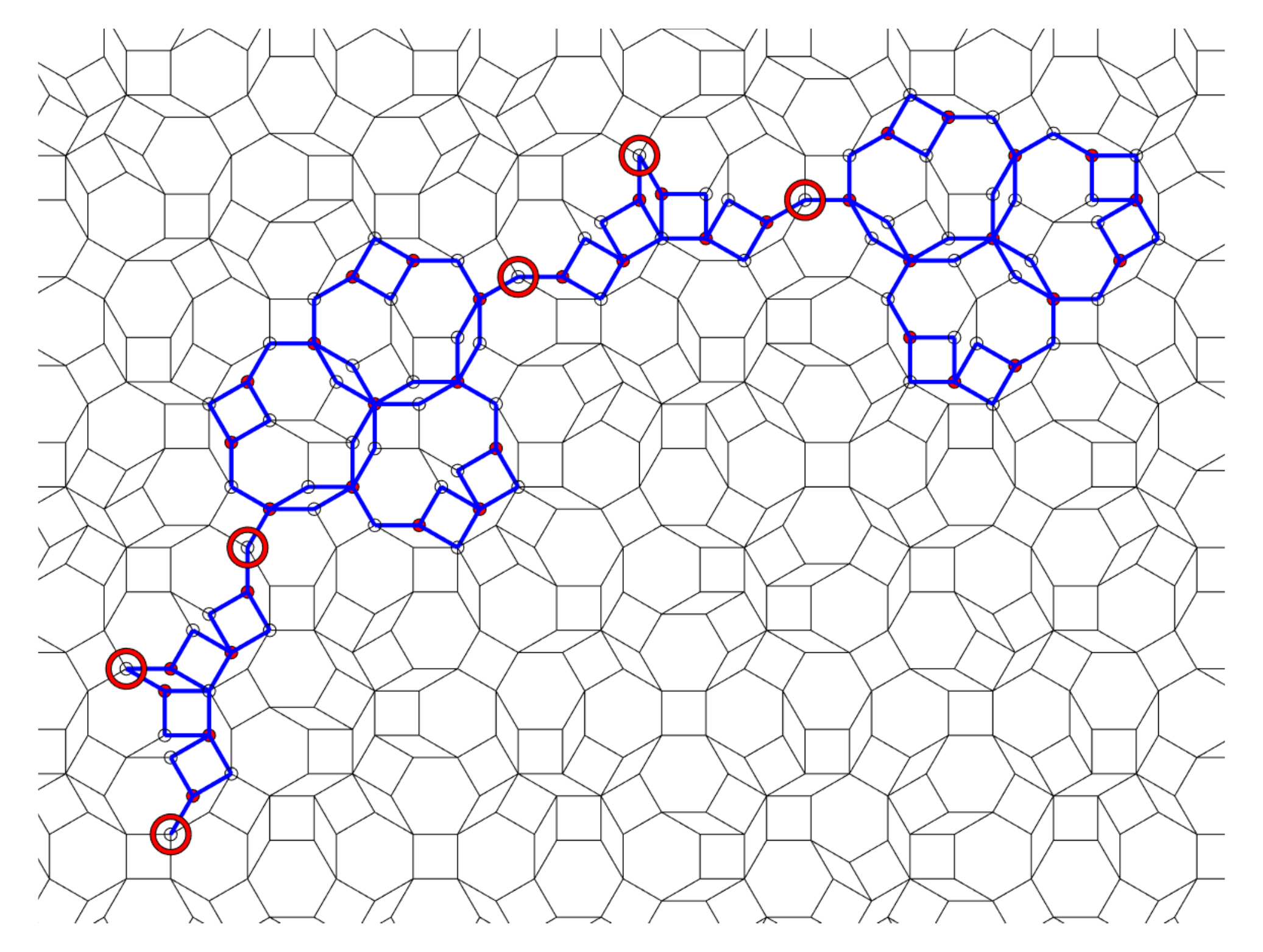}
    \caption{(a) Another local configuration forces the LS wavefunction to vanish on the encircled site. Five more sites (circled in red) in (b) can follow this site as forbidden sites.}
    \label{fig:ForbiddenRealSpace4}
\end{figure}

\begin{figure}[!htb]
    \centering
    \includegraphics[clip,width=0.48\textwidth]{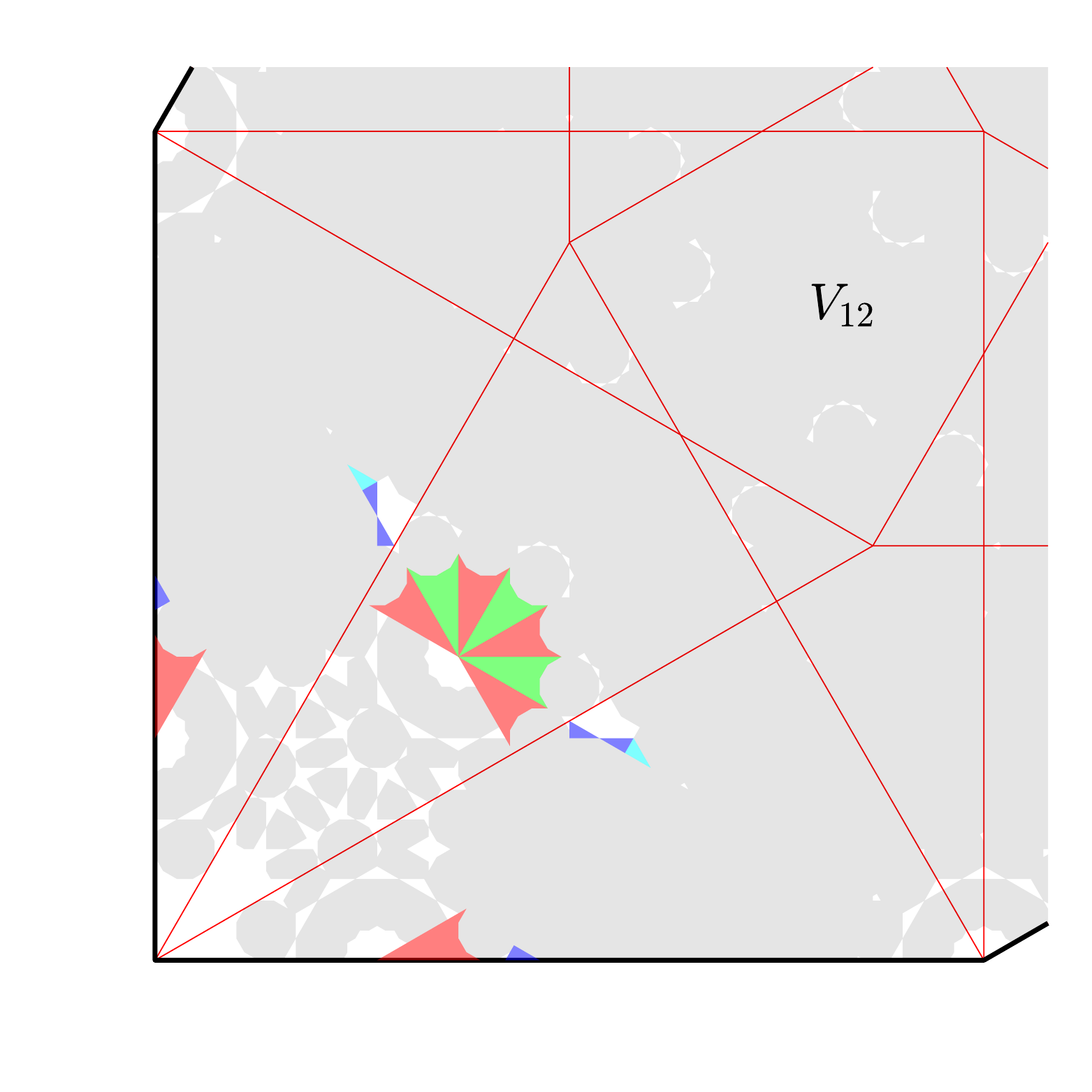}
    \caption{Perpendicular space regions forbidden by the four arguments given in the paper. Only the regions in one corner of $V_{12}$ are shown. Other regions can be deduced by rotation. Red regions follow from the argument in Fig.\ref{fig:ForbiddenRealSpace1}, green regions from Fig.\ref{fig:ForbiddenRealSpace3}. The local configurations causing the blue and cyan regions are given in the appendix.}
    \label{fig:ForbiddenPerpSpace}
\end{figure}
One of the most critical properties of LS on the PL is that there are large regions of the lattice where the LDOS is zero. More specifically, the lattice is split by strings of forbidden sites, and each part has LS only in one of the sublattices \cite{ara88}. Any site in the PL is either a forbidden site or in the support of at least one LS \cite{mok20}. Contrary to the PL, large-scale numerical results for the ABL show no forbidden sites \cite{kog20,okt21}. 

The LDOS picture we obtained for the SDL differs from the PL and the ABL results. First, LDOS of a sublattice is not confined to any regions like the PL but is spread somewhat uniformly throughout the lattice. However, there are isolated sites that have zero LDOS. We identified four ways local connectivity can prevent a site from hosting an LS. A single forbidden site can also forbid a finite string of sites. We give two of these arguments in this section, and two more are in the appendix.

First consider the local environment shown in Fig.\ref{fig:ForbiddenRealSpace1}, an $F $ site in the odd sublattice encircled by 5 $C$ sites and an $E$ site. All the neighbors of the central $F$ are in the even sublattice. Hence, they give an equation relating the wavefunctions of their neighbors. First, consider the equations provided by the sites labeled 1 and 2 in the figure. Subtracting these two equations give us $\Psi_1=\Psi_2$. The same operation on equations 3 and 4 similarly gives $\Psi_1=\Psi_3$. Finally, equations on sites 5 and 6 give $\Psi_2=\Psi_3+\Psi_4$. The only possible solution is $\Psi_4=0$, making this site a forbidden site. 
Any time a site is labeled as forbidden, we look for next-nearest neighbors that must have the same wavefunction due to local connections. There may be a string of attached sites to any forbidden site. There may be up to 11 more linked forbidden sites for the forbidden site identified above, as shown in Fig.\ref{fig:ForbiddenRealSpace1}.

As the second argument, consider the configuration in Fig.\ref{fig:ForbiddenRealSpace4}. First, using logic similar to the above argument, we can subtract equations of sites that share two neighbors to identify three pairs marked with the wavefunctions $\Psi_1,\pm\Psi_2$, and $\pm\Psi_3$. We identify two more sites with the wavefunctions $\Psi_4,\Psi_5$. Considering the sum of equations 2,3, and 4 and subtracting the equation from the central site 1, we obtain $\Psi_1+\Psi_4+\Psi_5=0$. A new forbidden site can be identified when this equation is combined with the equation on site 5. There may be six attached sites to this forbidden site, as shown in \ref{fig:ForbiddenRealSpace4}.

Two more independent arguments for forbidden sites are given in appendix B. Once the real space structure of a forbidden site is known, it is possible to count the frequency of their occurrence by calculating the acceptance domains for the sites that give the necessary equations. We do this for all the four arguments and their attached sites. The forbidden perpendicular space regions resulting from all four arguments are shown in Fig.\ref{fig:ForbiddenPerpSpace}. The total fraction of sites forbidden through the four arguments is  $f_{\mathrm{Forbid}}\simeq0.038955$.

\begin{figure}[!htb]
    \centering
    \includegraphics[clip,width=0.48\textwidth]{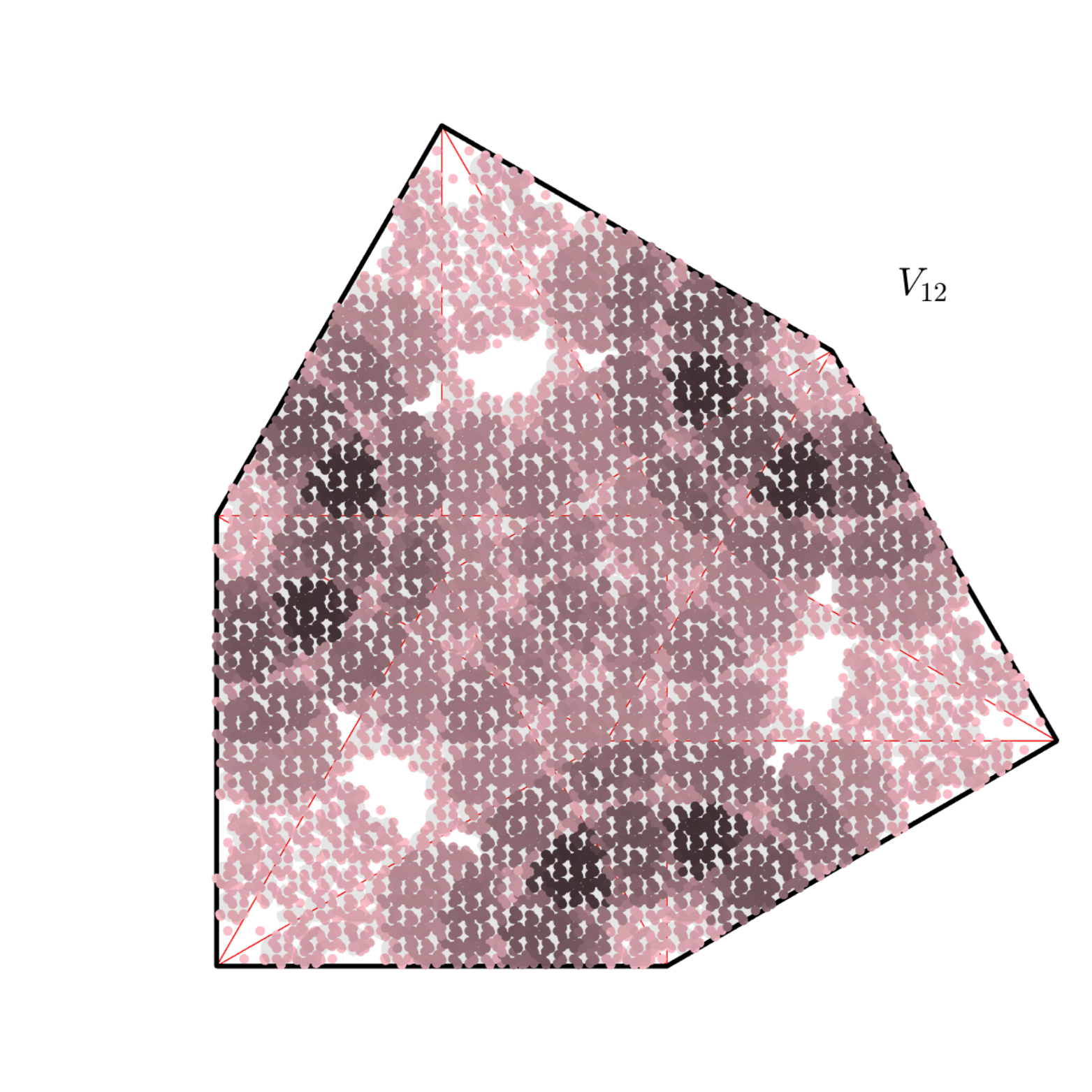}
    \caption{LDOS plotted in perpendicular space. Finite neighborhoods with a depth of 80 are used, so perpendicular space is not uniformly sampled. Notice that most of the empty regions correspond to forbidden sites. Also, LDOS follows the outline of the LS type allowed areas.}
    \label{fig:Fig05_PerpSpaceLDOS}
\end{figure}

We identified that large regions of perpendicular space belong to the support of LS and a much smaller area corresponds to the forbidden sites. Still, there are regions in perpendicular space that are not identified as forbidden or allowed for an LS vertex. Repeating the LDOS calculation in perpendicular space gives us at least a suggestion about the properties of the unidentified regions. The result of such a calculation on 80-deep neighborhoods of 5 random starting perpendicular space points is given in Fig.\ref{fig:Fig05_PerpSpaceLDOS}. Identifying small regions in perpendicular space is challenging as no neighborhood uniformly samples the perpendicular space.
Nonetheless, we expect that all F, D, and E vertices are in the support of some LS. There seem to be forbidden regions of C vertices next to the identified forbidden areas. We cannot be sure about the areas close to the tips of the C and the B regions; LS may cover those regions entirely. Overall, the lower bound we provide for the LS fraction and the identified forbidden sites seem to give a reasonably comprehensive picture of the perpendicular space.

\section{Conclusion}
\label{sec:Conclusion}

Socolar dodecagonal lattice is a quasicrystal closely related to the ABL and PL. A recent paper\cite{kog21} considered the vertex tight-binding model on this lattice and numerically found that $f_{\mathrm{Num}}\simeq0.076$ of the states are in the zero-energy manifold. The same paper claims that considering a small family of LS types does not adequately explain the observed LS fraction. We study the same model and used a scaling argument to find the LS fraction using smaller lattices with open boundary conditions. Our result agrees with Ref.\cite{kog21} for the LS fraction of $f_{\mathrm{Num}}\simeq0.0761$. However, we find different results for the frequencies of the LS types. We believe our method based on perpendicular space images of LS types is less error-prone than methods based on inflation-deflation. We find 18 LS types, which provide a lower bound  $f_{\mathrm{LS}}=\frac{10919-6304\sqrt{3}}{2}\simeq0.07585$ for the LS fraction.  

Beyond finding the LS types that explain the $99.7 \%$ of the numerical value, we also calculate the zero energy LDOS for this lattice. We find that zero-energy LDOS is non-zero throughout most of the lattice, yet we also observe that some sites have zero LDOS. We show that some local structures in the SDL prohibit specific sites from hosting LS. We give four arguments leading to forbidden sites and using their perpendicular space images, and we find that at least $f_{\mathrm{Forbid }}\simeq0.038955$ of the sites cannot host any LS. 

The LS type structure of the SDL is more similar to the ABL than the PL. All LS types found for the SDL and the ABL have wavefunctions of constant density and alternating sign. Both lattices require extensive LS types with very low frequency, while just six LS types span PL zero energy manifold. In a forthcoming paper, we also study the robustness of the zero-energy manifold with respect to an applied magnetic field. The SDL loses most of its zero energy LS with an applied field similar to the ABL and in direct contrast with the PL.
One property of SDL closer to the PL than the ABL is the presence of forbidden sites. 

All three lattices, the ABL, the PL, and the SDL, are obtained by a similar projection procedure from four, five, and six-dimensional simple cubic lattices. All three have perpendicular spaces that reduce to two-dimensional polygons and have scaling symmetries. Our results indicate that the zero energy manifold of the SDL and the ABL may be more generic compared to the PL. However, we cannot pinpoint what property of the quasicrystal controls the frequency, robustness, or the required number of LS types. Similarly, we cannot predict the existence or frequency of forbidden sites beyond finding specific instances of local environments. We hope these questions will stimulate further research into elementary excitations in quasicrystals.

\appendix*
\section{LS types with low frequency }

\begin{figure}[!htb]
    \centering
    \includegraphics[trim=8mm 8mm 8mm 8mm,clip,width=0.48\textwidth]{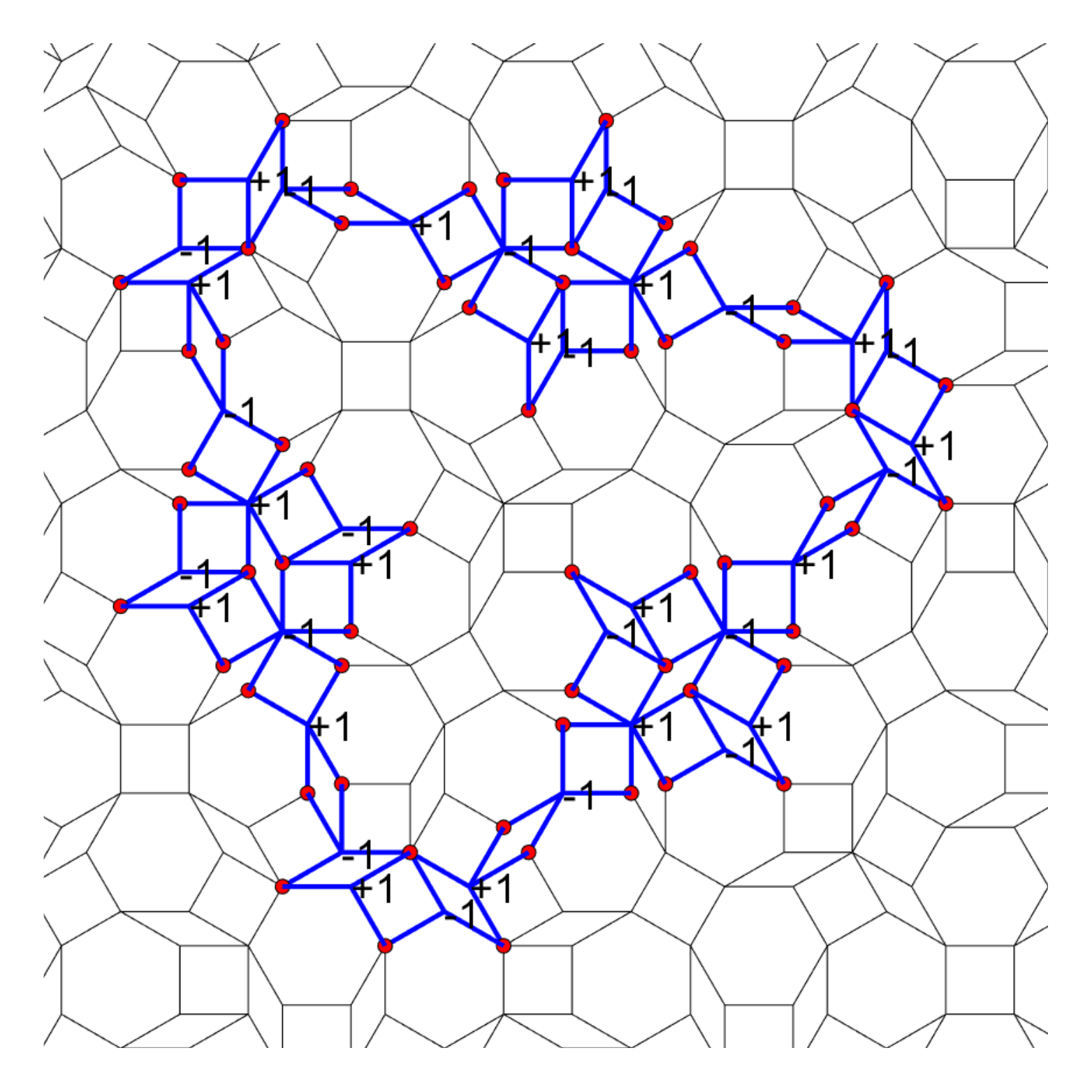}
    \includegraphics[trim=8mm 8mm 8mm 8mm,clip,width=0.48\textwidth]{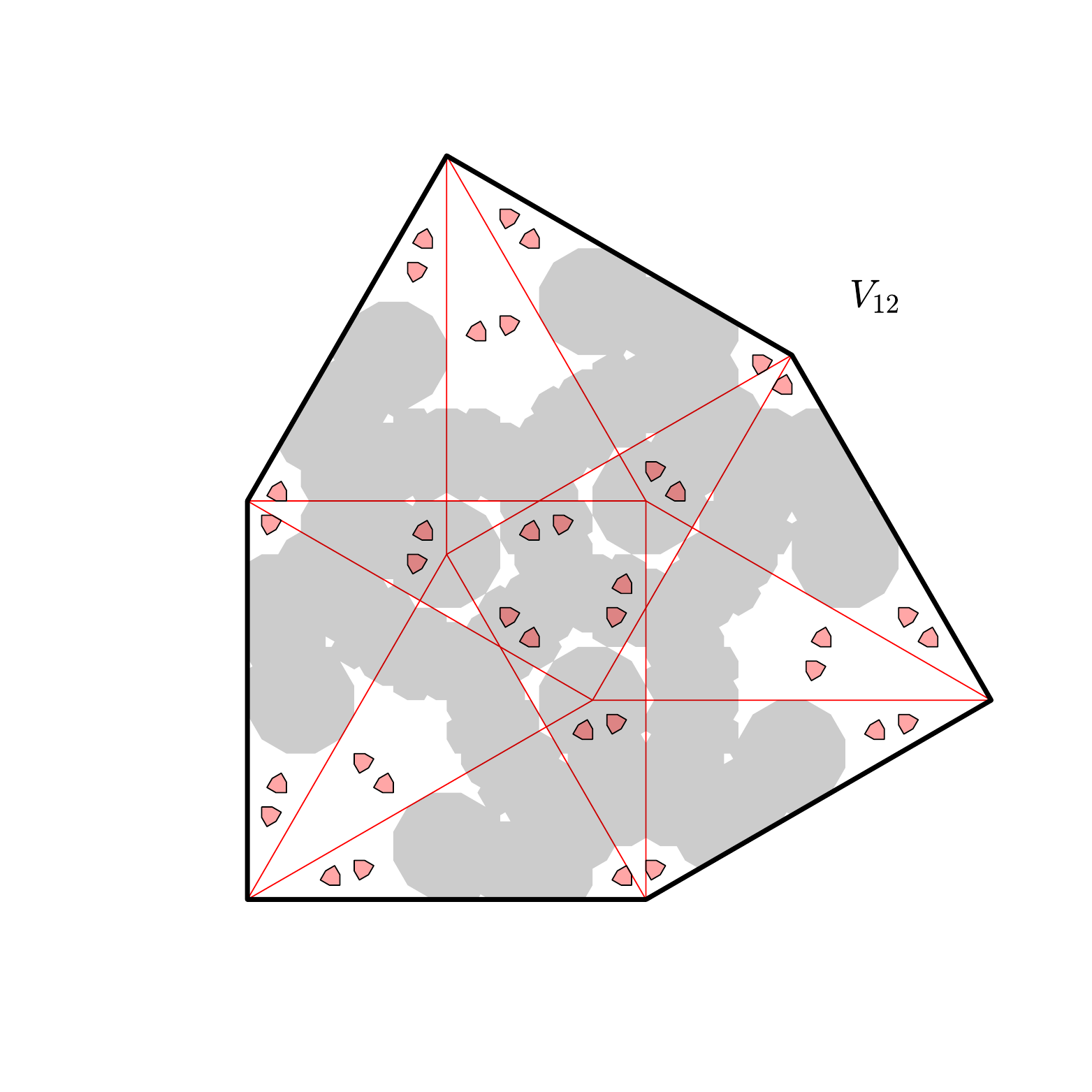}
    \caption{LS type-A3. The allowed region for each vertex is a hexagon similar to $V_{ij}$, but scaled with $\sqrt{\frac{\xi^{-5}}{2}}$. The frequency of this LS type is $\frac{\xi^{-5}}{2}\simeq0.00069$ }
    \label{fig:TypeA3_RealSpace}
\end{figure}

We give the real space configurations of ten LS types in addition to the eight in the main text. For each type in Fig.\ref{fig:TypeA3_RealSpace},...,\ref{fig:TypeD6_RealSpace}, we plot the real space configuration and the allowed areas for each vertex in $V_{12}.$ The areas covered by all previous LS types are indicated as a gray background so that the independence of the new LS type can be visually established.

\begin{figure}[!htb]
    \centering
    \includegraphics[trim=8mm 8mm 8mm 8mm,clip,width=0.48\textwidth]{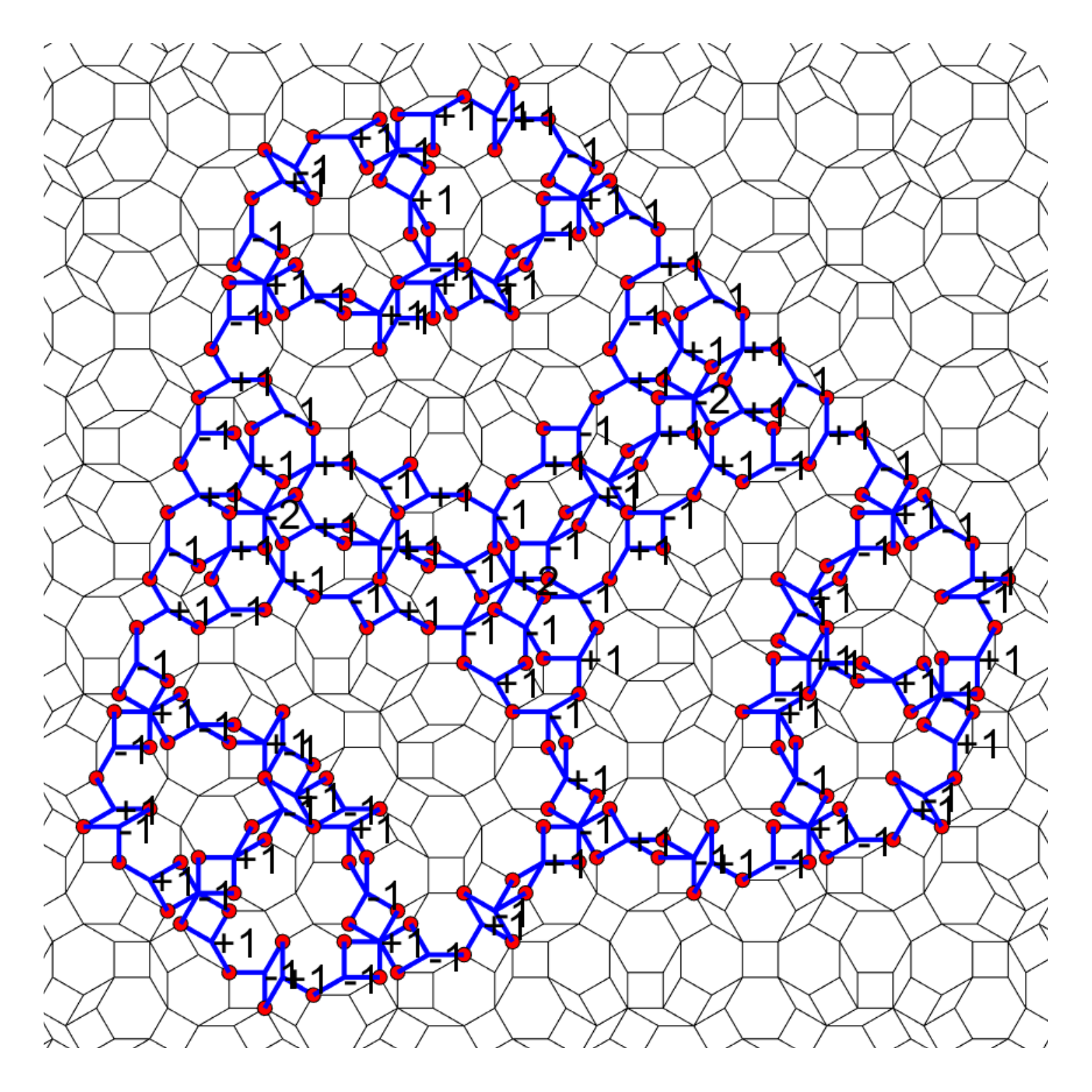}
    \includegraphics[trim=8mm 8mm 8mm 8mm,clip,width=0.48\textwidth]{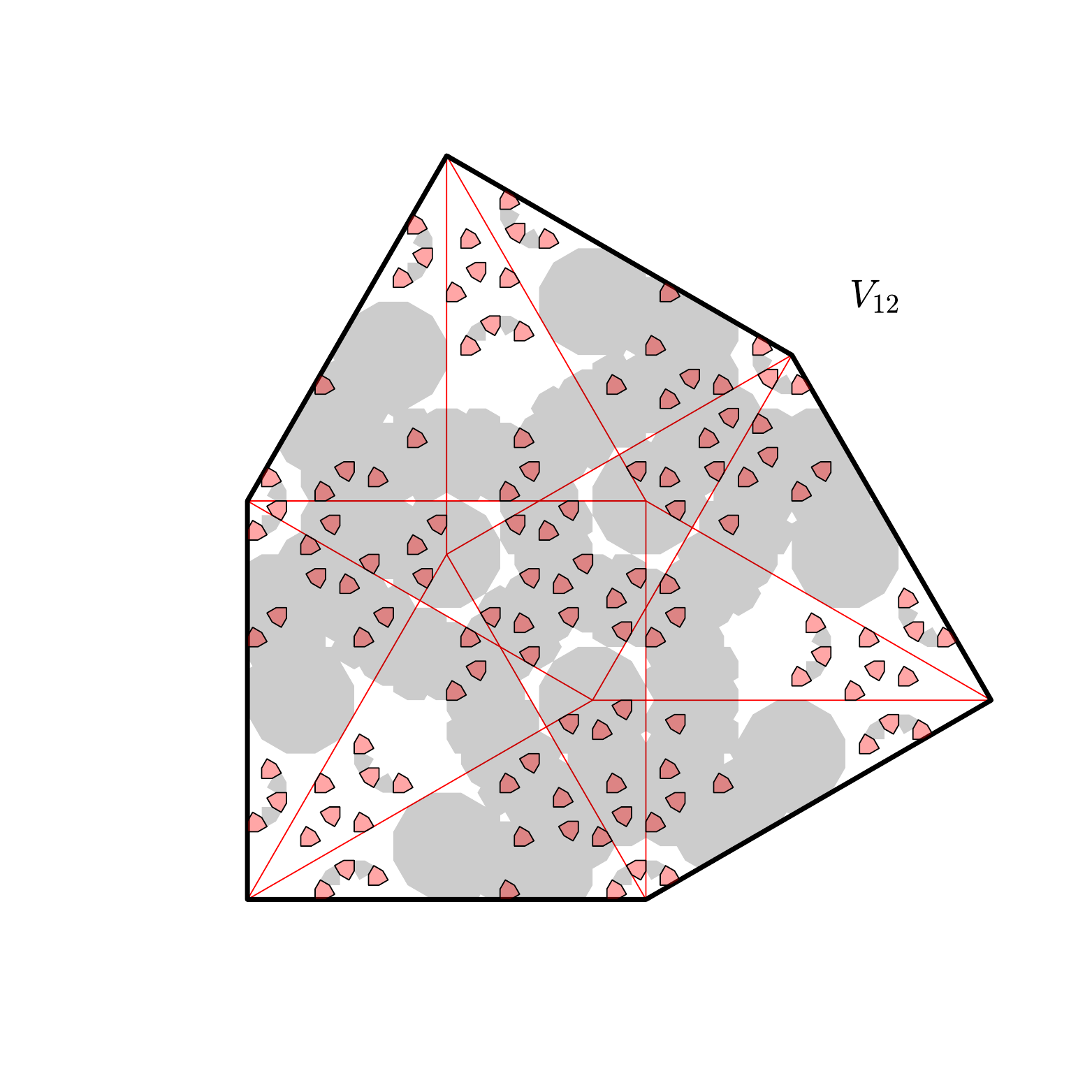}
    \caption{LS type-A4. The allowed region area and frequency is the same as type-A3 $f_{A4}\simeq0.00069$ although there are many more vertices in the support. }
    \label{fig:TypeA4_RealSpace}
\end{figure}
\begin{figure}[!htb]
    \centering
    \includegraphics[trim=8mm 8mm 8mm 8mm,clip,width=0.48\textwidth]{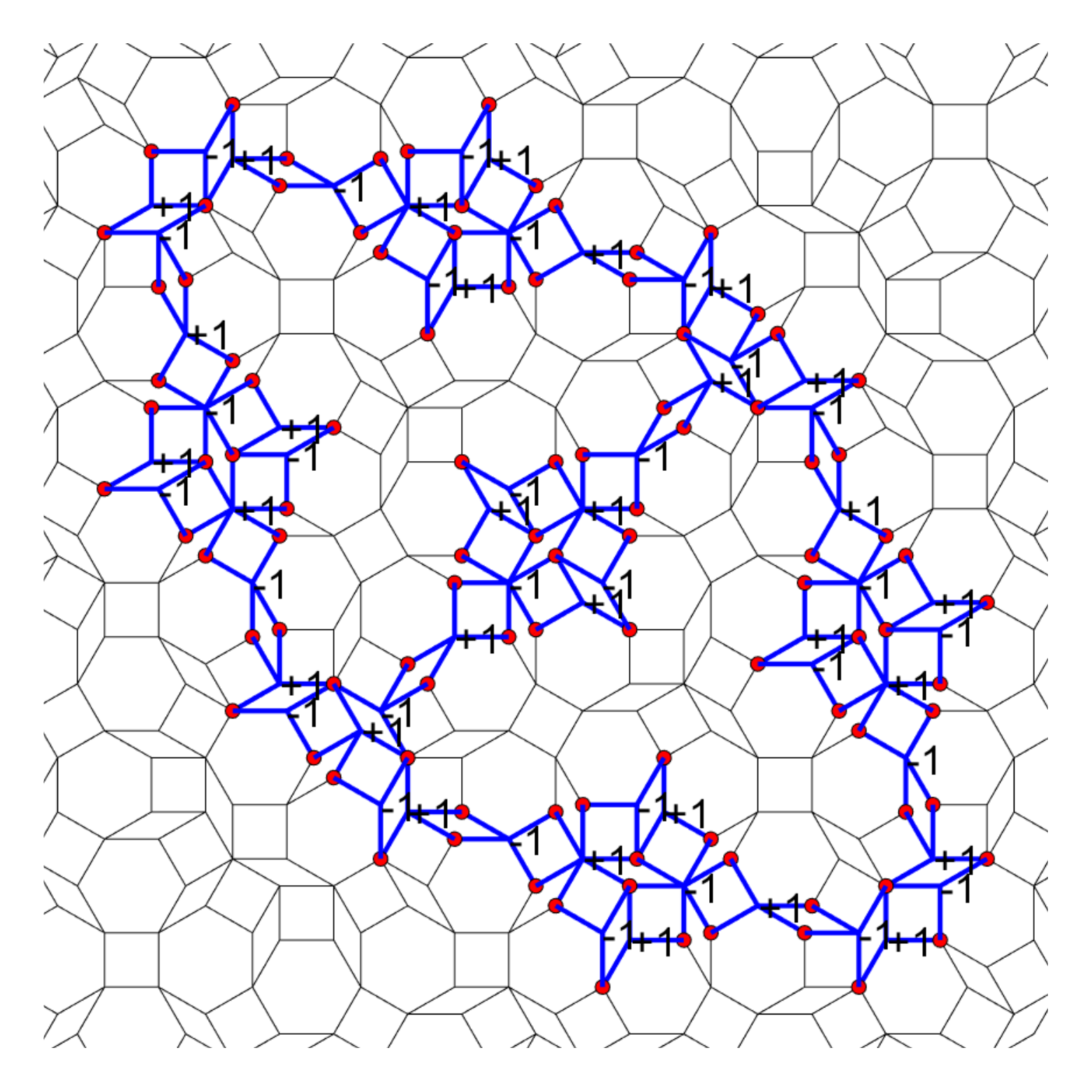}
    \includegraphics[trim=8mm 8mm 8mm 8mm,clip,width=0.48\textwidth]{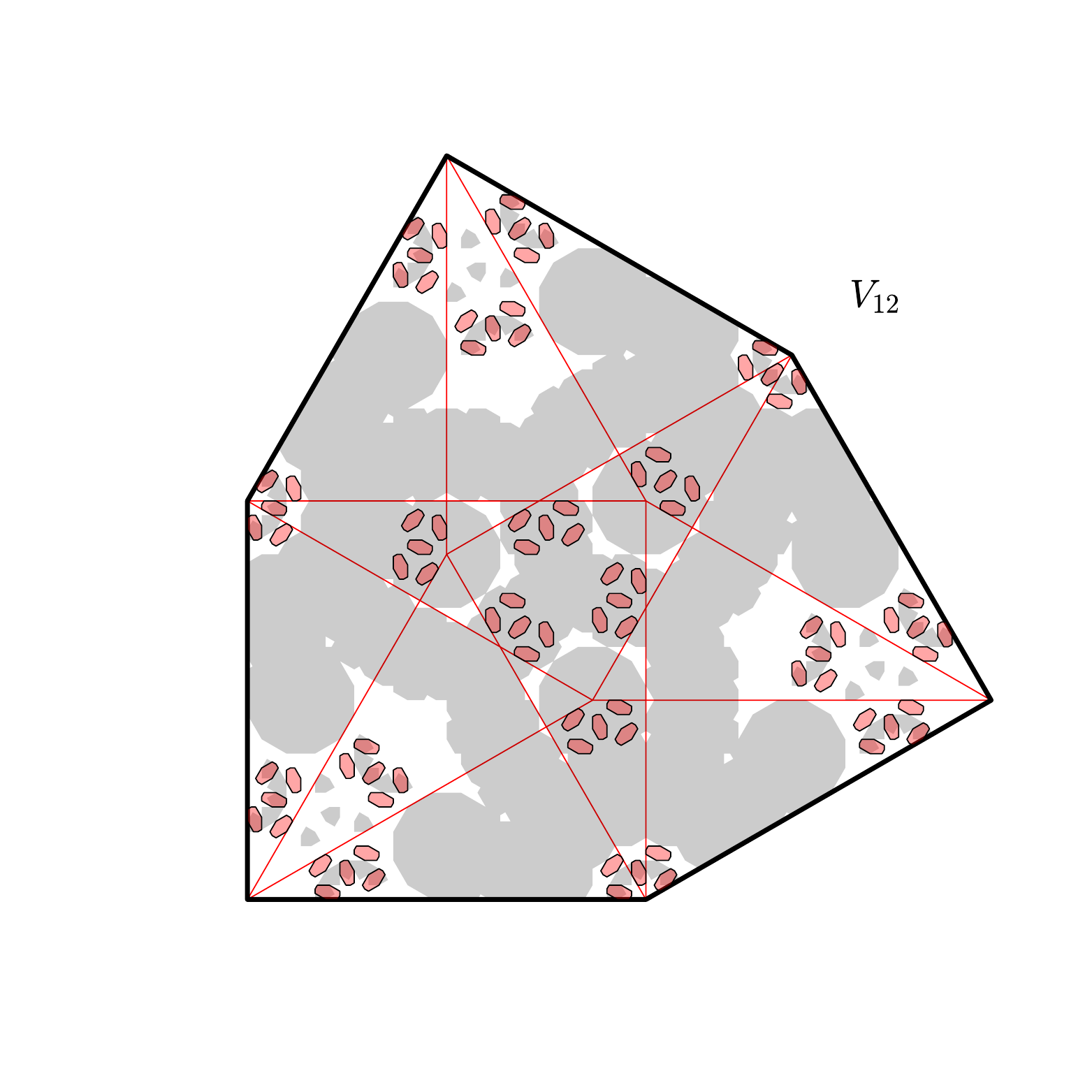}
    \caption{LS type-B3. The allowed region for each vertex is an irregular hexagon with $\pi/2$ rotational symmetry. Total frequency for this type is $f_{B3}=\frac{3\xi^{-5}+\xi^{-6}}{4}\simeq0.00113$}
    \label{fig:TypeB3_RealSpace}
\end{figure}
\begin{figure}[!htb]
    \centering
    \includegraphics[trim=8mm 8mm 8mm 8mm,clip,width=0.48\textwidth]{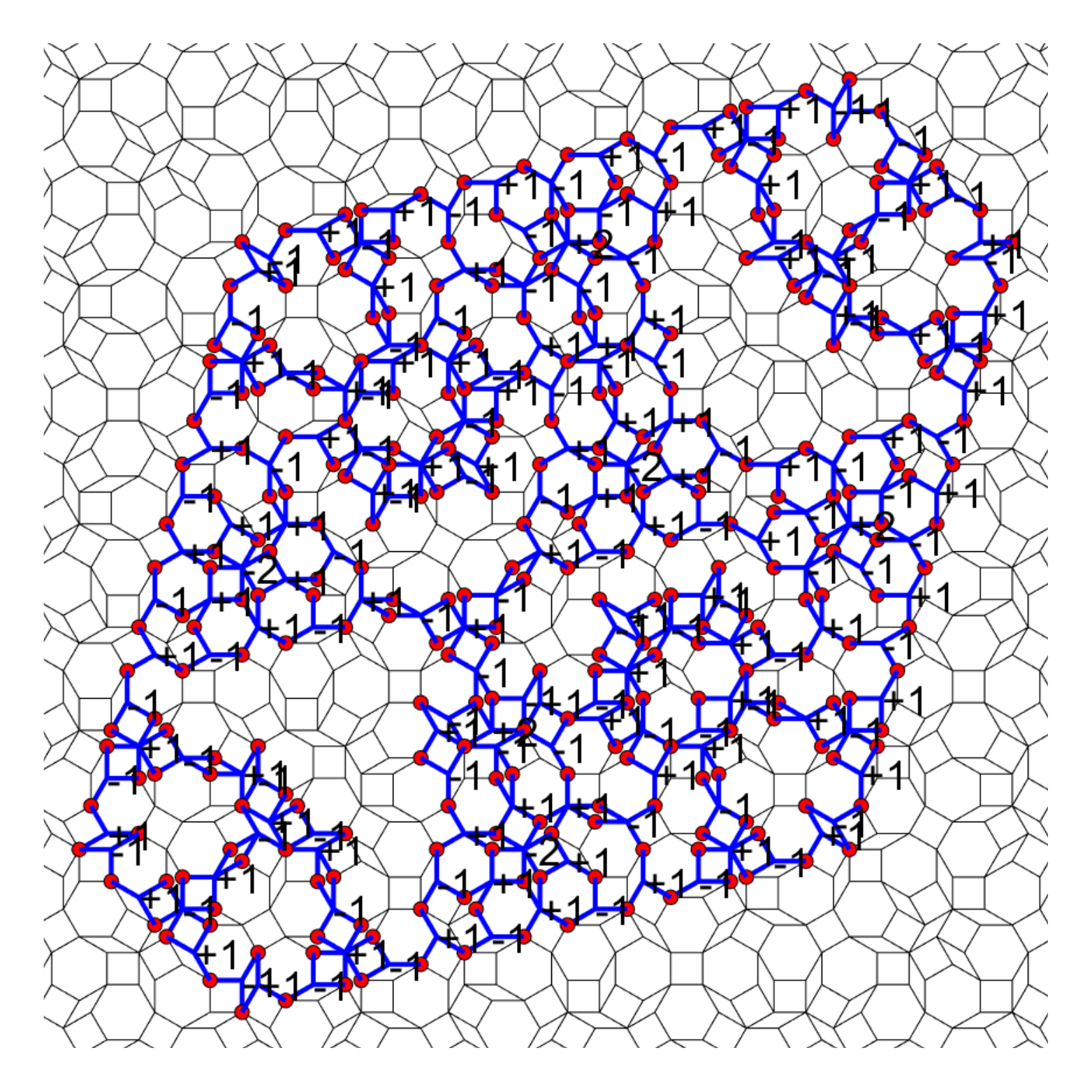}
    \includegraphics[trim=8mm 8mm 8mm 8mm,clip,width=0.48\textwidth]{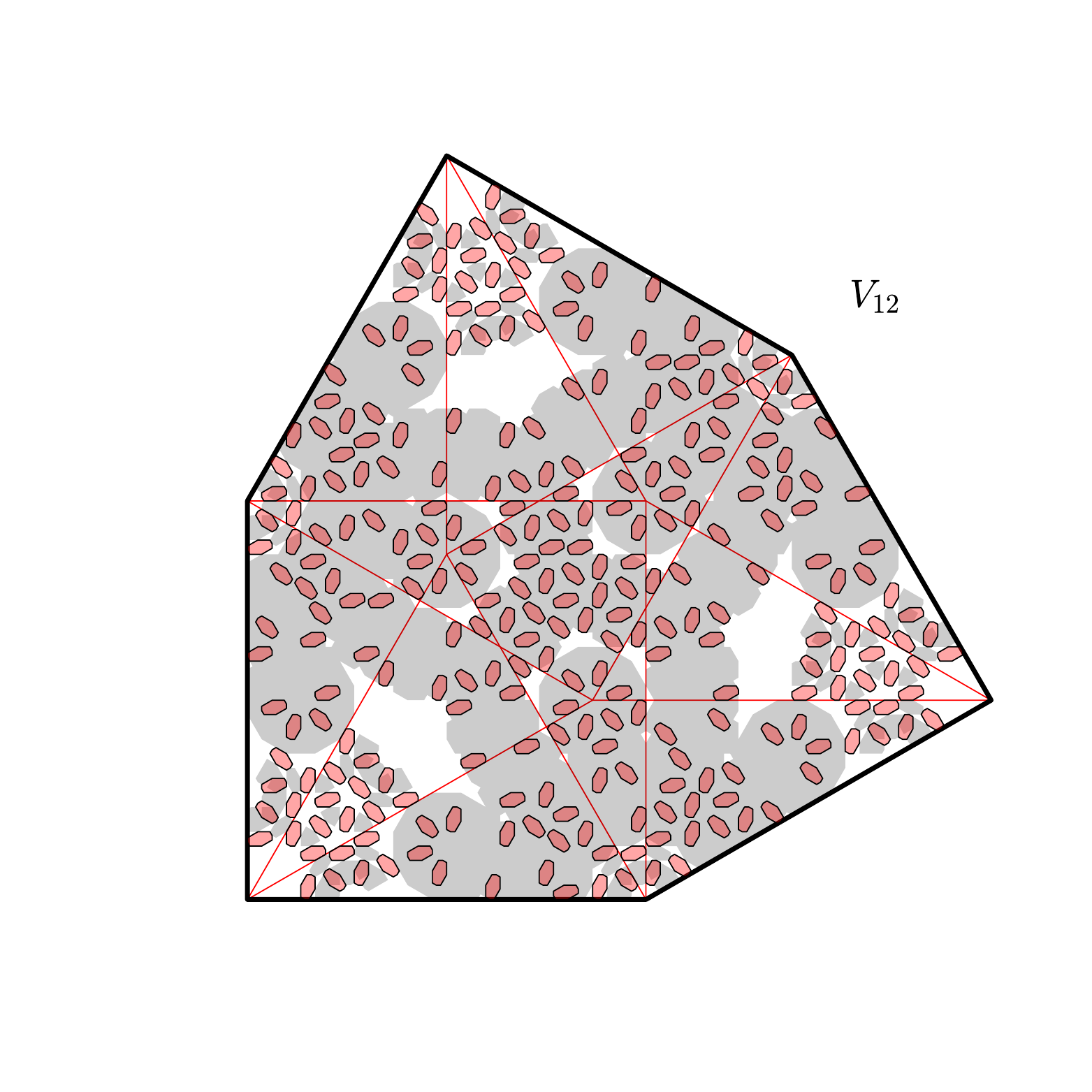}
    \caption{LS type-B4 has the same frequency as type-B3 $f_{B4}\simeq0.00113$.}
    \label{fig:TypeB4_RealSpace}
\end{figure}
\begin{figure}[!htb]
    \centering
    \includegraphics[trim=8mm 8mm 8mm 8mm,clip,width=0.48\textwidth]{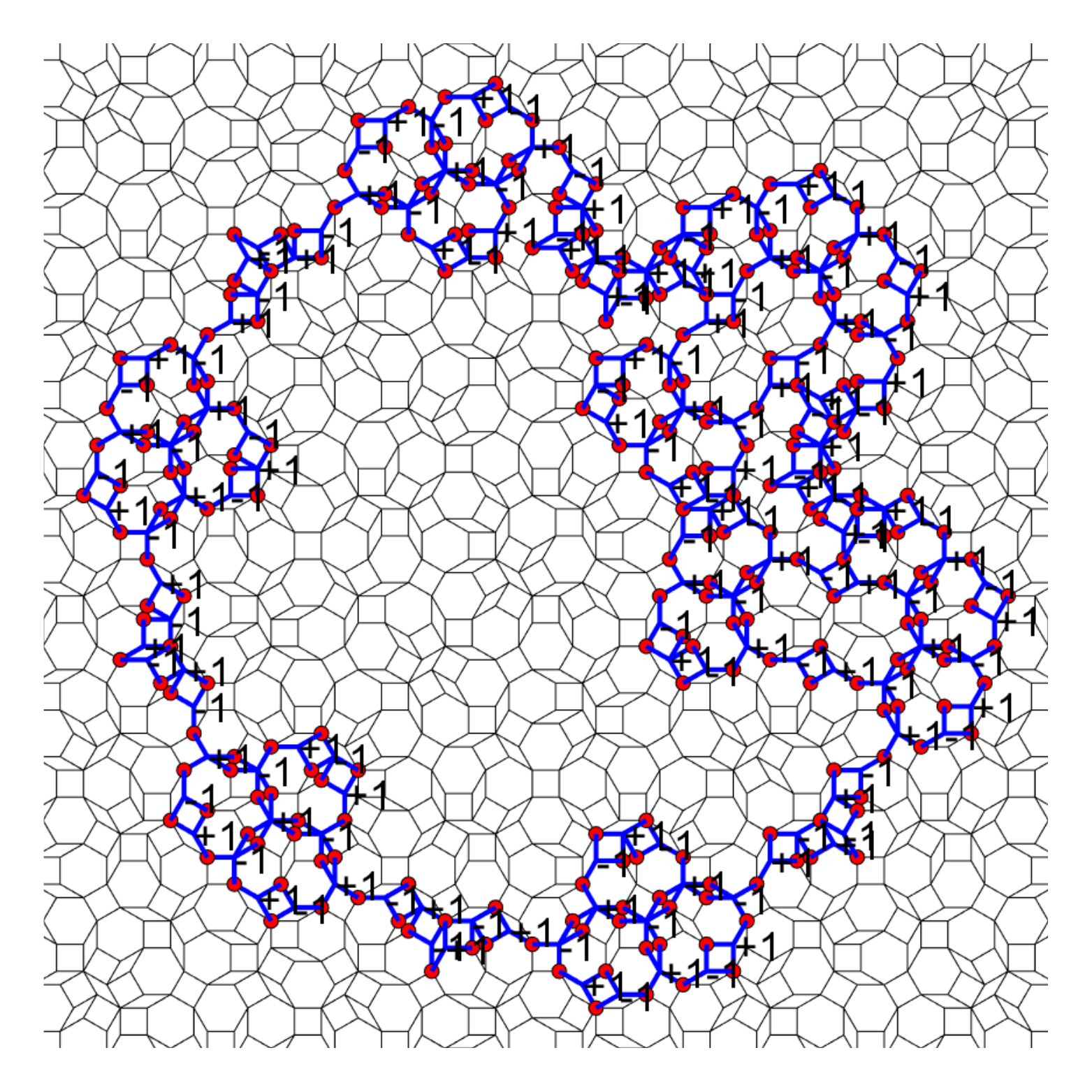}
    \includegraphics[trim=8mm 8mm 8mm 8mm,clip,width=0.48\textwidth]{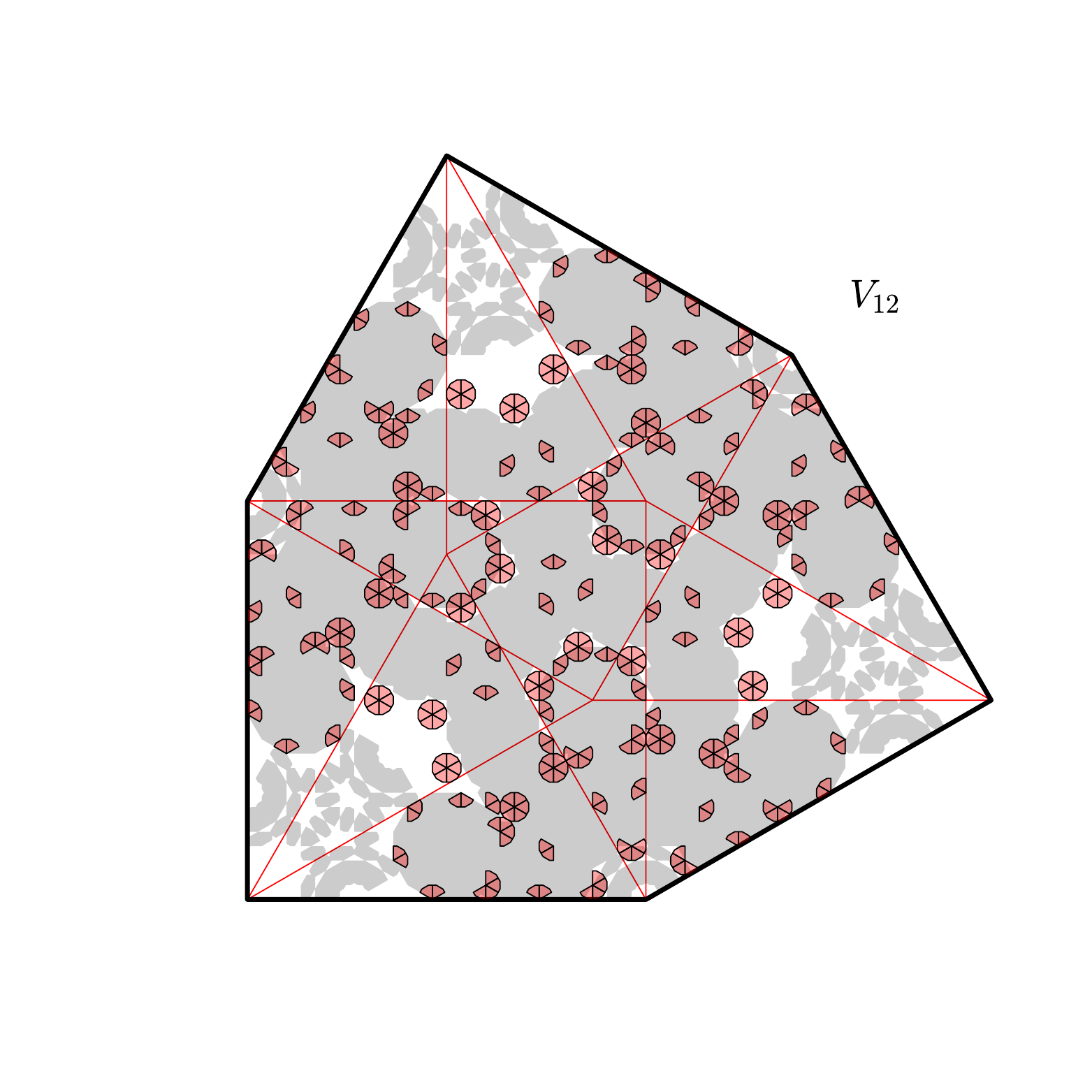}
    \caption{LS type-C5 has an allowed region which is an irregular pentagon with area $\frac{3\xi^{-5}}{2}\simeq0.00207$. The frequency is $\frac{\xi^{-5}+\xi^{-6}}{2}\simeq 0.00088$.}
    \label{fig:TypeD1_RealSpace}
\end{figure}
\begin{figure}[!htb]
    \centering
    \includegraphics[trim=8mm 8mm 8mm 8mm,clip,width=0.48\textwidth]{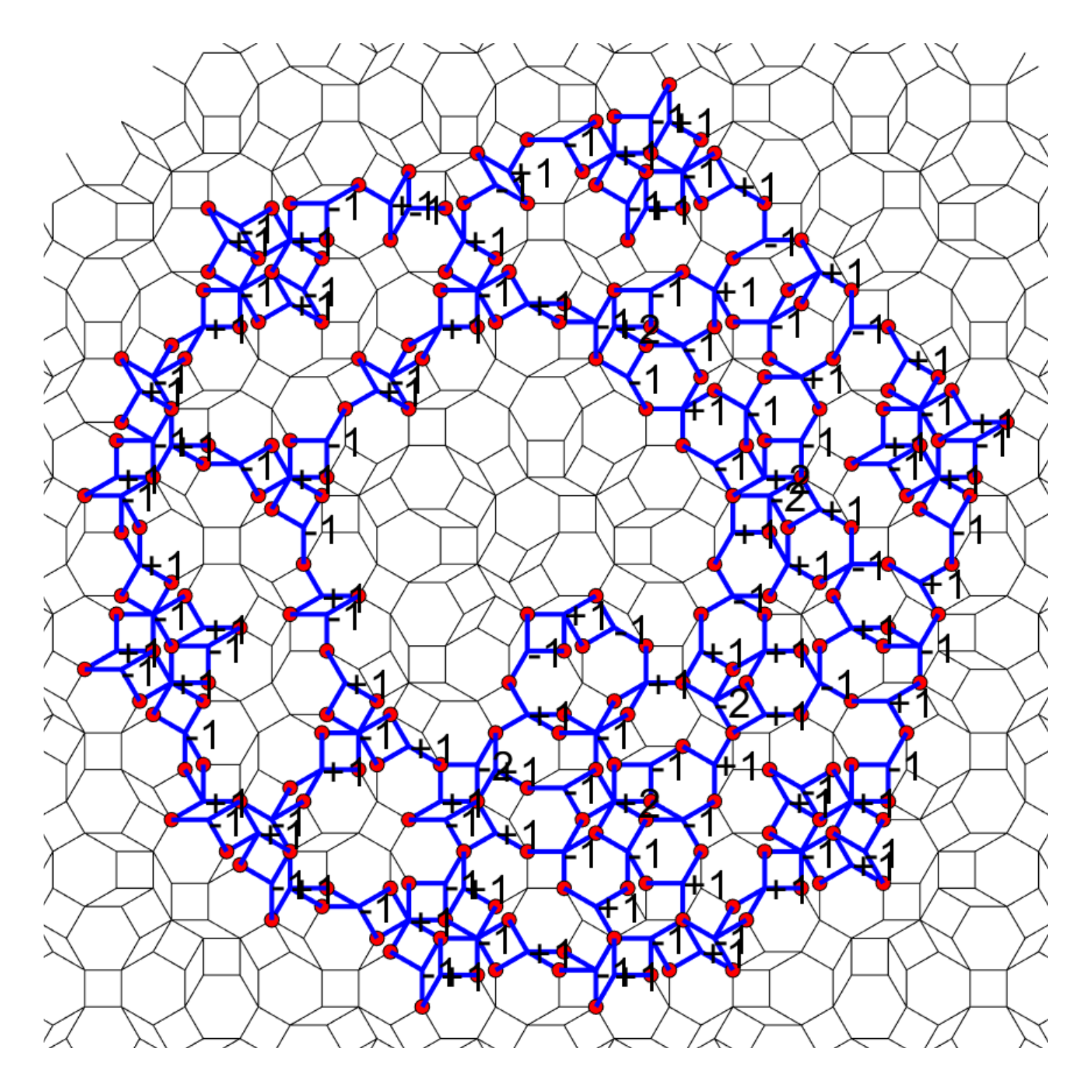}
    \includegraphics[trim=8mm 8mm 8mm 8mm,clip,width=0.48\textwidth]{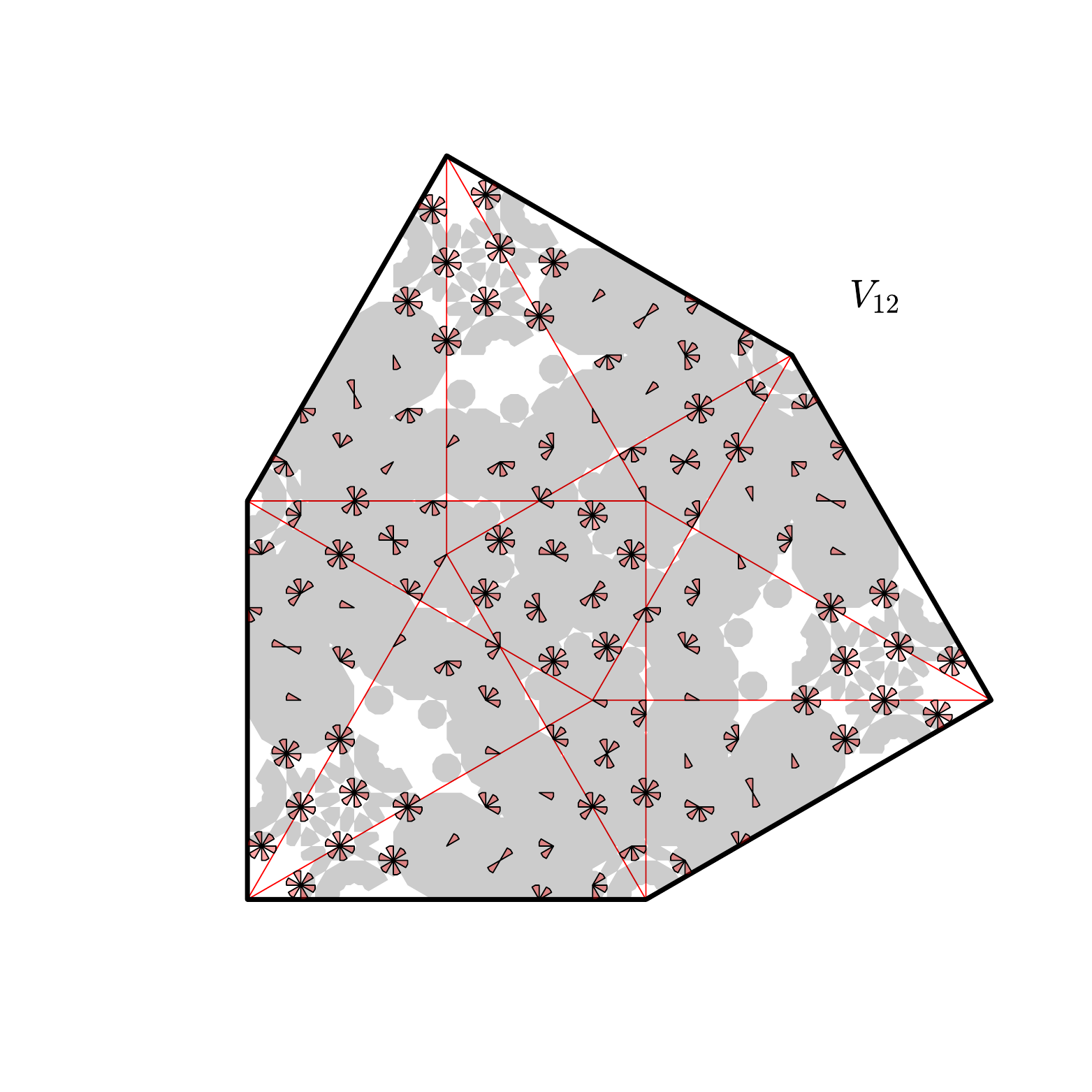}
    \caption{LS type-C6 has a frequency of $\frac{\xi^{-5}+\xi^{-6}}{4}\simeq 0.00044$.}
    \label{fig:TypeD2_RealSpace}
\end{figure}
\begin{figure}[!htb]
    \centering
    \includegraphics[trim=8mm 8mm 8mm 8mm,clip,width=0.48\textwidth]{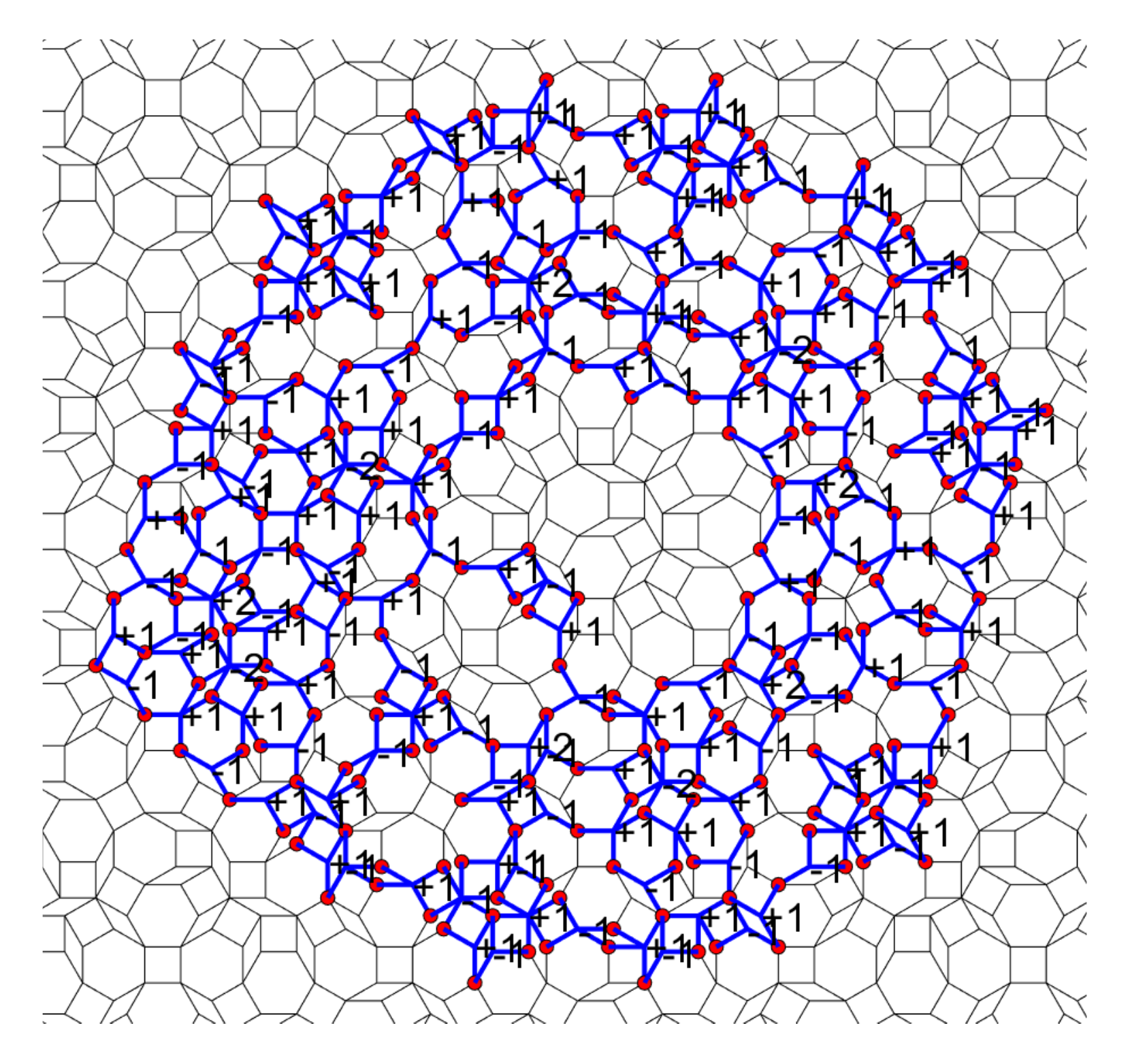}
    \includegraphics[trim=8mm 8mm 8mm 8mm,clip,width=0.48\textwidth]{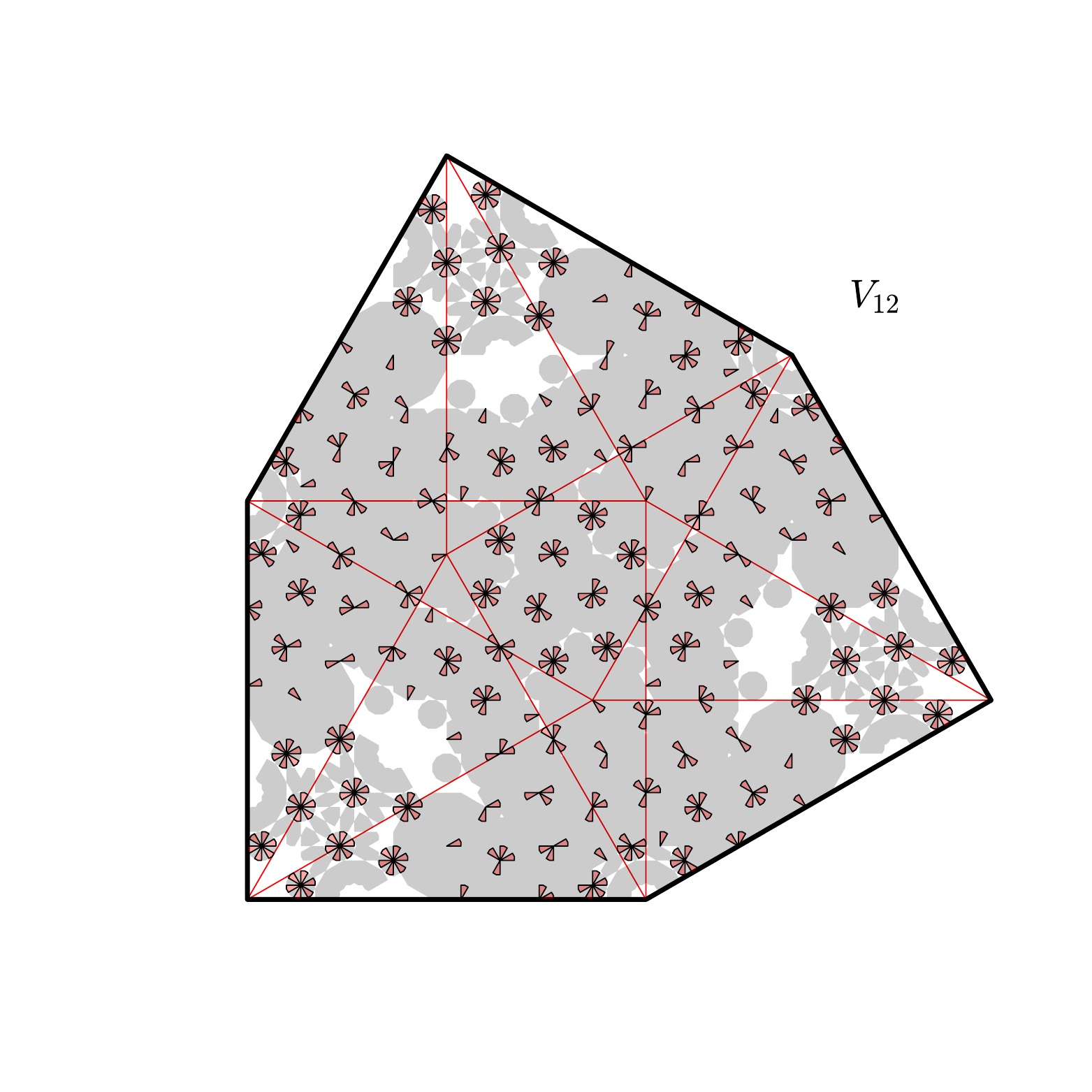}
    \caption{LS type-C7 fills perpendicular regions adjacent to type-C6. Its frequency is also the same with type-C6.}
    \label{fig:TypeD3_RealSpace}
\end{figure}
\begin{figure}[!htb]
    \centering
    \includegraphics[trim=8mm 8mm 8mm 8mm,clip,width=0.48\textwidth]{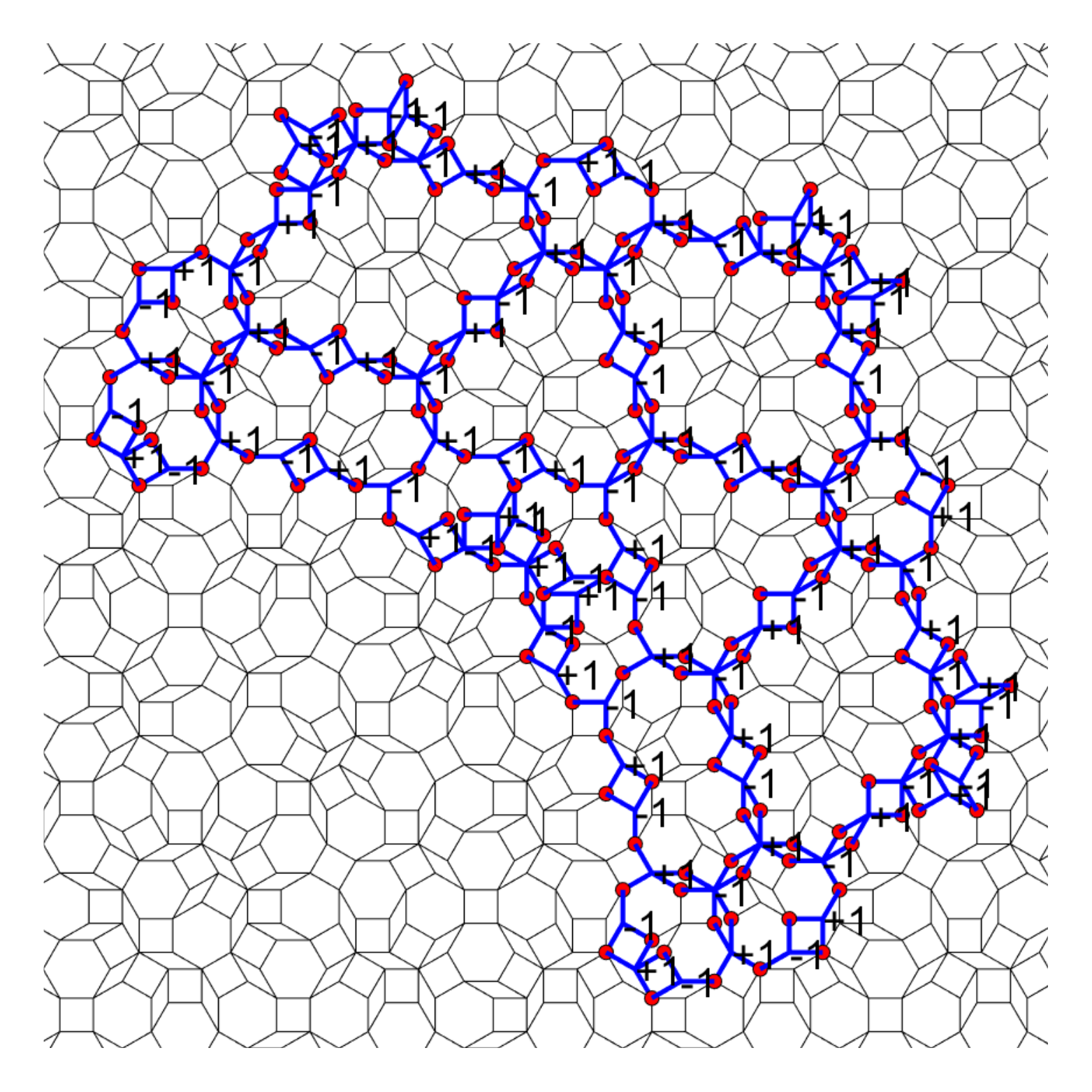}
    \includegraphics[trim=8mm 8mm 8mm 8mm,clip,width=0.48\textwidth]{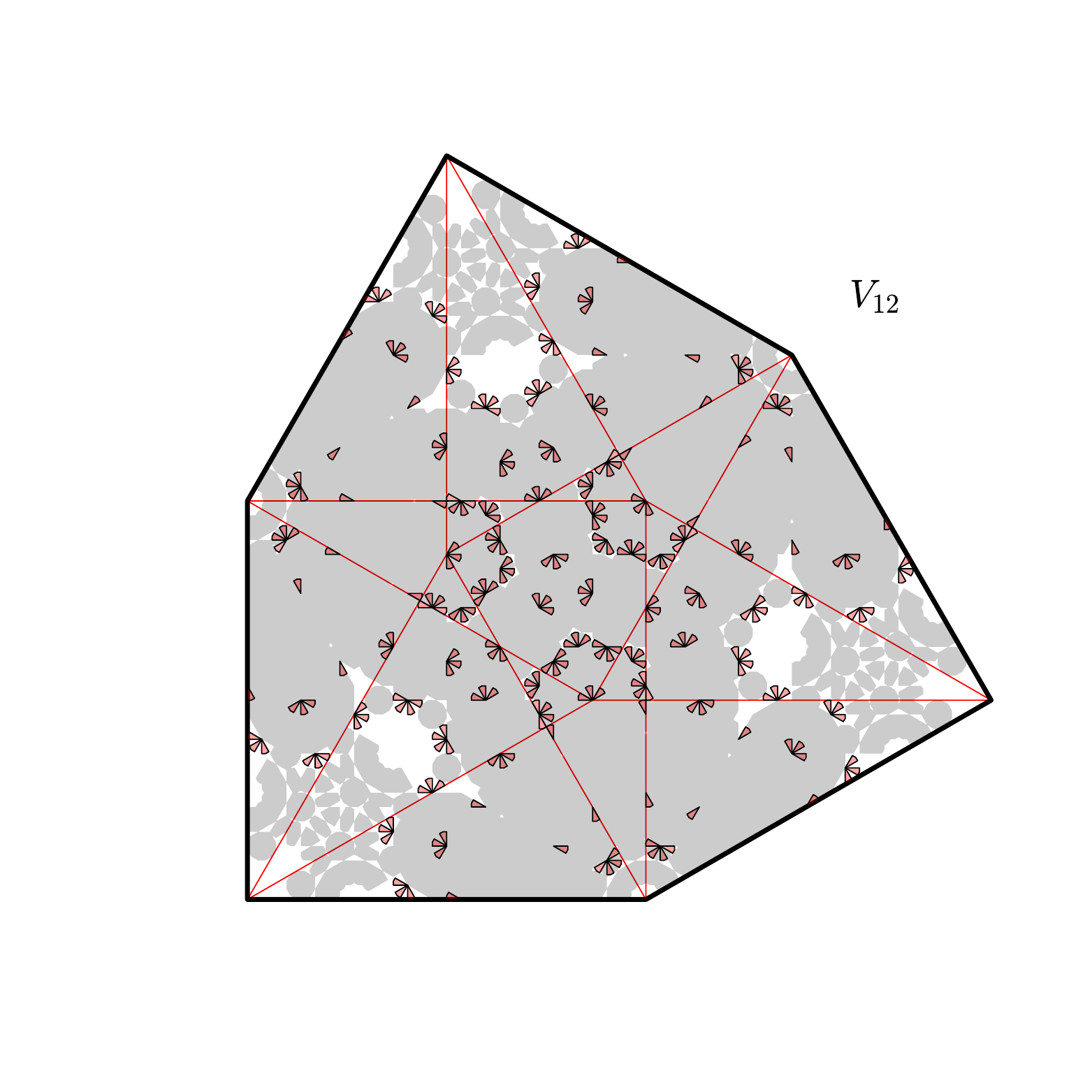}
    \caption{LS type-C8 has the same allowed region area and frequency as type-C6.}
    \label{fig:TypeD4_RealSpace}
\end{figure}
\begin{figure}[!htb]
    \centering
    \includegraphics[trim=8mm 8mm 8mm 8mm,clip,width=0.48\textwidth]{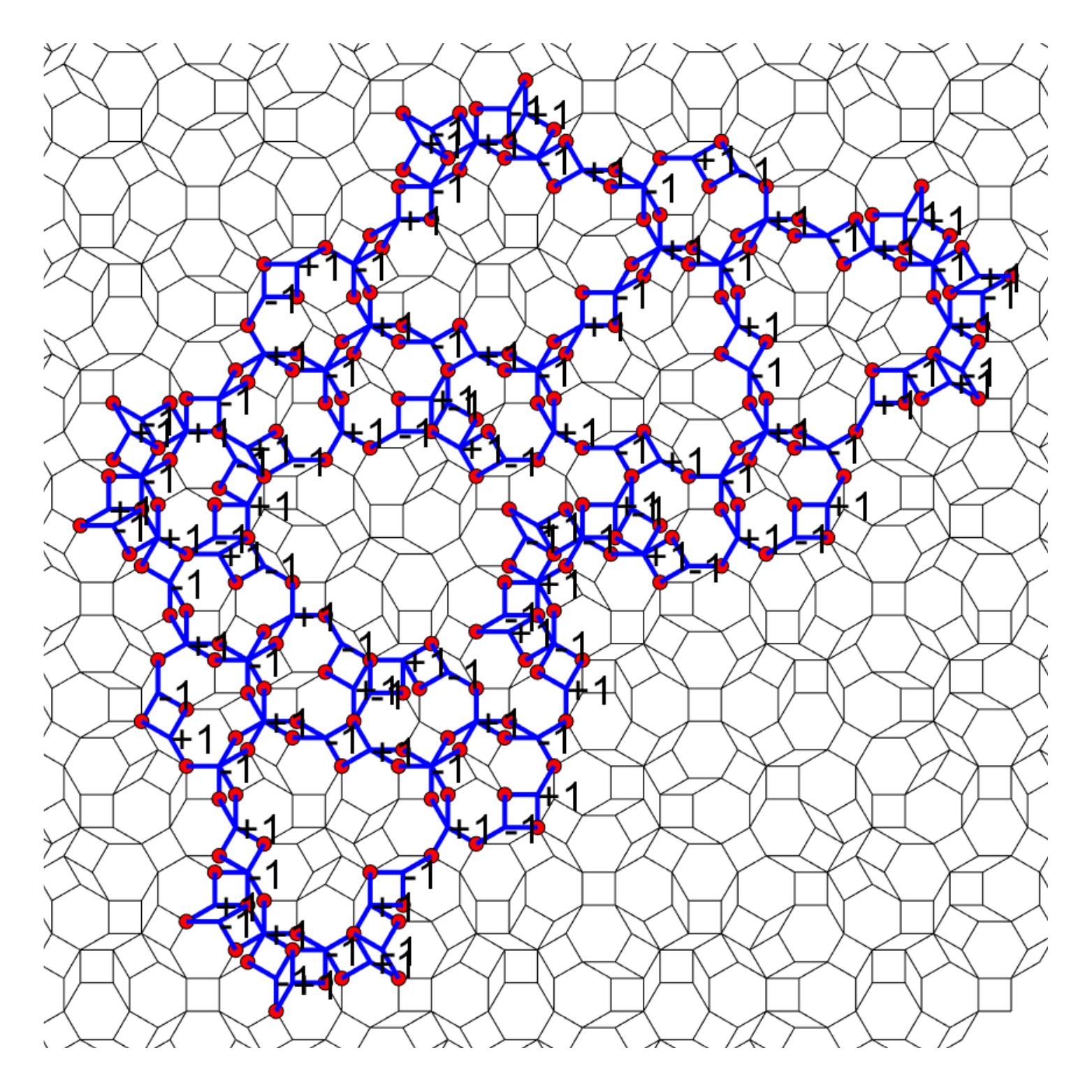}
    \includegraphics[trim=8mm 8mm 8mm 8mm,clip,width=0.48\textwidth]{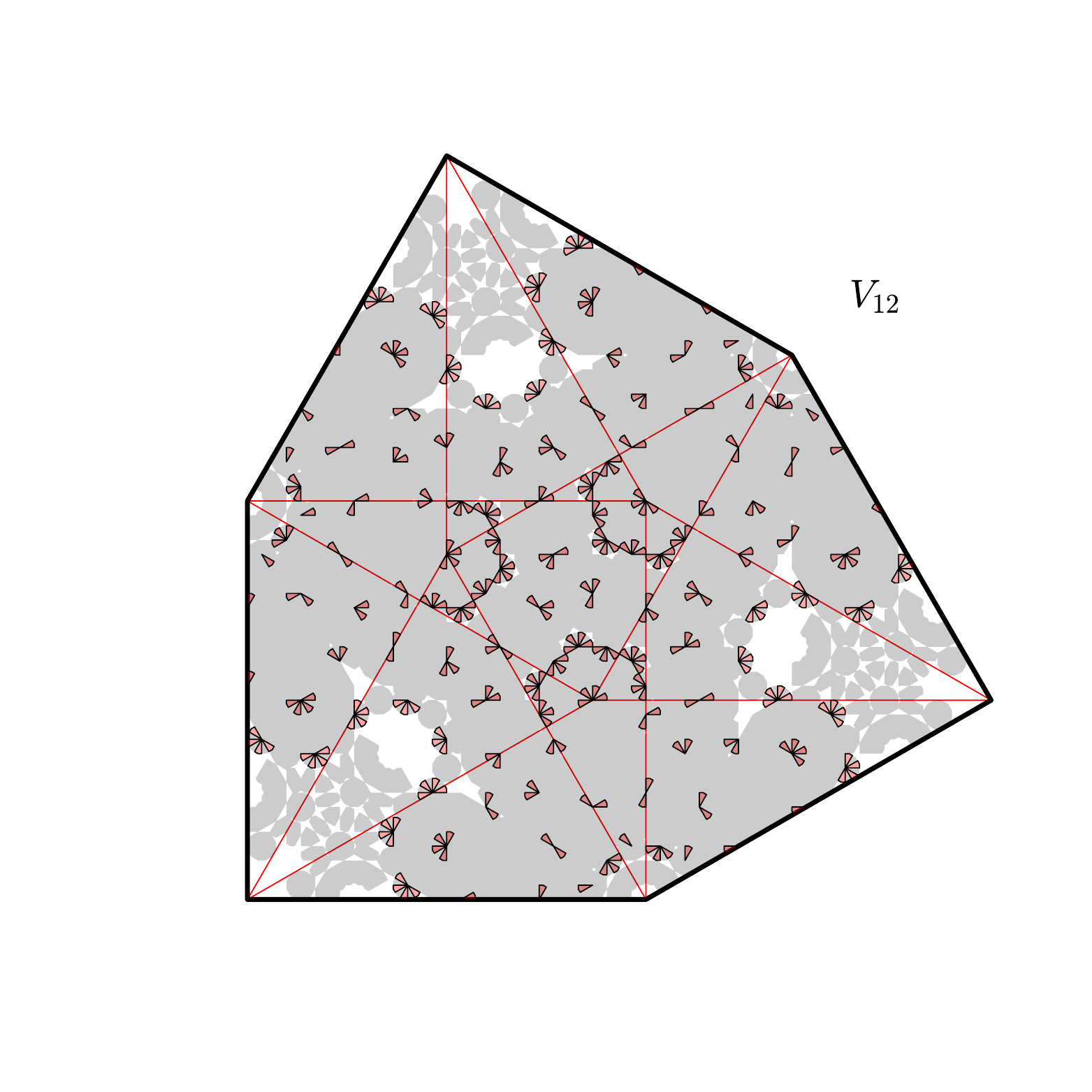}    
    \caption{LS type-C9 fills areas adjacent to type-C8, and has the same frequency as type-C6}
    \label{fig:TypeD5_RealSpace}
\end{figure}
\begin{figure}[!htb]
    \centering
    \includegraphics[trim=8mm 8mm 8mm 8mm,clip,width=0.48\textwidth]{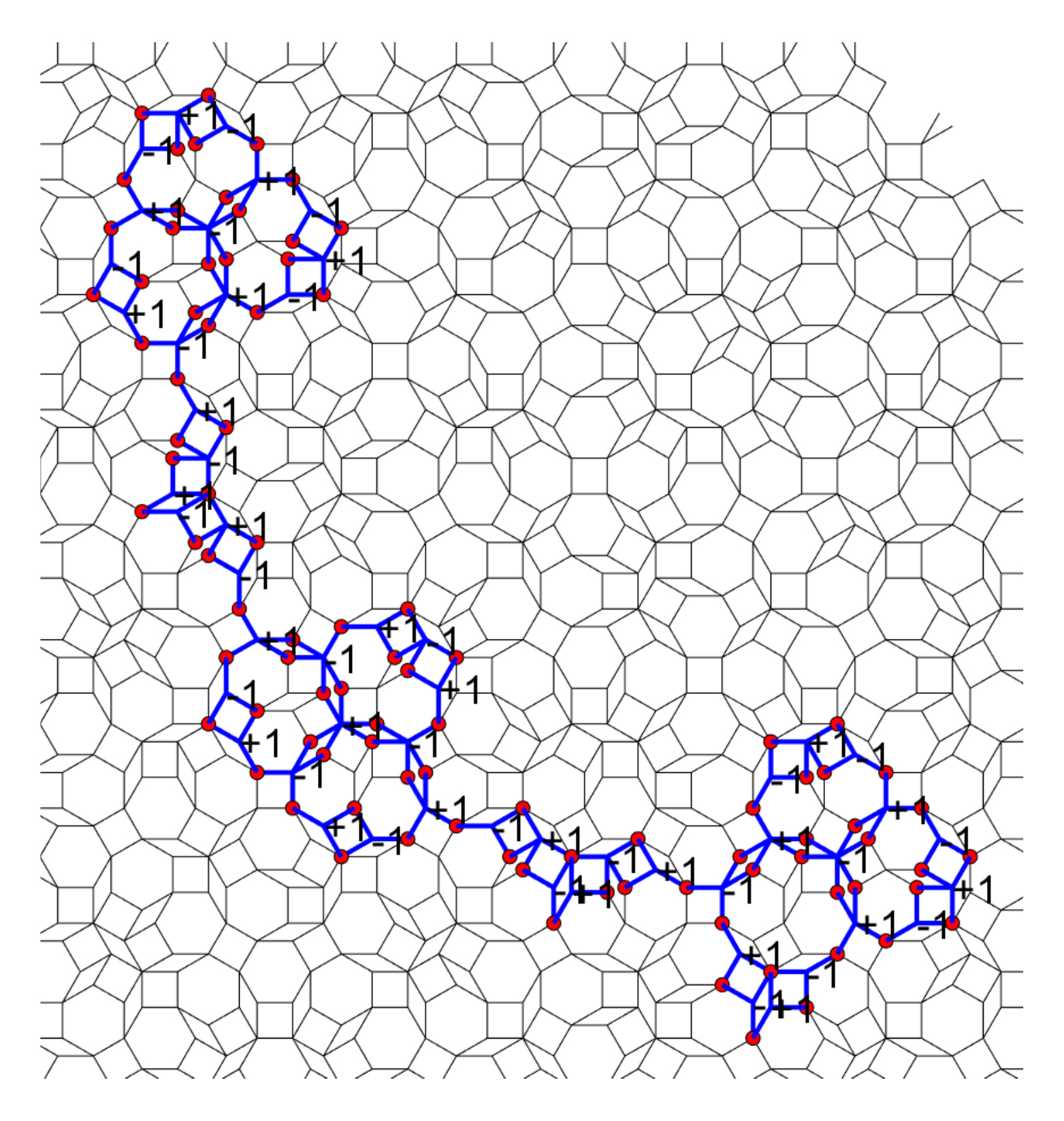}
    \includegraphics[trim=8mm 8mm 8mm 8mm,clip,width=0.48\textwidth]{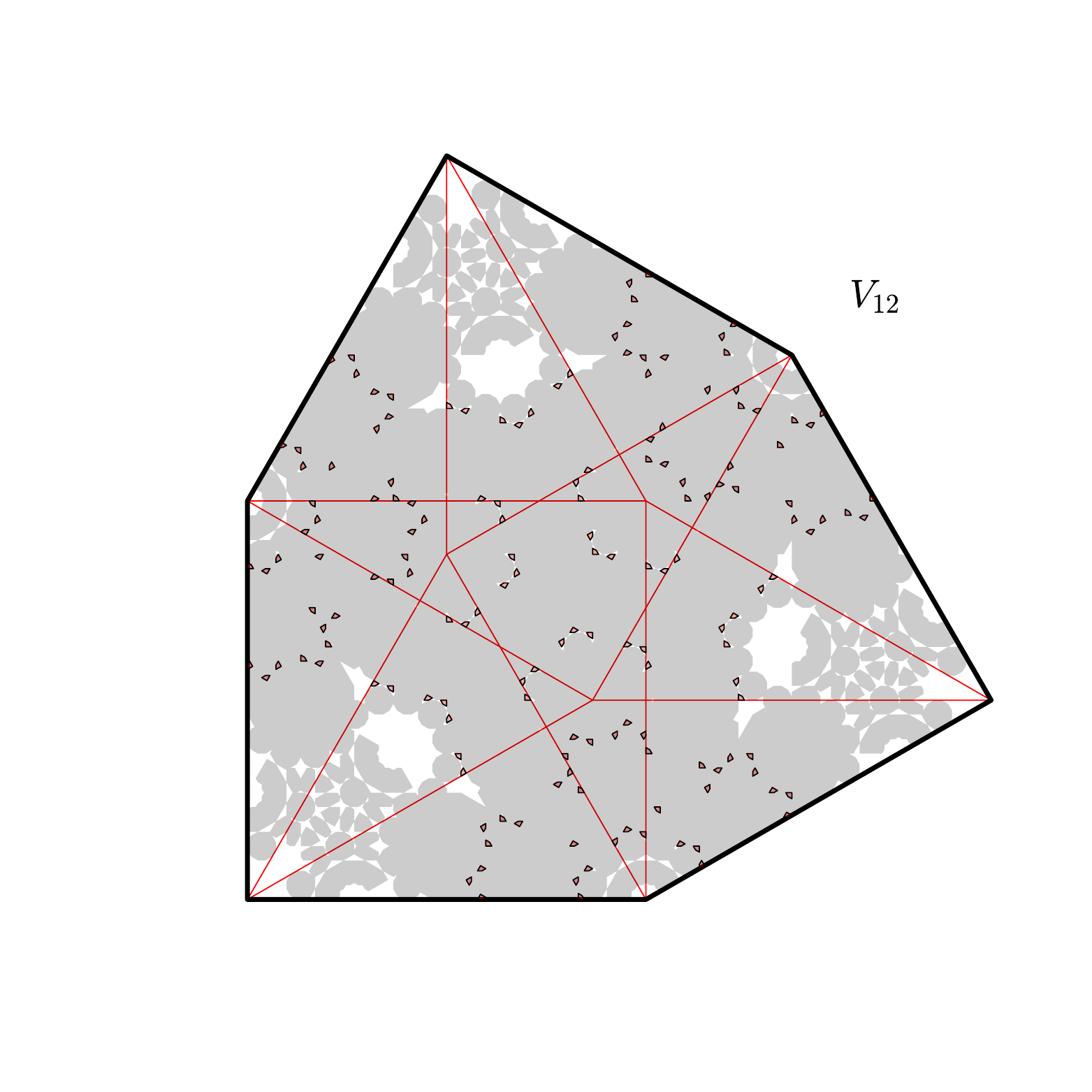}
    \caption{LS type-C10 has the lowest frequency of the LS types reported in this paper $f=\frac{\xi^{-6}}{2}\simeq0.00019$. There are further LS type which have lower frequencies which are not reported here.}
    \label{fig:TypeD6_RealSpace}
\end{figure}

\section{Forbidden sites with low frequency}

\begin{figure}[!htb]
    \centering
    \includegraphics[clip,width=0.43\textwidth]{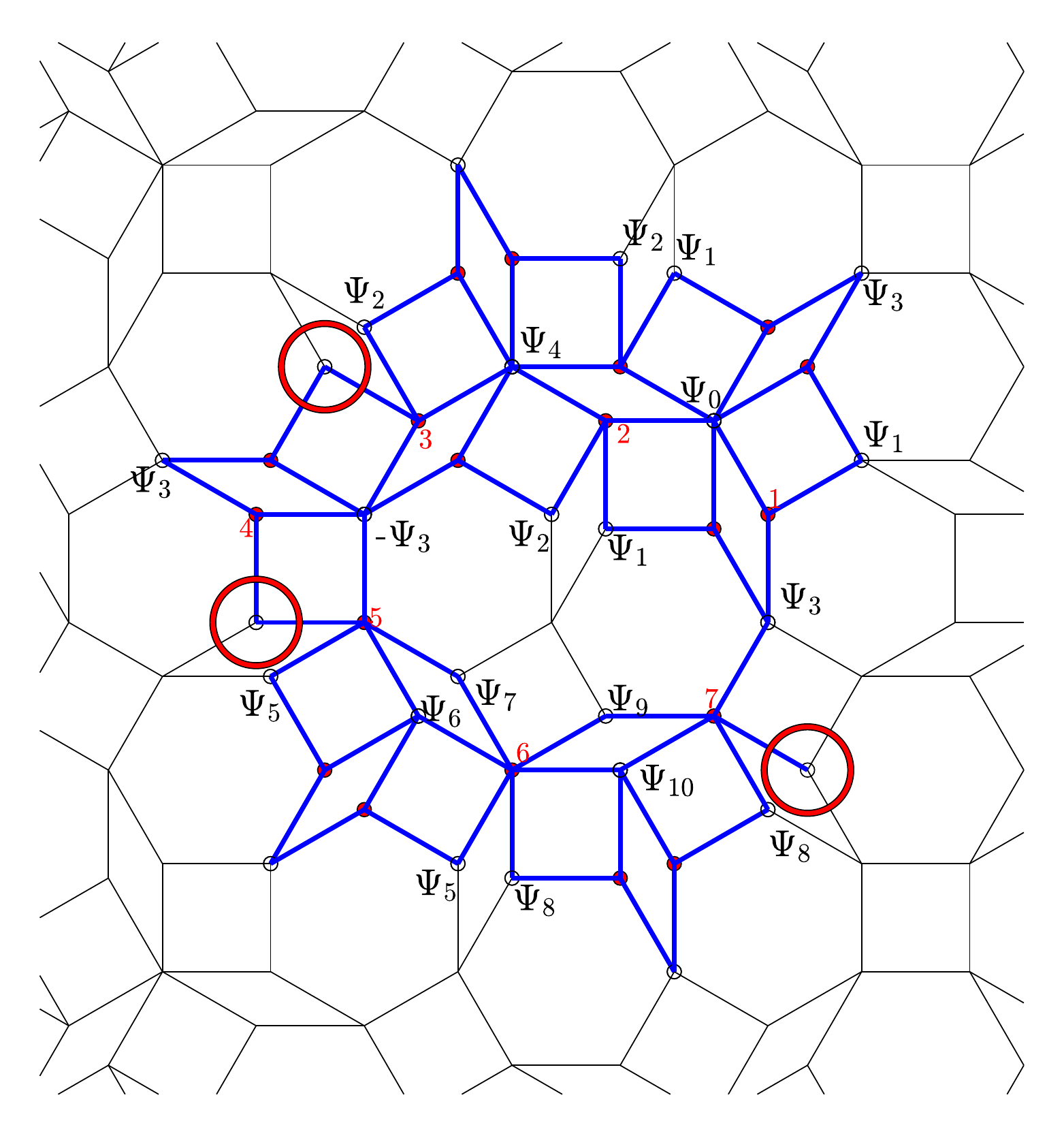}
    \includegraphics[clip,width=0.43\textwidth]{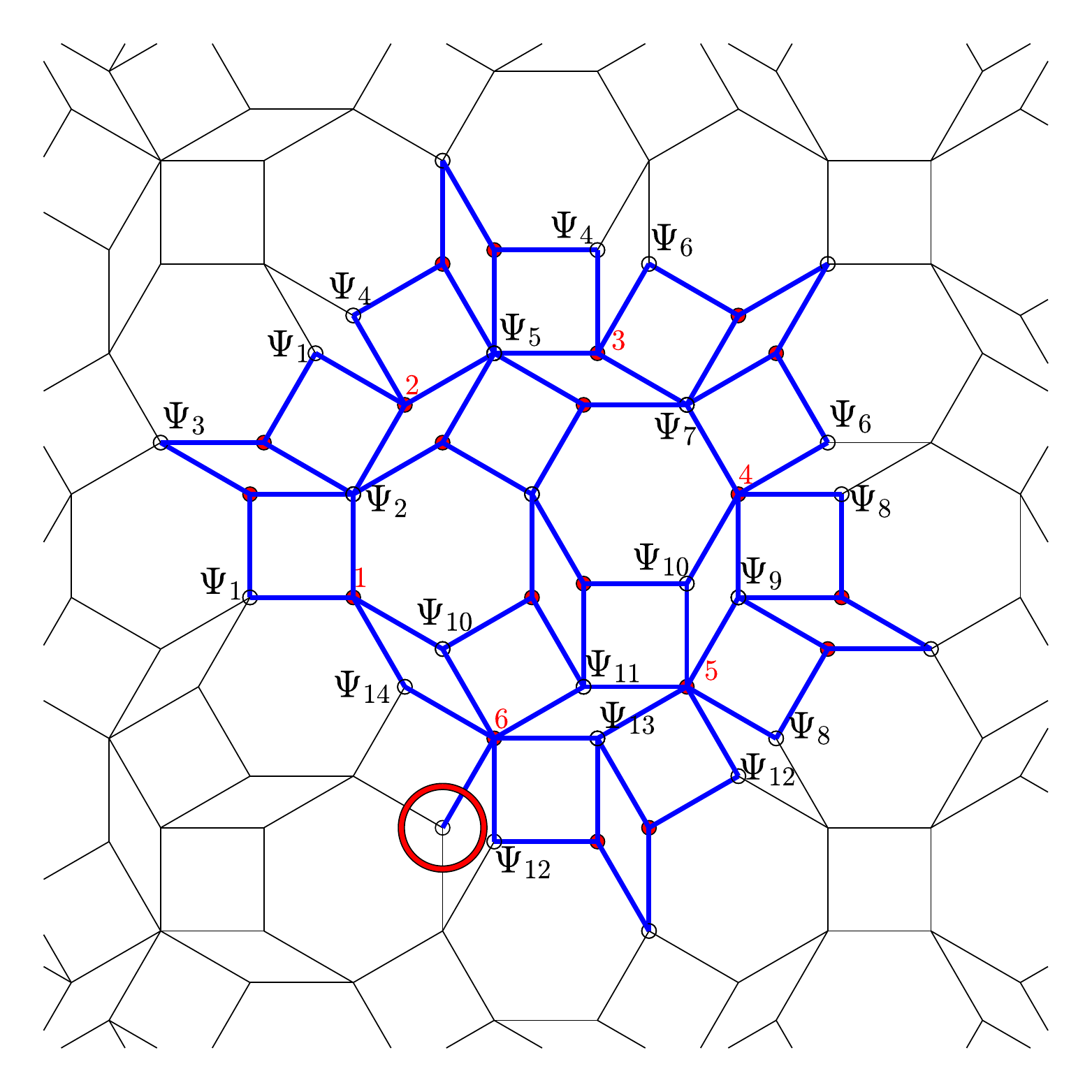}
    \caption{Local configurations which lead to new forbidden sites. The sites encircled in red cannot be in the support of an LS.}
    \label{fig:ForbiddenRealSpace3}
\end{figure}
The configurations in Fig.\ref{fig:ForbiddenRealSpace3} allow us to find new forbidden sites. We start  with the left configuration. Equations on site 1 and site 2 are $\Psi_0+\Psi_1+\Psi_3=0$ and $\Psi_0+\Psi_1+\Psi_2+\Psi_4=0$, respectively, and, here, we get $\Psi_2+\Psi_4=\Psi_3$. Thus, the equation on site 3 gives us the first forbidden site shown by the red circle for this configuration. The equation on site 4 gives the second forbidden site. We can easily get $\Psi_5+\Psi_6+\Psi_7=\Psi_3=-(\Psi_8+\Psi_9+\Psi_{10})$ by equations on sites 5 and 6. Finally, the equation on site 7 gives us the last forbidden site for this configuration.

For another forbidden site, consider the right configuration in  Fig.\ref{fig:ForbiddenRealSpace3}. Here equations on site 1, $\Psi_1+\Psi_2+\Psi_{10}+\Psi_{14}=0$, and on site 2, $\Psi_1+\Psi_2+\Psi_{4}+\Psi_{5}=0$, give us the following eaquaion; $\Psi_4+\Psi_5=\Psi_{10}+\Psi_{14}$. From equations on site 3, $\Psi_4+\Psi_5+\Psi_{6}+\Psi_{7}=0$, and on site 4, $\Psi_6+\Psi_7+\Psi_{8}+\Psi_{9}+\Psi_{10}=0$, we obtain following relations;$\Psi_1+\Psi_2=\Psi_{6}+\Psi_{7}=-(\Psi_8+\Psi_9+\Psi_{10})=-(\Psi_{10}+\Psi_{14})$. Equation on site 5, $\Psi_8+\Psi_9+\Psi_{10}+\Psi_{11}+\Psi_{12}+\Psi_{13}=0$, give us $-(\Psi_{10}+\Psi_{14})=\Psi_{11}+\Psi_{12}+\Psi_{13}$. Now if we consider equation on site 6, we find that the site shown by red circle is forbidden.


%

\end{document}